\documentclass{aa}  

\usepackage{graphicx}
\usepackage{pdflscape}
\usepackage{txfonts}
\usepackage{placeins} 
\usepackage{hyperref}
\hypersetup{colorlinks=true,citecolor=blue,linkcolor=black, urlcolor=black, linktoc=all,filecolor=black}
\newcommand{\cyc}[1]{c-C$_3$H$_2$} 
\newcommand{\meth}[1]{CH$_3$OH}
\newcommand{\prop}[1]{CH$_3$CCH}
\newcommand{\nht}[1]{$N_{\mathrm{H}_2}$}
\newcommand{\ngrad}[1]{$\nabla N_{\mathrm{H}_2}$}

\begin{document}

   \title{Chemical segregation analysed with unsupervised clustering}

   \author{K. Giers
          \inst{1},
          S. Spezzano\inst{1}, 
          Y. Lin\inst{1},
          T. Valdivia-Mena\inst{1,2},
          P. Caselli\inst{1},
          O. Sipil\"{a}\inst{1}
          }

   \institute{Max-Planck-Institute for Extraterrestrial Physics, Giessenbachstrasse 1, D-85748 Garching, Germany\\
              \email{kgiers@mpe.mpg.de}
         \and European Southern Observatory, Karl-Schwarzschild-Strasse 2, D-85748 Garching, Germany
             }

  \abstract 
   { Molecular emission is a powerful tool for studying the physical and chemical structures in cold and dense cores. The distribution and abundance of different molecular species provide information on the chemical composition and physical properties in these cores.}
   {We study the chemical segregation of three molecules -- \cyc\,, \meth\,, and \prop\, -- in the two starless cores B68 and L1521E, and the prestellar core L1544. }
   {We applied the density-based clustering algorithms DBSCAN and HDBSCAN to identify chemical and physical structures within these cores. To enable cross-core comparisons, the clustering input samples were characterised based on their physical environment, discarding the two-dimensional spatial information.
   }
   {Clustering analysis showed significant chemical differentiation across the cores. The clustering successfully reproduces the known molecular segregation of \cyc\, and \meth\, in all three cores. Furthermore, it identifies a segregation between \cyc\, and \prop\,, which is not apparent from the emission maps. Key features driving the clustering are integrated intensity, velocity offset, H$_2$ column density, and H$_2$ column density gradient. Different environmental conditions are reflected in the variations in the feature relevance across the cores.
   }
   {This study shows that density-based clustering provides valuable insights into chemical and physical structures of starless cores. It demonstrates that already small datasets covering only two or three molecules can yield meaningful results.
   In fact, this new approach revealed similarities in the clustering patterns of \meth\, and \prop\, relative to \cyc\,, suggesting that \cyc\, traces more outer layers or lower-density regions than to the other two molecules. 
   This allowed for insight into the \prop\, peak in L1544, which appears to trace a landing point of chemically fresh gas that is accreted to the core, highlighting the impact of accretion processes on molecular distributions. 
   }


\authorrunning{Giers et al.}

   \maketitle
%

\section{Introduction}

To understand star and planetary-system formation, it is crucial to understand the physics and chemistry in star-forming regions. Particularly important are starless dense cores, as they represent the earliest stages of star formation and set its initial conditions. 
In starless cores, the chemistry and physics can be studied without the complications caused by protostellar feedback.
Molecular line emission is a powerful tool for the study of the structure of these dense cores. The distribution and abundance of different molecular species provide information on their chemical composition and physical properties \citep[e.g.][]{Crapsi2007,Redaelli2021,Lin2022}. 
In addition, the line emission can help recover the chemical evolution of molecules. 
This has been shown, for example, with the inheritance of water \citep{Cleeves2014} and methanol \citep{Drozdovskaya2021,Drozdovskaya2022} in the Solar System, which come from the prestellar phase.
Prestellar cores are a subset of starless cores that are gravitationally bound and on the verge of star formation \citep[e.g.][]{Andre2000,KetoCaselli2008}. This makes prestellar cores dynamically evolved, with higher central densities and more pronounced temperature and velocity gradients compared to unbound starless cores \citep[e.g. see][]{Crapsi2007}.

A well-studied example is the prestellar core L1544 in the Taurus Molecular Cloud. Its molecular emission maps have led to the understanding of its evolutionary status, which is close to gravitational collapse \citep{Williams1999,Ohashi1999}, as well as its physical structure \citep[volume density, velocity, and the dust and gas temperature profiles; e.g. ][]{Crapsi2007,KetoCaselli2008,Keto2015,ChaconTanarro2019}.
\cite{Spezzano2016} observed a striking chemical differentiation between the carbon-bearing molecules \cyc\, and \meth\, in L1544, driven by differences in the external illumination onto the core. This chemical segregation has also been observed in other starless cores, linking it to their environments \citep{Spezzano2020}. In \cite{Spezzano2017}, the authors identified four molecular families in L1544, classified by the location of their emission peaks (carbon-chain peak, \meth\, peak, dust emission peak, and HNCO peak). Using principal component analysis, they found correlations between the different families and with the physical properties of the core. A similar analysis of the starless core L1521E by \cite{Nagy2019} also reported chemical differentiation between the \cyc\,, the \meth\,, and the dust emission peak.

In the big data era of astronomy, statistical methods are essential to analysing and interpreting the vast amounts of observational and simulated data. The rapid advancements in machine learning provide novel approaches to study the molecular complexity during the early stages of star formation. 
Unsupervised learning algorithms (clustering techniques in particular) are increasingly applied to identify hidden patterns, structures, and relationships in multidimensional datasets (e.g. see review by \citealt{Fotopoulou2024}). 
By grouping together data points based on similarities in various physical and chemical parameters, these methods help visualise subtle trends that might not be detected otherwise.

In astrochemistry, clustering methods have primarily been applied to large-scale surveys, for instance, to identify molecular clouds in an unbiased and systematic way \citep[e.g.][]{Colombo2015,Bron2018,Yan2022}. 
By isolating distinct structures in position-position-velocity space, these methods segment clouds into regions with similar physical or chemical properties. 
Furthermore, \cite{ValdiviaMena2023} have connected filament scales ($<$0.1\,pc) with envelope scales ($>$100\,au) by identifying (velocity-) coherent structures of inflowing material, known as streamers, through the clustering of molecular emission. 
Meanwhile, \cite{Okoda2020,Okoda2021} applied principal component analysis to molecular line emission, characterising velocity structures and distinct features surrounding a protostar.
Additionally, this method has also been used to disentangle overlapping kinematic components and to understand spectral variations \citep[e.g.][]{YunLee2023}, providing deeper insight into the dynamics of star-forming regions.

In this work, we apply the unsupervised clustering algorithms Density-Based Spatial Clustering of Applications with Noise (DBSCAN) and Hierarchical DBSCAN (HDBSCAN) to investigate the chemical segregation and differences between the molecular emission of \cyc\,, \meth\,, and \prop\, towards the starless cores B68 and L1521E and the prestellar core L1544.
The molecules were chosen as representatives of three of the molecular families found in L1544 \citep[see][]{Spezzano2017}. 
Our goal is to study the chemical segregation previously observed for \cyc\, and \meth\, from a different perspective using clustering techniques. 
Additionally, we aim to investigate the less understood differentiation between the two carbon chains, \cyc\, and \prop\,.
In our analysis, we take a novel approach by dropping the two-dimensional spatial information and instead characterising each pixel of the emission maps based on its physical parameters.
By concentrating on the core scale and incorporating both starless and prestellar cores, we study the influence of different evolutionary stages on the molecular emission and distribution as well as the effect of different environments.

In Sect.~\ref{sec:data}, we describe the data and the sources analysed in this work. In Sect.~\ref{sec:methodology}, we explain the preprocessing we applied to our dataset and what methods we used for the unsupervised clustering with DBSCAN and HDBSCAN.
The results of the dataset analysis and the clustering are presented in Sect.~\ref{sec:results}. 
We discuss the results and their implications in Sect.~\ref{sec:discussion} and present our conclusions in Sect.~\ref{sec:conclusion}.

\section{Observations and data reduction}\label{sec:data}

\subsection{Data}
The data presented in this work were taken with the IRAM 30\,m single-dish radio telescope on Pico Veleta in the Sierra Nevada, Spain. 
The observations were carried out between October 2013 and April 2018 (PIs: Silvia Spezzano, Zofia Nagy).
The data have also been used in \cite{Spezzano2017}, \cite{Nagy2019}, and \cite{Spezzano2020}.
The on-the-fly (OTF) maps were observed in position switching mode, using the EMIR E090 receiver and the Fourier transform spectrometer (FTS) backend with a spectral resolution of 50\,kHz.
The map sizes, sources and coordinates are listed in Table~\ref{Tab:SourceSample}.
The observed transitions are summarised in Table~\ref{Tab:ObservedLines}.

The data processing was done using the GILDAS software \citep{Pety2005} and the python packages \texttt{pandas} \citep{pandas} and \texttt{spectral-cube} \citep{Spectralcube}.
All emission maps were gridded to a pixel size of 8" with the CLASS software in the GILDAS packages; this corresponds to one-third to one-quarter of the actual beam size, depending on the frequency. To create a uniform dataset, we additionally resampled the data to a spectral resolution of 0.18\,km\,s$^{-1}$, corresponding to the resolution of the lowest frequency observation (82\,GHz).
The antenna temperature $T_A^*$ was converted to the main beam temperature $T_\mathrm{mb}$ using the relation $T_\mathrm{mb}=F_\mathrm{eff}/B_\mathrm{eff}\cdot T_A^*$. The corresponding values for the 30\,m forward ($F_\mathrm{eff}$) and main-beam efficiencies ($B_\mathrm{eff}$) are given in Table~\ref{Tab:ObservedLines}.

\begin{table*}[h]
    \centering
    \caption{Source sample.}
    \begin{tabular}{l c l c c c c}
    \noalign{\smallskip}
    \hline\hline 
    \noalign{\smallskip}
    & Right ascension & Declination & $V_\mathrm{sys}$\tablefootmark{b} & Distance\tablefootmark{c} & H$_2$ column density\tablefootmark{d} & map size \\
    & (J2000) & (J2000) & (km\,s$^{-1}$) & (pc) & ($10^{22}$\,cm$^{-2}$) & ($'\times'$) \\
    \noalign{\smallskip}
    \hline
    \noalign{\smallskip}
    B68 (s)\tablefootmark{a}   & 17:22:38.9 & -23:49:46.0   & 3.4 & 150 & 1.6 & $3.3\times2.5$ \\
    L1521E (s) & 04:29:15.7 & +26:14:05.0 & 6.9 & 140 & 2.3 & $2.5\times2.5$ \\
    L1544 (p)  & 05:04:17.2 & +25:10:42.8 & 7.2 & 170 & 2.8 & $2.5\times2.5$ \\
    \noalign{\smallskip}
    \hline
    \end{tabular}
    \label{Tab:SourceSample}
    \tablefoot{
    \tablefoottext{a}{Starless (s) or prestellar (p), following the definition given in \cite{Crapsi2005}.}
    \tablefoottext{b}{B68: \cite{Spezzano2020}; L1521E: this work; L1544: \cite{Caselli2002a}.}
    \tablefoottext{c}{B68: \cite{AlvesFranco2007}. L1521E: \cite{Galli2018}. L1544: \cite{Galli2019}.
    }
    \tablefoottext{d}{Values derived from \textit{Herschel}/SPIRE observations towards the dust peak \citep{Spezzano2016,Spezzano2020}.}
    }
\end{table*}

\begin{table*}[h]
    \centering
    \caption{Spectroscopic parameters of the observed lines. }
    \begin{tabular}{l c c c c c c c}
    \hline\hline 
    \noalign{\smallskip}
    Molecule & Transition & Frequency\tablefootmark{a} & $F_\mathrm{eff}/B_\mathrm{eff}$ & $E_\mathrm{up}$\tablefootmark{a} & $g_\mathrm{up}$\tablefootmark{a} & $A$\tablefootmark{a} & Reference \\
     &  & (MHz) &  & (K) &  & (s$^{-1}$) & \\
    \noalign{\smallskip}
    \hline
    \noalign{\smallskip}
    c-C$_3$H$_2$ & $2_{0,2}-1_{1,1}$ & 82093.544(1) & 0.95/0.81 & 6.4 & 5 & 1.89$\times10^{-5}$ & 1 \\
    c-C$_3$H$_2$ & $3_{2,2}-3_{1,3}$ & 84727.688(2) & 0.95/0.81 & 16.1 & 7 & 1.04$\times10^{-5}$ & 1 \\
    CH$_3$OH & $2_{1,2}-1_{1,1}\,(E_2)$ & 96739.358(2) & 0.95/0.80 & 12.5 & 20 & 2.56$\times10^{-6}$ & 2 \\
    CH$_3$OH & $2_{0,2}-1_{0,1}\,(A^+)$ & 96741.371(2) & 0.95/0.80 & 7.0 & 20 & 3.41$\times10^{-6}$ & 2 \\
    CH$_3$CCH & $5_1-4_1$    & 85455.6667(1) & 0.95/0.81 & 19.5 & 22 & 1.95$\times10^{-6}$ & 3 \\
    CH$_3$CCH & $5_0-4_0$    & 85457.3003(1) & 0.95/0.81 & 12.3 & 22 & 2.03$\times10^{-6}$ & 3 \\
    CH$_3$CCH & $6_{1}-5_{1}$  & 102546.0242(1) & 0.95/0.79 & 24.5 & 26 & 3.46$\times10^{-6}$ & 3 \\
    CH$_3$CCH & $6_{0}-5_{0}$  & 102547.9844(1) & 0.95/0.79 & 17.2 & 26 & 3.56$\times10^{-6}$ & 3 \\
    \noalign{\smallskip}
    \hline
    \end{tabular}
    \label{Tab:ObservedLines}
    \tablefoot{
    \tablefoottext{a}{Extracted from the Cologne Database for Molecular Spectroscopy \citep{Mueller2001}.}}
    \tablebib{
    (1) \cite{Thaddeus1985}; (2) \cite{XuLovas1997}; (3) \cite{Bauer1969}.
    }
\end{table*}

\subsection{Sources}\label{sec:data:sources}

The source sample is listed in Table~\ref{Tab:SourceSample}.
The pre-stellar core L1544 (following the definition of \citealt{Crapsi2005}) and the two starless cores (L1521E, B68) are located in different star-forming regions (Taurus, Ophiuchus) and therefore cover different evolutionary stages and environmental conditions.

L1521E and L1544 are located in the Taurus molecular cloud. Both cores are located at the edge of their filament, which exposes their southern sides to the local interstellar radiation field (ISRF). The higher illumination leads to an increase of C atoms in the gas phase and subsequently enhanced abundances of carbon chains such as \cyc\,, as discussed in \cite{Spezzano2020}.
L1521E was classified as a very young core because of its high abundances of carbon-chain molecules and low level of CO depletion \citep{Hirota2002,TafallaSantiago2004,Nagy2019}. 
The well-studied prestellar core L1544, on the other hand, is more evolved and shows signs of contraction, suggesting that it is on the verge of  star formation \citep{Williams1999,Ohashi1999,Lee2001,Caselli2002a,Caselli2012}. 
Within the central 1000\,au, L1544 exhibits an almost total (99.99\%) freeze-out of all species heavier than Helium \citep{Caselli2022}.

The isolated starless core B68 is located in the south of the Ophiuchus molecular cloud. 
It shows kinematic features of oscillation, indicating a stage prior to contraction \citep{Lada2003,KetoCaselli2008}. Due to it being a Bok globule (and therefore isolated), it can be assumed to be exposed to uniform external illumination.

\section{Methodology}\label{sec:methodology}

\subsection{Preprocessing}
The integrated intensity maps were generated by calculating the zeroth moment of the emission data cubes. 
We used the resulting map when the emission is extended over at least one telescope beam, as our focus in this work is primarily on the spatial distribution of the molecules. 
To facilitate comparisons between the molecules, all maps were convolved to an angular resolution of 32", corresponding to the half-power beam width (HPBW) of the largest observed beam.
The resulting integrated intensity maps of \cyc\,, \meth\, and \prop\, are shown in Figs.~\ref{fig:intmapsC3H2CH3OH} and \ref{fig:intmapsCH3CCH}.

To compare the distribution of molecular emission across three different cores with varying map sizes and distances, we treated the individual pixels of each map as input samples, and discarded their two-dimensional spatial information. 
Instead, we described the environment of each pixel using the projected distance to the dust peak, the H$_2$ column density, and the H$_2$ column density gradient at that position. 
Additionally, we used the Gaussian fit parameters -- velocity and linewidth -- obtained from fitting each pixel's spectrum with a one-dimensional Gaussian profile. 
This approach is valid as all cores in our sample display only one velocity component along the line of sight. 
For the analysis, we selected only the fits with a signal-to-noise ratio greater than or equal to three and an error below 70\%. 

We used six input features in the clustering dataset.
The following features characterise one input sample or emission pixel:

\paragraph{Integrated intensity:} The intensity was integrated over a range of $\pm$0.7\,km\,s$^{-1}$ around the centroid velocity. 
While the expected linewidth for optically thin emission in our cores is around 0.5\,km\,s$^{-1}$, we used a wider interval here to account for line shifts due the velocity gradients across the cores. For the clustering, we selected only pixels with a signal-to-noise ratio of at least three.
Each integrated intensity map was then normalised using the \texttt{MinMaxScaler} from the \texttt{scikit-learn preprocessing} package, scaling the values from zero to one before being added to the clustering dataset. This standardisation removes information about the absolute brightness of the molecular emission and the intensity ratios between different molecules, focusing instead on the distribution of emission across the core. In the analysis, this feature is referred to as `intensity'.

\paragraph{Velocity offset with respect to source's $V_\mathrm{LSR}$:} Here, we applied a selection criterium where the uncertainty in the fitted velocity position, $V_\mathrm{LSR}$, has to be smaller than $0.08\,$km\,s$^{-1}$ (which is channel width/2.355). To calculate the relative velocity offset of the emission line, we subtracted the systemic velocity from the velocity position: $V_\mathrm{offset}= V_\mathrm{LSR} - V_\mathrm{sys}$. The systemic velocities of the cores are listed in Table~\ref{Tab:SourceSample}. In the analysis, this feature is referred to as $V_\mathrm{offset}$.

\paragraph{Linewidth:} The minimum linewidth was set to be the spectral resolution of the spectra (0.18\,km\,s$^{-1}$). In the analysis, this feature is referred to as `linewidth'.

\paragraph{H$_2$ column density:} We used the H$_2$ column density maps derived from $Herschel$ SPIRE maps \citep{Spezzano2016,Spezzano2020}, which are shown in Fig.~\ref{fig:NH2maps}. Similar to the integrated intensity maps, each column density map was normalised individually before being added to the clustering dataset. In the analysis, this feature is referred to as \nht\,.

\paragraph{Distance to the dust continuum peak:} This feature was calculated as the distance between the equatorial coordinates of each pixel and the dust emission peak of the corresponding core. 
To be consistent across all cores, the dust emission peak was approximated by the emission peak in the H$_2$ column density map.
In the analysis, this feature is referred to as `dist2dust'.

\paragraph{H$_2$ column density gradient: } 
To calculate this feature, we convolved the H$_2$ column density maps with a Gaussian derivative kernel using a standard deviation of two telescope beams, as described in \cite{Soler2013}. Details of the derivation and the resulting gradients can be found in Sec.~\ref{sec:NH2gradient} and Appendix~\ref{app:NH2gradient}, where the gradient maps are presented in Fig.~\ref{fig:NH2maps}.
As with the integrated intensity and H$_2$ column density maps, each gradient map was normalised individually before being added to the clustering dataset. In the analysis, this feature is referred to as \ngrad\,.\\

The features $V_\mathrm{offset}$, linewidth, and dist2dust are normalised only when chosen for a specific sub-dataset (see Tables~\ref{Tab:FeatureCombinations} and \ref{tab:casestudylines}). This ensures that the feature values of all selected molecules are standardised together rather than individually, as is done for the emission.

\subsection{Clustering}
We used two density-based clustering algorithms in this analysis: DBSCAN \citep{DBSCAN} and HDBSCAN \citep{HDBSCAN}. 
For DBSCAN, we used the \texttt{scikit-learn} implementation \citep{scikit-learn} and for HDBSCAN the \texttt{hdbscan} Python package \citep{HDBSCANpython}.
Density-based clustering separates high-density areas from low-density areas by grouping points that are closer than a given distance threshold, which is defined by the hyperparameter \textsc{epsilon}. The minimum number of points required to form a cluster is set by the hyperparameter \textsc{min\_samples}; groupings smaller than this are considered as noise.
Unlike \textsc{k-means} clustering, which defines clusters as spherical, density-based clustering allows for clusters of arbitrary shape.
DBSCAN classifies points into three types: 
core points, which have at least \textsc{min\_samples} of neighbouring points within the \textsc{epsilon} distance; 
border points, which have less than \textsc{min\_samples} of neighbouring points, but are within \textsc{epsilon} distance to at least one core point; 
noise points, which have less than \textsc{min\_samples} of neighbouring points and are not within \textsc{epsilon} distance to any core point, so they are not part of any cluster.
HDBSCAN extends DBSCAN by converting it into a hierarchical clustering algorithm, allowing for clusters with varying densities.
Clusters are defined by the hyperparameter \textsc{min\_cluster\_size}, which sets the minimum number of points required for a group to be considered a cluster, and the hyperparameter \textsc{min\_samples}, which sets the minimum number of neighbouring points needed for a point to be considered a core point.

We used three features at a time as inputs to both DBSCAN and HDBSCAN, which helped to simplify the output interpretation.
The integrated intensity is kept as a fixed input feature to retain the information about the molecular emission, while the other two features are varied. 
The ten resulting feature combinations are listed in Table \ref{Tab:FeatureCombinations}.
To optimise the clustering, we perform a systematic grid search for the respective hyperparameters: \textsc{epsilon} (ranging from 0.05 to 0.155, in steps of 0.005) and \textsc{minsamples} (ranging from 10 to 100, in steps of 5) for DBSCAN; and \textsc{minclustersize} (ranging from 5 to 20) and \textsc{minsamples} (ranging from 1 to \textsc{minclustersize}) for HDBSCAN.

To determine the best tuning, we use the density-based clustering validation (DBCV) score \citep{DBCV}. This score is particularly well-suited to validating density-based clustering methods because it accounts for outliers and noise points (unlike cross-validation, for example). 
To calculate the score, a kernel density function first estimates the local density of data points around each object. Then, the cluster quality is evaluated by comparing the minimum density within clusters (which represents cluster cohesion) to the maximum density between clusters (which represents cluster separation).
We use the HDBSCAN function \texttt{validity\_index}, which is a fast approximation of the original DBCV score. Although this function is provided by the \texttt{hdbscan} package, it only requires the data points and the corresponding cluster labels as input, and therefore we also applied it to the DBSCAN results.
For HDBSCAN, we additionally used the intrinsic attribute \texttt{relative\_validity}, which is another fast approximation of the DBCV score. Since this intrinsic score gives slightly different results from the \texttt{validity\_index} function, we consider both scores in our evaluation. 
Finally, we select the best result based on the number of noise points and the relative sizes of the clusters.
However, since both the \texttt{validity\_index} function and the \texttt{relative\_validity} attribute are only approximations of the DBCV score, we treat them as relative measures and use them only to compare results across different hyperparameters choices, not across different datasets.

In addition to the DBCV score, we applied two postprocessing criteria to determine the optimal clustering results for each method, feature combination, and dataset:
(1) the number of clusters found by the algorithm has to be between two and five (inclusive), and (2) the found clusters must collectively cover at least 50\% of the data points.
Tests with more than five clusters showed that in this case, larger clusters are merely subdivided into smaller ones covering the same total amount of data and provide no additional insights. 
Conversely, when only one cluster is found, it typically contains over 90\% of the input data points, offering little value to understanding the data distribution.
Most datasets show two to four prominent trends that are effectively captured by the clustering. 
When the resulting clusters cover less than 50\% of the data, they fail to represent the overall trends and instead highlight minor sub-clusters while overlooking the majority of the data.

\begin{table}[h]
    \centering
    \caption{Feature combinations used in the clustering.}
    \begin{tabular}{ c| c |c| c}
    \hline\hline    
    \noalign{\smallskip}
    Combination & Feature 1 & Feature 2 & Feature 3 \\
    \noalign{\smallskip}
    \hline
    \noalign{\smallskip}
     1 & intensity & $V_\mathrm{offset}$ & linewidth \\
     2 & intensity & $V_\mathrm{offset}$ & dist2dust \\
     3 & intensity & $V_\mathrm{offset}$ & \nht\, \\
     4 & intensity & $V_\mathrm{offset}$ & \ngrad\, \\
     5 & intensity & linewidth & dist2dust \\
     6 & intensity & linewidth & \nht\, \\
     7 & intensity & linewidth & \ngrad\, \\
     8 & intensity & dist2dust & \ngrad\, \\
     9 & intensity & dist2dust & \nht\, \\
     10 & intensity & \nht\, & \ngrad\, \\
    \noalign{\smallskip} 
    \hline
    \end{tabular}
    \label{Tab:FeatureCombinations}
\end{table}

\subsection{Case studies}

In addition to \cyc\,, \meth\,, and \prop\,, the dataset used for this study covers molecules such as CCS, HC$_3$N, HC$^{18}$O$^+$, C$_4$H, HNCO, and CS, with between 8 to 20 detected molecules per core. 
However, to enhance the interpretability and extract meaningful patterns, we focused our analysis on the three key molecules \cyc\,, \meth\,, and \prop\,. This targeted approach allowed for a clear understanding of the relationships between these molecules, and helped to maintain clarity in the complex clustering output. 
During the clustering process, the features of the molecular maps were combined without giving the algorithm any prior information about which data point corresponds to each molecule.

We used the following four datasets as input to the clustering:

\paragraph{Case 1:} \cyc\, vs \meth\, \\
The two molecules show a well-known and well-studied segregation in the sources of our dataset \citep{Spezzano2016,Spezzano2020}, which is driven by environmental effects.
Through clustering, we analyse how this chemical differentiation is represented in the six features and how it influences the clusters identified by the algorithm.
This approach allowed us to assess the effectiveness of the clustering technique.
The transitions used in each core are listed in Table \ref{tab:casestudylines}, along with the initial ratio between the data points of each molecule in this sub-dataset.

\paragraph{Case 2:} \cyc\, vs \prop\,\\
These two molecules display a chemical differentiation in the prestellar core L1544 that is not yet understood \citep{Spezzano2017}. While \prop\, is a carbon chain such as \cyc\, and is therefore expected to peak in the carbon-chain rich south-east of the core, similar to \cyc\,, it instead peaks in the north-west of the core.
To ensure a balanced number of data points between the two molecules, we use two transitions of \prop\, for B68, and three transitions for L1521E, as shown in Table \ref{tab:casestudylines}.

\paragraph{Case 3:} \meth{} vs \prop{} \\
The two molecules show a chemical differentiation in the prestellar core L1544 \citep{Spezzano2017}. Both peak in the northern part of the core, \meth{} in the north-east, and \prop{} in the north-west. 
We use this combination to rule out biases that might arise from a clustering with \cyc{}.

\paragraph{Case 4:} \cyc\, vs \meth\, vs \prop\, \\
We use the combination of all three molecules to validate the results from the other two case studies.
This approach also helps to eliminate biases in the algorithm that might arise from comparing only two molecules.
The dataset for each core contains the combined data of Case 1 and Case 2 (see Table \ref{tab:casestudylines}).

\begin{table}
    \centering
    \caption{Molecular transitions used in Case 1, Case 2, and Case 3.}
    \begin{tabular}{l|ccc} 
        \hline\hline
    \noalign{\smallskip}
    \multicolumn{4}{c}{Case 1} \\
    \noalign{\smallskip}
    \hline
    \noalign{\smallskip}
         & \cyc\, & \meth\, & ratio\\
        B68 & $2_{02}-1_{11}$ & $2_{02}-1_{01}$ & 50/50\\
        L1521E & $2_{02}-1_{11}$ & $2_{02}-1_{01}$ & 45/55 \\
        L1544 & $3_{2,2}-3_{1,3}$ & $2_{02}-1_{01}$ & 42/58 \\
    \noalign{\smallskip}
    \hline
    \noalign{\smallskip}
        \multicolumn{4}{c}{Case 2} \\
    \noalign{\smallskip}
    \hline
    \noalign{\smallskip}
         & \cyc\, & \prop\, & ratio\\
        B68 & $2_{02}-1_{11}$ & $5_0-4_0$, $5_1-4_1$ & 57/43\\
        L1521E & $2_{02}-1_{11}$ & $5_0-4_0$, $5_1-4_1$, $6_0-5_0$ & 45/55 \\
        L1544 & $3_{2,2}-3_{1,3}$ & $6_1-5_1$ & 46/54 \\
        \noalign{\smallskip}
    \hline
    \noalign{\smallskip}
        \multicolumn{4}{c}{Case 3} \\
    \noalign{\smallskip}
    \hline
    \noalign{\smallskip}
         & \meth\, & \prop\, & ratio\\
        B68 & $2_{02}-1_{01}$ & $5_0-4_0$, $5_1-4_1$ & 57/43\\
        L1521E & $2_{02}-1_{01}$ & $5_0-4_0$, $5_1-4_1$, $6_0-5_0$ & 50/50 \\
        L1544 & $2_{02}-1_{01}$ & $6_1-5_1$ & 54/46 \\
    \noalign{\smallskip}
    \hline
    \end{tabular}
    \tablefoot{For Case 4, the transitions from Case 1 and Case 2 are combined for each core. Additionally, the ratio of data points for \cyc\, to \meth\, (Case 1), \cyc\, to \prop\, (Case 2), and \meth\, to \prop\, (Case 3) in each dataset is given as percentages.}
    \label{tab:casestudylines}
\end{table}

\section{Results}\label{sec:results}

\subsection{H$_2$ column density gradient}\label{sec:NH2gradient}

To derive the H$_2$ column density gradients for our three cores, we apply the method presented in \cite{Soler2013}. Therefore, the gradient is calculated by convolving the H$_2$ column density maps with a Gaussian derivative kernel in the x and y direction. To derive the total gradient, we combine the two directions. We use the method \texttt{gaussian\_filter} from the Python package \texttt{scipy.ndimage}. To depict the filament environments of the cores, we choose a Gaussian kernel with standard deviation equivalent to two telescope beams (=2x 32" or 2x 4 pixels).

The derived H$_2$ column density gradient maps are presented in Fig.~\ref{fig:NH2maps} in the Appendix, alongside the corresponding H$_2$ column density maps.
The observed gradients agree with the different levels of exposure to the ISRF of the different cores (see Sec.~\ref{sec:data:sources}).
The isolated starless core B68 has a mostly uniform N(H$_2$) gradient, forming a ring-like structure at the edges and a \ngrad\,$\approx0$ in the centre, representing a uniform external illumination.
For L1521E, the larger illumination towards the south is represented by an increased N(H$_2$) gradient along the south and lower values in the protected centre.
Similar to L1521E, the N(H$_2$) gradient of the prestellar core L1544 depicts how the south of the core is more exposed to the ISRF, resulting in larger values of \ngrad\,.
In contrast, the N(H$_2$) gradient is much lower in the more protected centre and north of the core, where both \meth\, and \prop\, peak.

\subsection{Comparison of dataset features} \label{sec:correlationplots}

\begin{figure*}[h]
    \centering
    \includegraphics[width=0.7\textwidth]{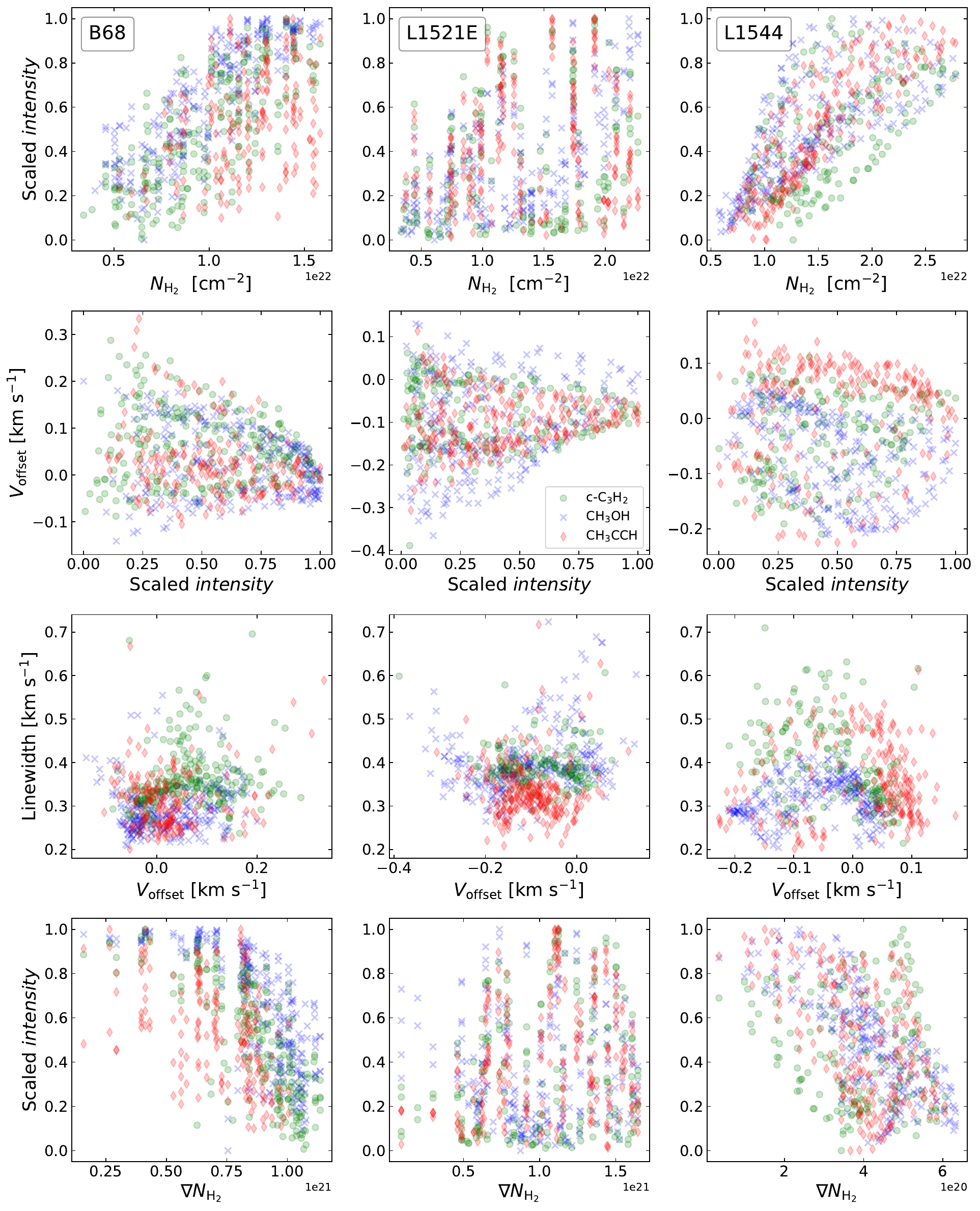}
    \caption[alt={Comparison of different features for B68, L1521E, and L1544. Compared are scaled intensity and \nht\,, $V_\mathrm{offset}$ and scaled intensity, linewidth and $V_\mathrm{offset}$, and scaled intensity and \ngrad\,.}]{Comparison of different features for B68 (\textit{left}), L1521E (\textit{middle}), and L1544 (\textit{right}), with \cyc\, given in green (circle), \meth\, in blue (cross), and \prop\, in red (diamond).}
    \label{fig:Correlationplots}
\end{figure*}

Figure~\ref{fig:Correlationplots} shows a comparison of selected features of the dataset, illustrating the data points for \cyc\, (green circles), \meth\, (blue crosses), and \prop\, (red diamonds) observed towards B68 (left), L1521E (centre), and L1544 (right).
The plots reveal distinct patterns and behaviours that vary depending on the core, molecule, and feature.

The distributions of intensity over \nht\, and intensity over $V_\mathrm{offset}$, shown in the top two rows of Fig.~\ref{fig:Correlationplots}, reflect the distribution of the molecular emission across the cores. 
In B68, all molecules exhibit similar distributions, with peak intensity at the highest H$_2$ column density. 
In L1521E, \cyc\, and \prop\, peak in the south-eastern part of the core at lower \nht\, compared to \meth\,, which peaks at the dust peak.
In contrast, in L1544, all molecules peak in different locations across the core with varying projected distances to the dust peak.
Velocity-wise, \meth\, in L1521E shows a slightly different behaviour compared to the carbon chains: at high intensity it extends to higher $V_\mathrm{offset}$, while at lower intensity it spreads to lower $V_\mathrm{offset}$. 
Additionally, in L1544, \prop\, spans a broader velocity range than \cyc\, and \meth\,, suggesting that it traces a different layer of the core.

The \cyc\, emission shows a wide range of linewidths, which are in general broader and reach higher values compared to the other molecules (see third row in Fig.~\ref{fig:Correlationplots}), likely tracing more turbulent material (compare e.g. \citealt{Lin2022} for L1544). 
In L1521E, the linewidths of \cyc\, have a more compact distribution around a value of 0.4\,km\,s$^{-1}$.
For \meth\,, the linewidths are typically smaller in B68 and L1544, between  0.25-0.30\,km\,s$^{-1}$ and 0.30-0.35\,km\,s$^{-1}$, respectively, while in L1521E, they are slightly higher, around 0.4\,km\,s$^{-1}$, with some values reaching up to 0.65\,km\,s$^{-1}$. 
The linewidths of \prop\, are more compactly distributed at lower values in the two starless cores (averaging around 0.3\,km\,s$^{-1}$), but extend up to 0.5\,km\,s$^{-1}$ in the prestellar core L1544, possibly indicating the presence of more turbulent material.

The \ngrad\,-intensity distribution also reflects the morphology of the molecular emission across the cores (see bottom row in Fig.~\ref{fig:Correlationplots}).
In L1521E, all molecules display a similar behaviour, with the highest intensities occurring at high, though not maximum, \ngrad\, values.
In contrast, in L1544, the \cyc\, intensity peaks at a much higher \ngrad\, than \meth\, and \prop\,. This illustrates the active photochemistry in the south of the core caused by the external illumination of the core, leading to an increased abundance of carbon-chains such as \cyc\,. 
This is supported by a local peak in \prop\, emission in the south and a sharp decline in \meth\, intensity at a higher \ngrad\,.
In B68, which has a more spherical shape and is uniformly illuminated, the molecular emissions peak at lower \ngrad\, values in the protected centre of the core. The emission then shows a rather steep decline at a higher \ngrad\, in the outer parts of the core. 
Additionally, \prop\, exhibits lower intensity values even at a lower \ngrad\,, because the emission map is less spatially extended across this core compared to \cyc\, and \meth\, (see Figs.~\ref{fig:intmapsC3H2CH3OH} and \ref{fig:intmapsCH3CCH}).

In summary, the plots point out the wealth of chemical and physical differences between the cores.
In L1544, all molecules are clearly separated and behave differently, which indicates a more profound chemical segregation in this core.
In contrast, in B68 and L1521E, the molecules show a more similar behaviour and are less segregated, which could be linked to their earlier evolutionary stages compared to L1544.
Consequently, an unbiased approach such as clustering can provide valuable insights into the varying chemical environments across these cores.

\subsection{Clustering}\label{subsec:clustering}

\begin{figure*}[h!]
    \includegraphics[width=0.49\textwidth]{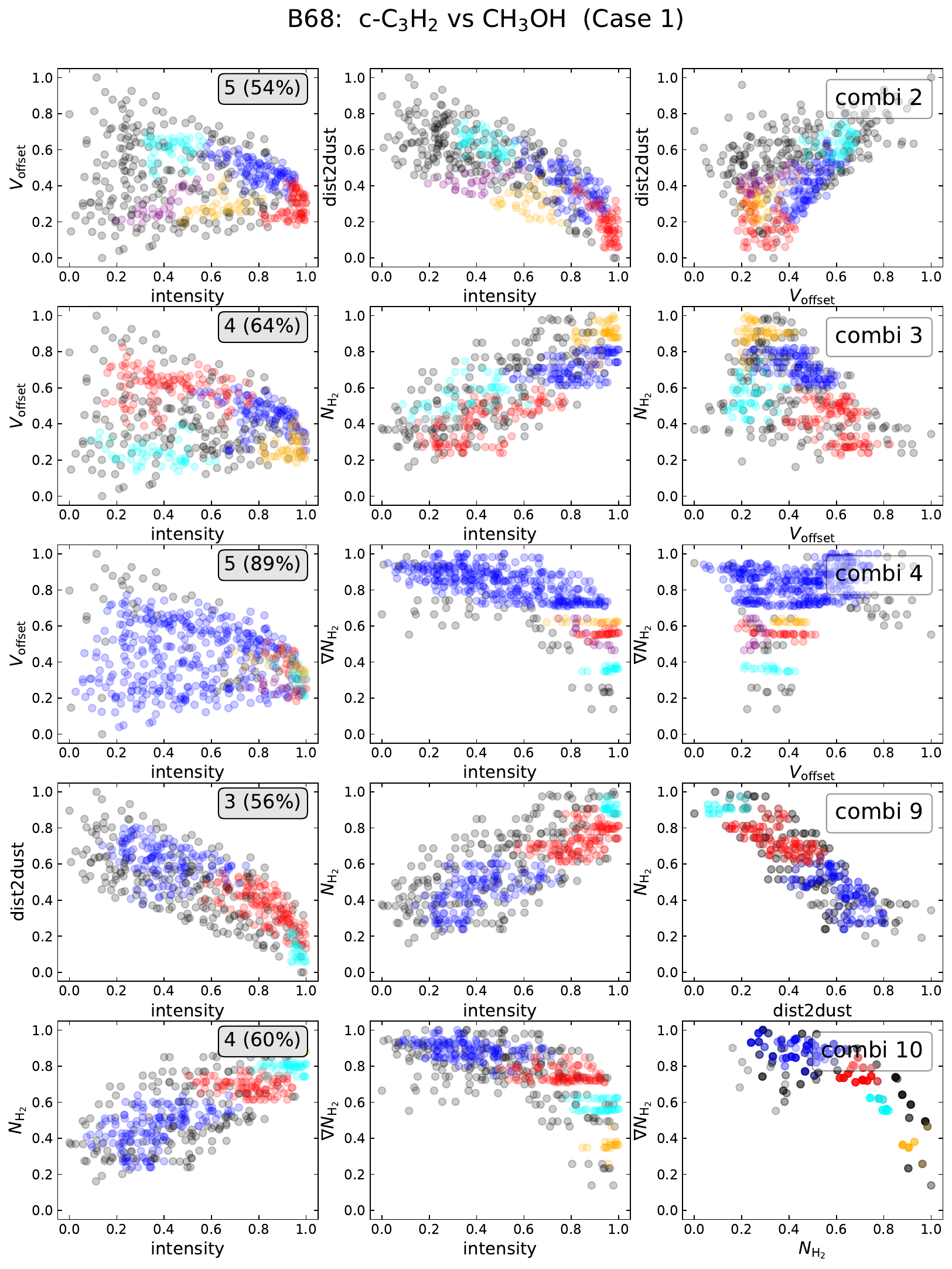}
    \includegraphics[width=0.49\textwidth]{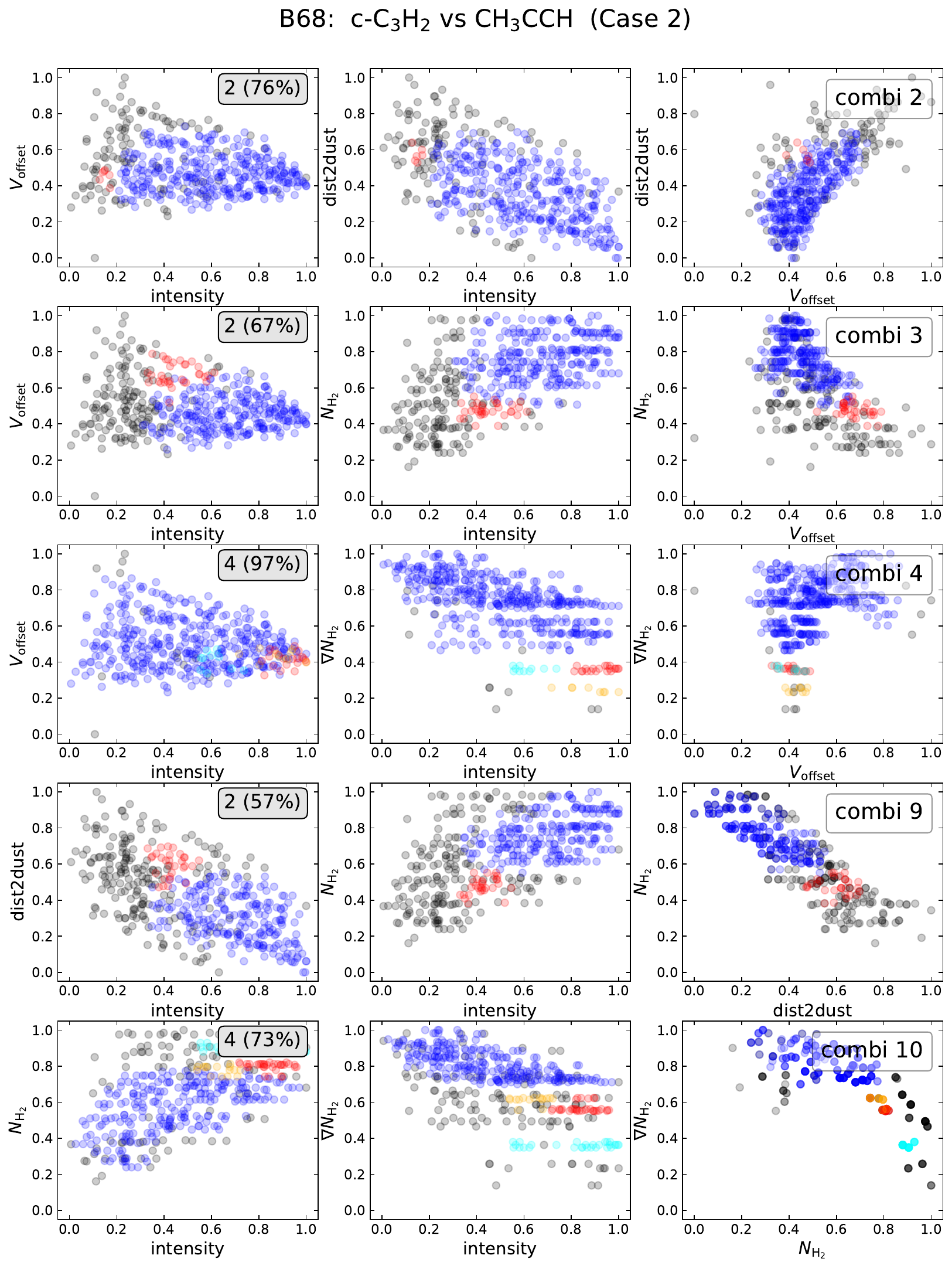}
    \includegraphics[width=0.47\textwidth]{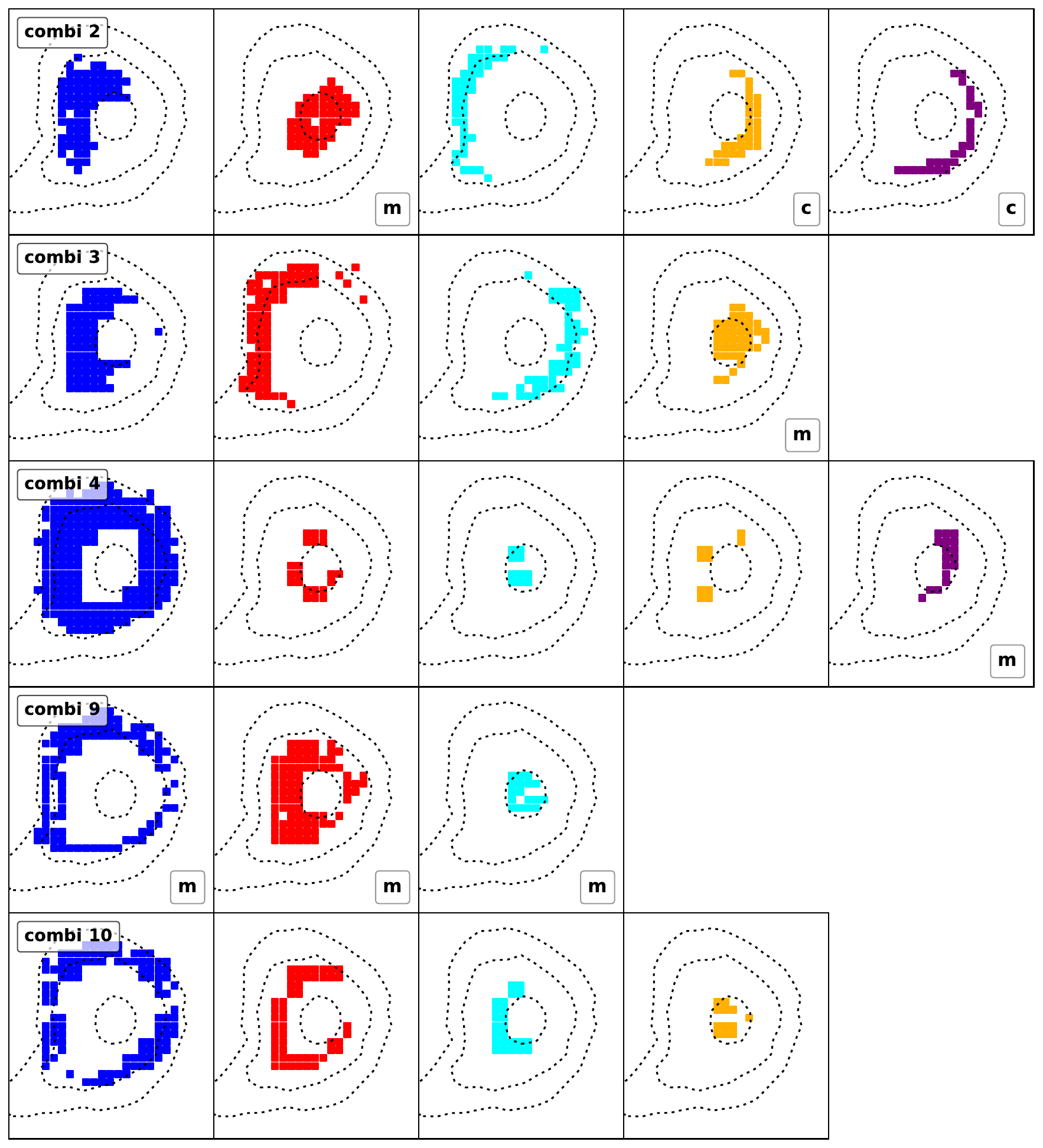}
    \includegraphics[width=0.38\textwidth]{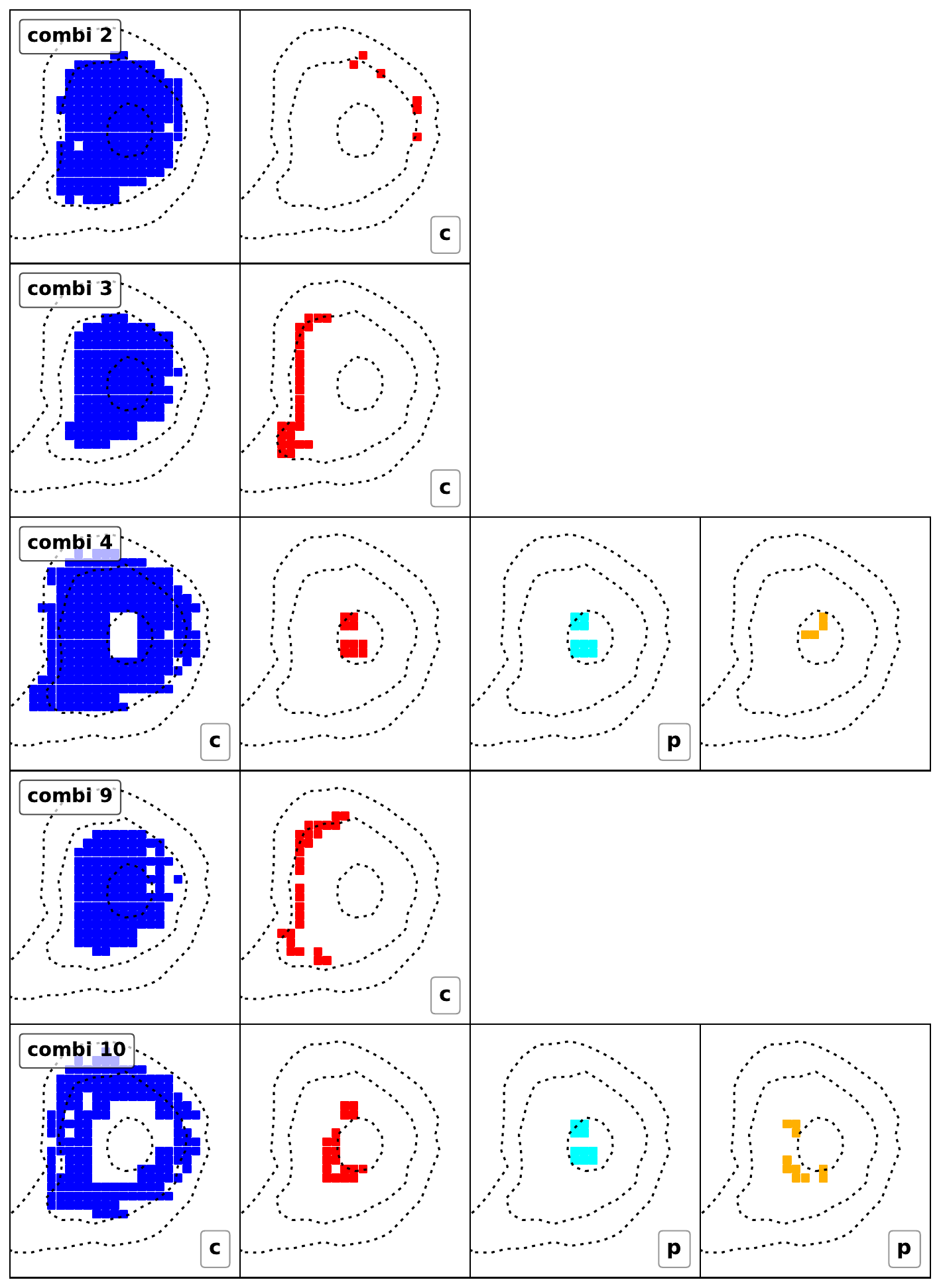}
    \caption[alt={Clustering results for the starless core B68 for the dataset of Case 1 and Case 2, for the feature combinations 2, 3, 4, 9, and 10. Each row represents a different combinaitons of features (see Table.~\ref{Tab:FeatureCombinations}). The top part shows the distribution of the resulting clusters in the input features. The bottom part shows the corresponding spatial distribution of each cluster across the core.}]{Clustering results for the starless core B68 for the dataset of Case 1 (\textit{left}) and the dataset of Case 2 (\textit{right}) for feature combinations 2, 3, 4, 9, and 10. Each row represents a different combination of features (see Table.~\ref{Tab:FeatureCombinations}): combination 2 (intensity, $V_\mathrm{offset}$, and dist2dust), combination 3 (intensity, $V_\mathrm{offset}$, and \nht\,), combination 4 (intensity, $V_\mathrm{offset}$, and \ngrad\,), combination 9 (intensity, dist2dust, and \nht\,), and combination 10 (intensity, \nht\,, and \ngrad\,).
    \textit{Top}: Distribution of the resulting clusters in the input features. The annotations provide information on how many clusters were found by the algorithm (two to five) and what percentage of data points are assigned to the clusters. Noise points (=points not assigned to any cluster) are plotted in black. The colours of the clusters are ordered by cluster size: the biggest cluster is given in blue, followed by red, cyan, yellow, and purple.
    \textit{Bottom}: Corresponding spatial distribution of each cluster across the core. The annotations indicate if a cluster contains more than $60\%$ of one molecule (c: \cyc\,; m: \meth\,; p: \prop\,). The dashed line contours represent 30\%, 50\%, 90\% of the H$_2$ column density peak derived from \textit{Herschel} maps \cite{Spezzano2020}.}
    \label{fig:ClusteringB68}
\end{figure*}

In Figs.~\ref{fig:ClusteringB68}, \ref{fig:ClusteringL1521E}, and \ref{fig:ClusteringL1544}, we show the clustering results for B68, L1521E, and L1544, respectively. They present the results for the feature combinations 2, 3, 4, 9, and 10 for Case 1 and Case 2 (for details see Table~\ref{Tab:FeatureCombinations}). 
The remaining results for Cases 1 and 2, together with the results for Cases 3 and 4 are published on Zenodo, in Figs.~\href{https://zenodo.org/records/15519030}{C.3-C.14}. 
The molecular ratio (i.e. the ratio of data points belonging to each of the molecules in a specific cluster) and the number of data points assigned to a cluster are given in Table~\ref{Tab:ClusterContent}.
In Sections~\ref{subsec:case1}-~\ref{subsec:case4}, we describe the results for each case individually.

Both the DBSCAN and the HDBSCAN results are included in the analysis. 
However, to improve readability, only one result is represented for each combination.
This was decided manually for each combination based on the amount of noise points and the number of clusters. 
This ensures that all scientific results discussed in this work are also presented visually.

To evaluate the possible chemical segregation within the clusters, we focused on imbalanced clusters, where the molecular ratio deviates from the initial ratio by at least 10\% (e.g. 37/63 instead of 47/53).
These imbalanced clusters are particularly insightful because they highlight regions where specific molecular abundances diverge from the average distribution, and they potentially indicate distinct physical or chemical processes at work.
The imbalanced clusters are marked in boldface in Table~\ref{Tab:ClusterContent}. 
Our primary focus is to extract meaningful chemical and scientific insights from the clustering patterns.
In fact, many clusters exhibit an excess of, or are dominated by, one molecule, indicating chemical differentiation across all three cases and all three cores in this study.
However, clusters that exclusively contain data points from a single molecule are rare and typically contain only a small number of points (N$\leq20$). Overall, we prioritised clusters that cover at least one telescope beam, which corresponds to a size of roughly 20 data points or pixels.

\subsubsection{Case 1: \cyc\, vs \meth\,}\label{subsec:case1}

In some results, the clusters vary greatly in size. 
The largest cluster typically maintains a balanced molecular ratio, similar to the input ratio (see Table~\ref{tab:casestudylines}), while the smaller clusters tend to show more variation (see Table~\ref{Tab:ClusterContent}). 
For L1521E and L1544, the input data has a slightly higher proportion of \meth\, compared to \cyc\, (see Table~\ref{tab:casestudylines}), causing the largest cluster to often show a slight excess in \meth\,. 
In the following section, we discuss the clustering results for each core individually.

\paragraph{B68:} 
Imbalanced clusters with an excess of \cyc\, or \meth\, are spatially separated into different regions of the core. 
\meth\,-dominated clusters are concentrated around the dust peak at the centre of the core, characterised by features such as high intensity and high \nht\, (e.g. red clusters in combs. 2, 4, and 7, or yellow cluster in comb. 3; see Figs.~\ref{fig:ClusteringB68} and \href{https://zenodo.org/records/15519030}{C.3}, and Table~\ref{Tab:ClusterContent}). 
In contrast, \cyc\,-dominated clusters are confined to the (south-)west region and associated with data points of moderate intensity (see yellow and purple clusters in comb. 2 in Fig.~\ref{fig:ClusteringB68}). 
In combination 9 and 10, \nht\, and \ngrad\, are structured in ring-like clusters around the dust peak (see Fig.~\ref{fig:ClusteringB68}), all dominated by \meth\,, and with no significant contribution from \cyc\,. 
Interestingly, in B68, the largest cluster of each combination (shown in blue) is slightly imbalanced towards \meth\, (on average 3\%, see Table~\ref{Tab:ClusterContent}), even though the initial molecular ratio is 50/50. The smaller clusters, however, display greater variation.

\paragraph{L1521E:} 
In this core, many features are clustered into separate structures dominated by either \meth\, or \cyc\,. This molecular segregation is evident across multiple combinations: 
The intensity-dist2dust distribution is clustered into two separate diagonals (see blue and red clusters in combs. 2 and 5 in Figs.~\ref{fig:ClusteringL1521E} and \href{https://zenodo.org/records/15519030}{C.4}). The lower intensity diagonal (blue) corresponds to regions in the north-west of the core and is imbalanced towards \cyc\, (see Table~\ref{Tab:ClusterContent}). 
Conversely, the upper intensity diagonal (red) is associated with the south-east of the core and is dominated by \meth\,.
The same cluster distribution, with similar molecular imbalance, appears in the intensity-\nht\, and the intensity-$V_\mathrm{offset}$ distributions: clusters dominated by \cyc\, (in the north-west) show high \nht\, and high  $V_\mathrm{offset}$ (see blue cluster in comb. 2, red cluster in comb. 3), while \meth\,-dominated clusters  (south-east) show low \nht\, and lower $V_\mathrm{offset}$ (see red clusters in comb. 2 and 9, and blue cluster in comb. 3).
Additionally, combination 2 (Fig.~\ref{fig:ClusteringL1521E}) shows that the \meth\,-dominated clusters appear as a narrow diagonal structure in dist2dust over $V_\mathrm{offset}$, while the \cyc\, clusters present a separate, more diffuse distribution.
To summarise, \cyc\, and \meth\, show a separation in \nht\, and $V_\mathrm{offset}$ in L1521E, and cluster into separate diagonals in the intensity-dist2dust distribution. 
Beyond the north-west/south-east separation, 
the $V_\mathrm{offset}$-\ngrad\, distribution is split into \cyc\, at high \ngrad\, and mid $V_\mathrm{offset}$ in the south of the core, and \meth\, at mid \ngrad\, and lower $V_\mathrm{offset}$, in the east (\cyc\,: blue cluster in comb. 4; \meth,: red cluster in comb. 4; see Figs.~\ref{fig:ClusteringL1521E} and \href{https://zenodo.org/records/15519030}{C.4}).
Additionally, \meth\, shows a cluster at the core centre, characterised by high intensity and high \nht\, (see yellow cluster in comb. 9 in Fig.~\ref{fig:ClusteringL1521E}).

\paragraph{L1544:} 
Similar to L1521E, many features in L1544 are clustered into separate structures ($V_\mathrm{offset}$ over intensity, dist2dust over intensity, \nht\, over $V_\mathrm{offset}$, dist2dust over $V_\mathrm{offset}$, linewidth over $V_\mathrm{offset}$, \ngrad\, over $V_\mathrm{offset}$, \nht\, over dist2dust), dividing the core into north/south, and on-centre/off-centre regions. 
However, unlike in L1521E, these structures do not consistently correspond to a specific molecular segregation. 
A separation of the two molecules is visible in $V_\mathrm{offset}$ and \ngrad\, in combination 4, where a \cyc\,-dominated cluster (red) is concentrated in the northern part of the core and a \meth\,-dominated cluster (blue) is found in the south, both with lower intensity (see Fig.~\ref{fig:ClusteringL1544}).
Additionally, \cyc\, clusters appear at the core centre with high intensity, high linewidth, high \nht\, and low \ngrad\, (see cyan cluster in comb. 3 in Fig.~\ref{fig:ClusteringL1544}, and red clusters in combs. 7 and 10 in Figs.~\ref{fig:ClusteringL1544} and \href{https://zenodo.org/records/15519030}{C.5}). 
\meth\,, on the other hand, forms a cluster on the \prop\, peak in the north-west, characterised by low $V_\mathrm{offset}$ and higher intensity (see red cluster in comb. 2 in Fig.~\ref{fig:ClusteringL1544}). It also appears across the northern part of the core with lower intensity and low $V_\mathrm{offset}$ (red cluster in comb. 3, see Fig.~\ref{fig:ClusteringL1544}). 
In combination 10, \cyc\, (red) is clustered on the dust peak, while \meth\, (blue) is off-peak, with separation visible in \nht\,, and \ngrad\, (see Fig.~\ref{fig:ClusteringL1544}). A similar split appears in \nht\, in combination 6 (see \href{https://zenodo.org/records/15519030}{C.5}), though both corresponding clusters (blue and red) are imbalanced towards \meth\, without a molecular separation.

\paragraph{Summary of Case 1:} 
All cores display cluster structures with a clear separation between \cyc\, and \meth\,. 
However, the molecules are not necessarily clustered on their respective peaks.
The molecular separation is primarily visible in the features intensity,  $V_\mathrm{offset}$, and \nht\,, and for L1521E and L1544 also in \ngrad\,.
In general, the clustering reveals recurring structures in several feature combinations, where one molecule dominates over the other.
Additionally, B68 and L1544 show structures in \nht\, and \ngrad\, that divide the core into on-centre and around-centre. These divisions, however, are not always linked to a molecular separation.

\FloatBarrier
\subsubsection{Case 2: \cyc\, vs \prop\,}\label{subsec:case2}

Similar to Case 1, \cyc\, is slightly underrepresented in the datasets for L1521E and L1544, resulting in clusters that are more imbalanced towards \prop\,. For B68, the opposite is true, resulting in a slight excess of \cyc\, in many clusters. 
In the following, we discuss the clustering results for each core individually:

\paragraph{B68:}
The clusters dominated by either \cyc\, or \prop\, are spatially separated, showing behaviour similar to the segregation of \cyc\, and \meth\, in Case 1. 
High intensity \prop\, is clustered at the core centre, where the molecule peaks (see red and cyan clusters in comb. 4, red cluster in comb. 8, and cyan and yellow clusters in comb. 10 in Figs.~\ref{fig:ClusteringB68} and \href{https://zenodo.org/records/15519030}{C.6}).
In contrast, \cyc\, is clustered off its emission peak, along the east side of the core, with lower intensity (see red clusters in combs. 1,3, and 9 in Fig.~\ref{fig:ClusteringB68}). 
Beyond intensity, the molecular separation is also evident in $V_\mathrm{offset}$ and \nht\,: \cyc\, is associated with high $V_\mathrm{offset}$ and low \nht\,, while \prop\, is associated with low $V_\mathrm{offset}$ and high \nht\,.
Interesting to note is that this \cyc\, cluster is shaped along high values of \ngrad\, (see Fig.~\ref{fig:NH2maps}), which is not included in the mentioned feature combinations (1, 3, 9).
In combination 10, a \cyc\,-dominated cluster forms a broad ring (blue), corresponding to high values of \ngrad\, (see also Fig.~\ref{fig:NH2maps}), surrounding \prop\,-dominated clusters in the core centre (cyan and yellow). This separation is visible in \ngrad\, and \nht\, (see Fig.~\ref{fig:ClusteringB68}).

\paragraph{L1521E:} 
Similar to Case 1, clusters dominated by \cyc\, and \prop\, are spatially separated, forming separate structures in various features (e.g. $V_\mathrm{offset}$ over intensity, dist2dust over intensity, dist2dust over $V_\mathrm{offset}$, \nht\, over intensity, \nht\, over linewidth).
As in Case 1, \cyc\, is clustered in the north and north-west, characterised by high \nht\,, high $V_\mathrm{offset}$, and appearing as lower diagonal in the intensity-dist2dust distribution (see red clusters in combs. 2, 3, and 6, and blue clusters in combs. 5 and 9 in Figs.~\ref{fig:ClusteringL1521E} and \href{https://zenodo.org/records/15519030}{C.7}).
In contrast, \prop\,-dominated clusters are found at low \nht\, and low $V_\mathrm{offset}$, similar to \meth\, in Case 1, building the upper diagonal in the intensity-dist2dust and located in the south and south-east of the core (see blue clusters in combs. 2, 3, and 9, and red cluster in comb. 5 in Figs.~\ref{fig:ClusteringL1521E} and \href{https://zenodo.org/records/15519030}{C.7}).
\prop\,-dominated clusters are concentrated around the carbon peak in the south-east and are associated with higher \nht\, (see cyan and yellow clusters in comb. 4, cyan cluster in comb. 6, and cyan and purple clusters in comb. 10 in Figs.~\ref{fig:ClusteringL1521E} and \href{https://zenodo.org/records/15519030}{C.7}).
Additionally, a low intensity \cyc\, cluster is located along the sharp edge of the core in the south-west (see cyan cluster in comb. 9 in Fig.~\ref{fig:ClusteringL1521E}). Similar to B68, the shape of this cluster follows values of high \ngrad\,, even though this feature is not included in combination 9.

\paragraph{L1544:}
For \cyc\, and \prop\,, the clustered features reveal structures similar to those found in Case 1 (e.g. in $V_\mathrm{offset}$ over intensity, linewidth over $V_\mathrm{offset}$, \nht\, over $V_\mathrm{offset}$, \ngrad\, over $V_\mathrm{offset}$, \nht\, over linewidth), which again divide the core into north/south and on-centre/off-centre regions. 
As in Case 1, \cyc\, is predominantly associated with the northern part of the core, while \prop\, is dominant in the south. This separation is visible in $V_\mathrm{offset}$ and \ngrad\, (\cyc\,: red clusters in combs. 1, 2, and 4; \prop\,: blue clusters in combs. 1 and 4; see Figs.~\ref{fig:ClusteringL1544} and \href{https://zenodo.org/records/15519030}{C.8}). 
In addition, both molecules cluster in the core centre, exhibiting high intensity, higher $V_\mathrm{offset}$ and higher linewidth (\cyc\,: cyan cluster in comb. 1; \prop\,: yellow cluster in comb. 1, and red cluster in comb. 6; see Figs.~\ref{fig:ClusteringL1544} and \href{https://zenodo.org/records/15519030}{C.8}). 
In combination 3, a \prop\,-dominated cluster also covers the \cyc\, peak in the south-east of the core, characterised by high intensity, high \nht\,, and high $V_\mathrm{offset}$ (see red cluster in Fig.~\ref{fig:ClusteringL1544}).

\paragraph{Summary of Case 2:}
Clusters dominated by \cyc\, or \prop\, show spatial separation across all cores, similar to the segregation observed between \cyc\, and \meth\, in Case 1, revealing distinct structures in various features combinations.
The molecular segregation is evident in the same features as in Case 1, intensity, $V_\mathrm{offset}$, and \nht\,, with B68 and L1544 also showing separation in \ngrad\,.
Overall, the clustering shows the same strong divisions of the cores into north/south, east/west, on-centre/off-centre regions as seen in Case 1, highlighting structural and chemical themes across the cores.

\subsubsection{Case 3: \meth\, vs \prop\,}\label{subsec:case3}

For B68, \prop\, is slightly underrepresented in this dataset, resulting in clusters that are more imbalanced towards \meth\,.
In the following, we discuss the clustering results for each core individually.

\paragraph{B68 (Fig.~\href{https://zenodo.org/records/15519030}{C.9}):}
\meth\,-dominated clusters and \prop\,-dominated clusters are not clearly spatially separated. Instead, both are found in the central area of the core and on the core centre (\meth\.: blue cluster in combs. 1, 2, 6, 9, cyan cluster in comb. 6; \prop\,: red cluster in comb. 1 and cyan cluster in combs. 8 and 10), reproducing the clustering behaviour of Case 1 for \meth\, and Case 2 for \prop\,. 
A direct, feature-wise separation of the two molecules occurs only in combination 1 in intensity, with \meth\, at higher and \prop\, at lower values.
Apart from that, \meth\, is clustered along the east of the core with high $V_\mathrm{offset}$, and along the west of the core with low $V_\mathrm{offset}$ (east: cyan cluster in comb. 1, red cluster in combs. 2 and 3; west: yellow cluster in comb. 4).
The shapes of those clusters follow high values of \ngrad\,, even though this feature is only included in combination 4, resembling patterns of \cyc\, in Case 2 (east) and Case 1 (west). 
In combinations 8 and 10, the clustering creates ring-like structures around the dust peak, visible in \nht{}, dist2dust, and \ngrad\,, with \prop\, concentrated at the centre and \meth\, forming the outer rings - reflecting structures found in Case 1 (\meth\,) and Case 2 (\prop\,).

\paragraph{L1521E (Fig.~\href{https://zenodo.org/records/15519030}{C.10}):}
As in Cases 1 and 2, we see a spatial separation of the two input molecules, \meth\, and \prop\,, but here it is less pronounced. 
\prop\, is primarily clustered in the (south-) east of the core, similar to Case 2 (e.g. see blue and red clusters in comb. 1, cyan cluster in combs. 4, 5 and 10, yellow cluster in combs. 7 and 8). 
In contrast, \meth\, is clustered along the north of the core (see red cluster in combs. 3 and 8, blue cluster in comb. 9), showing patterns similar to \cyc\, in Case 1 and Case 2, but contrary to its own clustering behaviour in Case 1. 
In terms of features, we see an indirect separation in \nht\, values: \prop\, is associated with mid \nht\,, and \meth\, with higher \nht\, (\prop\,: cyan cluster in comb. 10; \meth\,: blue cluster in comb. 9).
Combination 5 shows a direct spatial separation of the two molecules into east (\prop\,) and west (\meth\,), visible as upper and lower diagonal in the intensity/dist2dust-distribution.
Apart from that, \prop\, is clustered at the sharp edge of the core in the south-west, with high \ngrad\, and mid $V_\mathrm{offset}$ (see red cluster in comb. 4, blue cluster in comb. 7), similar to \cyc\, in Case 2.
Both \meth\, and \prop\, are additionally clustered at the core centre, at high \nht\,, with \meth\, at high intensity (see red cluster in comb. 6), and \prop\, at lower intensity (see cyan cluster in comb. 3, purple cluster in comb. 6 and yellow cluster in comb. 9). This pattern was not observed for either \meth\, or \prop\, in Case 1 or Case 2.

\paragraph{L1544 (Fig.~\href{https://zenodo.org/records/15519030}{C.11}):}
For \meth\, and \prop\,, the clustered features reveal structures similar to those found in Case 1 and Case 2 (e.g. in $V_\mathrm{offset}$ over intensity, linewidth over $V_\mathrm{offset}$, \nht\, over $V_\mathrm{offset}$, \ngrad\, over $V_\mathrm{offset}$, \nht\, over linewidth).
As before, this leads to a division of the core into north/south and on-centre/off-centre regions.
\meth\, exhibits a north-south separation, visible in the features $V_\mathrm{offset}$, intensity, and \ngrad\, (see blue and red clusters in combs. 1-4), similar to the pattern seen in Case 1.
In contrast, \prop\, is predominantly clustered in the south and on the \cyc\, peak (see yellow cluster in combs. 3, 4, and 7), as well as at the core centre (see yellow cluster in comb. 2, cyan cluster in comb. 3, red cluster in combs. 5, 6 and 9), which were both observed in Case 2. 
Additionally, \prop\, is clustered at its molecule peak in the north-west, with high intensity, high linewidth, and low \ngrad\,, which was not seen in Case 2 (see cyan cluster in comb. 7).
Direct molecular segregation occurs only in comb. 3, where \prop\, is clustered at the core centre (cyan cluster) with high intensity and high \nht\,, while \meth\, is clustered around the centre at lower intensity and lower \nht\, (blue and red clusters).
Ring-like cluster structures are visible in \nht\, and \ngrad\, in combinations 6, 8, 9, and 10; however, they do not display any molecular segregation but instead a balanced ratio between \prop\, and \meth\,.

\paragraph{Summary of Case 3:}
All cores display cluster structures with feature-wise or spatial separation between \meth\, and \prop\,. However, the molecular segregation is less distinct than in Case 1 and Case 2, and it is mainly visible in the features intensity, $V_\mathrm{offset}$, \nht\,, and \ngrad\,.
In B68 and L1544, the clustering predominantly reproduces the structures and clusters found in Case 1 and 2.
All three cores show minor differences of cluster behaviour compared to Case 1 and Case 2, where \meth\, or \prop\, behave similar to \cyc\,. This is particularly evident in L1521E, where \meth\, and \prop\, show a more distinct separation, similar to Case 1 and 2.

\subsubsection{Case 4: \cyc\, vs \meth\, vs \prop\,}\label{subsec:case4}

With a dataset containing three molecules, it is more difficult and less clear to identify molecular segregation, as the ratios between the molecules mostly do not show big variations from the initial ratio of the dataset (the initial ratios for \cyc\,/\meth\,/\prop\, are 37/36/27 for B68, 29/35/36 for L1521E, and 28/39/33 for L1544).
In the following, we discuss the clustering results for each core individually.

In B68 (see Fig.~\href{https://zenodo.org/records/15519030}{C.12}), \cyc\, is clustered in a shell along the east, reproducing the structure of Case 1. \prop\, is clustered in the core centre, reproducing the behaviour in Case 2 and Case 3. However, \meth\, is not clustered in the core centre as in Case 1 but instead around the centre in shells along the east and along the west, similar to what was found in Case 3.
Additionally, \cyc\, and \meth\, show concentric clusters around the centre, visible in \nht\,, similar to before.

In L1521E (see Fig.~\href{https://zenodo.org/records/15519030}{C.13}), \cyc\, is clustered in the northern part of the core, reproducing the clusters in both Case 1 and 2. It also shows the small cluster along the sharp edge in the south-west of the core, seen in Case 2. \prop\, is clustered in the south of the core and the core centre, reproducing the cluster structures of Case 2 and Case 3.
\meth\,, on the other hand, is not clustered in the south as seen in Case 1 but instead in the core centre, as in Case 3 (where its emission peaks) and the north-west of the core.

In L1544 (see Fig.~\href{https://zenodo.org/records/15519030}{C.14}), the division of the core into north/south is visible in $V_\mathrm{offset}$ for \meth\,, similar to Cases 1 and 3. 
For \cyc\,, the association with the northern part seen in Case 1 and Case 2 cannot be reproduced. Instead, it is 
clustered only on the \meth\, peak in the north-east of the core. The cluster in the core centre can be reproduced.
Additionally, both \cyc\, and \meth\, are clustered on their respective molecular peaks, which was not seen in Case 1 or Case 2.
For \prop\,, both the cluster in the core centre and on the \cyc\, peak are reproduced. However, Case 4 does not recreate \prop\, as being associated with the southern part of the core in the north-south division as in Cases 2 and 3.
Also, the molecular separation in \ngrad\, and $V_\mathrm{offset}$ that was seen in combination 4 in Case 1 and 2 is not reproduced with this combined dataset.

To summarise, in B68 and L1521E, the clustering behaviour of \cyc\, and \prop\, seen in Case 2 can be reproduced, but \meth\, behaves now differently than in Case 1.
In L1544, however, the behaviour of \meth\, and part of \prop\, seen in Case 1 and Case 2 can be reproduced, but not the behaviour of \cyc\,.
We discuss this further in Sect.\ref{sec:discussion:clustering}

\subsection{\prop\, abundances} \label{sec:dustpeakabundances}

This section focuses on comparing the \prop\, abundances towards the dust peaks of the different cores. In addition to B68, L1521E, and L1544, we include data from the prestellar cores HMM-1, L429, L694-2, and OphD, which were observed but not analysed in the IRAM project of \cite{Spezzano2020}.
To calculate the abundances at the dust peaks, we divide the \prop\, column density by the H$_2$ column density.
The H$_2$ column density at the dust peak is extracted from the respective N(H$_2$) map (derived from $Herschel$ SPIRE observations, \citealt{Spezzano2020}) using a circular aperture with a diameter of 16" (matching the Herschel map pixel size).
We assume a 20\% uncertainty for the resulting values. 
To derive N(\prop\,), we convolve the \prop\, spectral cubes with the 40" beam of the $Herschel$ telescope and extract the spectrum at the dust peak using the same 16" aperture.
The column density is calculated using a one-dimensional Gaussian fit, and following \cite{Mangum2015}, with the respective spectroscopic parameters listed in Table~\ref{Tab:ObservedLines}.
Since we do not have sufficient data to determine a precise excitation temperature via a rotational diagram for all cores, we adopt a standard excitation temperature of 8\,K for consistency.
A lower (higher) excitation temperature only affects the abundances of \prop\, (and \cyc\, in L1544) by shifting them to higher (lower) values, but the overall trend stays the same.
For each core, we select the most optically thin transition with the smallest (propagated) error in column density. 
Specifically, we use the \prop\, ($5_1-4_1$) transition for most cores, except for L694-2 and L1544, where the ($5_0-4_0$) and ($6_1-5_1$) transitions are used, respectively.

\begin{figure}[h!]
    \centering
    \includegraphics[width=\hsize]{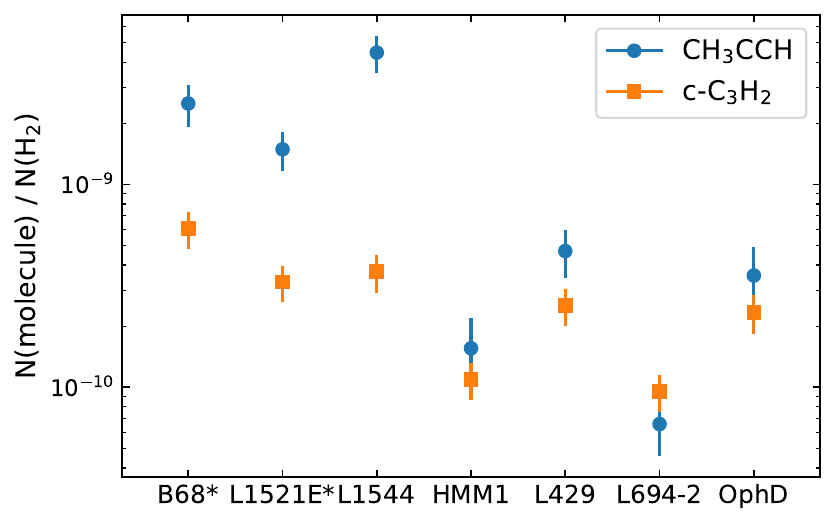}
    \caption[alt={Abundances of \prop\,, and \cyc\, at the dust peaks of the core sample.}]{Abundances of \prop\, and \cyc\, at the dust peaks of the cores. 
    The molecular column density was calculated assuming $T_\mathrm{ex}=8$\,K (see text for details). The starless cores are marked with an asterisk.}
    \label{fig:abundancesatdustpeak}
\end{figure}

For comparison, we also calculate the abundances of \cyc\,, using the ($2_{02}-1_{11}$) transition for all cores except L1544, where the ($3_{2,2}-3{1,3}$) transition is applied. Figure~\ref{fig:DustpeakSpectra} displays the extracted spectra along with their Gaussian fits.
Figure~\ref{fig:abundancesatdustpeak} presents the resulting \prop\, (blue circles) and \cyc\, (orange squares) abundances at the dust peaks of the different cores for an assumed excitation temperature of 8\,K.

The \prop\, abundances of the starless cores (see left part of Fig.~\ref{fig:abundancesatdustpeak}) are about one order of magnitude higher than the values of the prestellar cores (see right part of Fig.~\ref{fig:abundancesatdustpeak}), suggesting an evolutionary trend of \prop\, from the starless to the prestellar phase. 
Notably, the \prop\, abundances in L1544 are significantly higher compared to the other prestellar cores and even compared to the starless cores.
This suggests that the observed variations in \prop\, are probably influenced not only by the evolutionary stage but also by environmental factors. This is be discussed further in Sec.~\ref{sec:discussion:ch3cch}.
However, to further study the interplay between environmental and evolutionary or dynamical effects on the \prop\, abundance spread, additional work is necessary that goes beyond the scope of this paper.

The \cyc\, abundances show much less variation than those of \prop\, and do not exhibit a significant difference between the starless and the prestellar stages. The abundances are spread within one order of magnitude, suggesting that \cyc\, and \prop\, trace different layers within the core. This indicates that \cyc\, is largely unaffected by the evolution from the starless to the prestellar stage.

\FloatBarrier
\section{Discussion}\label{sec:discussion}

\subsection{Density-based clustering}\label{sec:discussion:clustering}

\paragraph{Segregation between \cyc\, and \meth\,.}
Density-based clustering is able to find molecular differentiation in our dataset.
In particular, the clustering with the dataset containing \cyc\, and \meth\, (Case 1) successfully reproduces the known molecular segregation between these two molecules in B68, L1521E, and L1544.
This segregation is attributed to uneven illumination across the cores, as discussed in \cite{Spezzano2016,Spezzano2020}. 
The differentiation mainly appears in the features intensity, $V_\mathrm{offset}$, \nht\,, and \ngrad\,. 
In addition, the following pairs of features frequently show segregation: 
intensity/$V_\mathrm{offset}$, intensity/dist2dust, intensity/\nht\,, intensity/\ngrad\,, $V_\mathrm{offset}$/\nht\,, $V_\mathrm{offset}$/\ngrad\,.\\

\paragraph{Segregation between \cyc\, and \prop\,.}
The clustering analysis of the dataset containing \cyc\, and \prop\, (Case 2) reveals molecular segregation between these two carbon chains in all three cores. 
Like in Case 1, the segregation appears in the features intensity, $V_\mathrm{offset}$, \nht\,, \ngrad\,, and similar pairs of features. 
However, in B68 and L1521E, a differentiation between these two molecules is not apparent in their emission maps (see Figs.~\ref{fig:intmapsC3H2CH3OH} and \ref{fig:intmapsCH3CCH}) and was therefore previously unrecognised.
The segregation between \cyc\, and \prop\, suggests that these molecules trace different layers in the cores, representing different physical conditions. A similar result was discussed in \cite{Lin2022}, where \cyc\, was found to trace lower density regions than for example \meth\,.

In B68 and L1521E, the emission of \prop\, is less spatially extended compared to \cyc\, (see Figs.~\ref{fig:intmapsC3H2CH3OH} and \ref{fig:intmapsCH3CCH}). In B68, the \prop\, emission is concentrated on the core centre, while in L1521E, it is primarily located in the eastern part of the core. Due to the lower number of available data points for \prop\, compared to \cyc\,, the clustering dataset in Case 2 includes two \prop\, transitions for B68 and three transitions for L1521E (as detailed in Table~\ref{tab:casestudylines}). This results in a higher density of data points in the core centre of B68 and the eastern part of L1521E, increasing the likelihood of forming clusters at these locations with density-based algorithms such as DBSCAN and HDBSCAN.
The incomplete coverage of \prop\, across the cores becomes more apparent in Case 3 (\meth\,, and \prop\,) and Case 4 (\cyc\,, \meth\,, and \prop\,).
In Case 4, \prop\, forms clusters similar to those in Case 2 (\cyc\, and \prop\,), but \meth\, shows different behaviour compared to Case 1 (\cyc\, and \meth\,). In contrast, in L1544 -- where the emission maps of \cyc\,, \meth\,, and \prop\, extend across the entire core -- the clustering results of Case 4 differ significantly from Case 1 and 2.
In Case 3, on the other hand, most clusters found in Case 1 and 2 are recreated. The only exception is \meth\, in L1521E, that mimics the clustering behaviour of \cyc\, instead.

\paragraph{Similarities between \meth\, and \prop\,.}
Our analysis reveals that in B68 and L1521E, the clusters dominated by \meth\, in Case 1 behave very similarly to those dominated by \prop\, in Case 2, despite the fact that the \meth\, emission is as spatially extended as \cyc\, in both cores and \prop\, is not.
The \prop\,- and \meth\,-dominated clusters are associated with the same features -- \nht\, and $V_\mathrm{offset}$ for L1521E, intensity for B68 -- and are located in the same regions within the cores (south-east for L1521E, and the core centre for B68).
Additionally, these clusters are spatially distinct from the \cyc\, clusters. 
As shown in Fig~\ref{fig:Correlationplots}, \cyc\, exhibits broader linewidths than \meth\, and \prop\, across B68, likely indicating that it traces a different, more turbulent layer. These clustering results may therefore reflect differences in the physical layers traced by the molecules.

In L1544, the similarity in cluster behaviour between \meth\, and \prop\, relative to \cyc\, is observed in only one feature pairing: \ngrad\, and $V_\mathrm{offset}$ (see combination 4). 
This could be linked to both \meth\, and \prop\, tracing gas influenced by slow accretion flows.
For \meth\,, such an association has been demonstrated and discussed, for instance, by \cite{Lin2022}. In Sec.~\ref{sec:discussion:ch3cch}, we further explore how the \prop\, peak in L1544 may also be impacted by inflowing gas.

\paragraph{Relevance of different features in the clustering analysis.}
In our clustering analysis, $V_\mathrm{offset}$ appears to be a dominant feature that drives the division of the cores into the different clusters, and in some cases, molecular separation.
In L1521E, the core is divided into north-west/south-east, while L1544 shows a north/south split, both reflecting the velocity structure of the cores. A similar split is indicated in B68, with a separation between the core centre and a shell along the eastern side, although this division is less pronounced.
The strong dependence of velocity structure with chemical prominence indicates that static chemical models might not be sufficient to predict observed features in full. While the overall physical structure of these cores is generally well-described by quasi-static models, 
understanding the anisotropic chemical structures requires a more dynamic approach \citep[see also][]{Lin2022}.

Both B68 and L1544 show clusters with concentric ring structures around their core centres, following the patterns of \nht\, and \ngrad\,. This ring-like pattern is a result of the rather spherical shape of these cores. In contrast, L1521E, which is more elongated and irregularly shaped, does not show this pattern. 
The clustering analysis of both L1521E (Case1) and L1544 (Case1/Case2) reveals molecular separation in the $V_\mathrm{offset}$-\ngrad\, distribution, which does not appear for B68. This difference may be due to environmental factors, as B68, a Bok globule, is exposed to relatively uniform external illumination.
In L1521E, the clusters with the highest \ngrad\, values in this feature pairing (see combination 4) are dominated by \cyc\,. These data points are located at the filamentary edge in the southern part of the core, where \cyc\, peaks and external illumination is strongest. 
In L1544, the data points with highest \ngrad\, values also come from the southern part of the core, near the filamentary edge with high external illumination. However, in this case they are dominated by \meth\, and \prop\, instead of \cyc\,. 
This suggests that the clustering results reflect the distinct environmental conditions within each core.
Overall, the clusters with prominent \nht\, and \ngrad\, features appear to represent the chemical patterns across the core structures, with differences in the clustering likely tied to varying environmental conditions.

The features dist2dust and linewidth appear to be less significant in our analysis, as they rarely exhibit distinct structures or molecular separations by themselves. However, when combined with other features, such as dist2dust/intensity or linewidth/$V_\mathrm{offset}$, they provide additional insights. The one-dimensional projected distance to the dust peak seems less relevant in the clustering analysis compared to \nht\, and \ngrad\,, as these two features better characterise the immediate environment of a data point or pixel.

\subsection{Evolution traced by \prop\,} \label{sec:discussion:ch3cch}

In the starless cores B68 and L1521E, the distribution of the \prop\, emission overlaps with that of \cyc\, (see Fig.~\ref{fig:intmapsC3H2CH3OH} and Fig.~\ref{fig:intmapsCH3CCH}).
In L1544, however, the peak of the \prop\, emission is not located in the south-east of the core, where the other carbon chains are found, but rather in the north-west. 
It is known that in the north-east of L1544, around the \meth\, peak, two filaments converge \citep[e.g. see][]{Andre2010,Spezzano2016}. This may result in slow accretion flows \citep{Punanova2018,Lin2022}, which could deliver fresh material to the core and help replenish \prop\,.

To investigate the formation and destruction routes of \prop\,, we conducted chemical simulations using the gas-grain chemical model \texttt{pyRate} \citep{Sipila2015}, applied to the standard physical model of L1544 \citep{Keto2015}. 
For the gas-phase chemical network we adopted the 2014 public release of the KIDA chemical network \citep[kida.uva.2014;][]{WakelamKIDA2015}, while the grain-surface network is an updated version of the one presented in \cite{Semenov2010}.
The simulation yields radial abundance profiles as a function of time; we checked the results at an evolutionary time of $10^5\,$yrs.
The model shows that at intermediate densities ($n\sim10^4$\,cm$^{-3}$),  \prop\, is mainly formed by the dissociative recombination of C$_3$H$_5^+$:
\begin{equation}
    \mathrm{C}_3\mathrm{H}_5^+ + e^- \longrightarrow \mathrm{CH}_3\mathrm{CCH} + \mathrm{H}\;,
\end{equation}
while it mainly gets destroyed by the reaction with free carbon:
\begin{equation}
    \mathrm{CH}_3\mathrm{CCH} + \mathrm{C} \longrightarrow \mathrm{C}_4\mathrm{H}_3 + \mathrm{H}\;.
\end{equation}

\begin{figure*}
    \centering
    \includegraphics[width=0.33\textwidth]{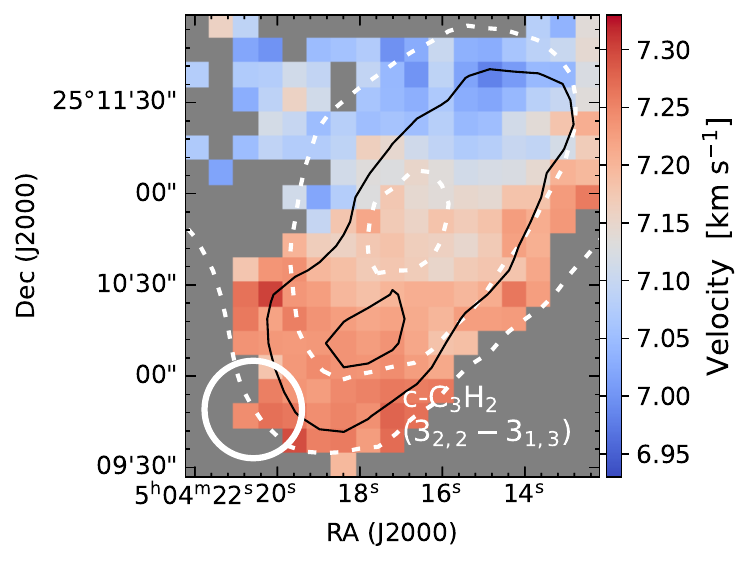}
    \includegraphics[width=0.33\textwidth]{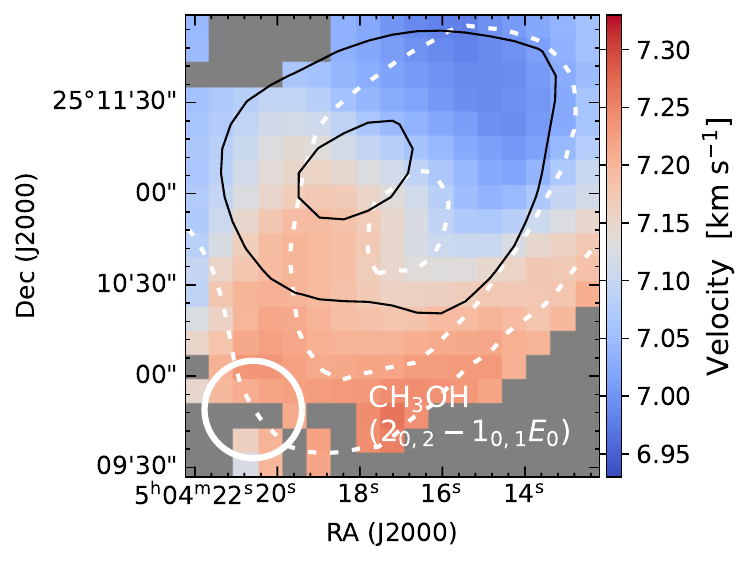}
    \includegraphics[width=0.33\textwidth]{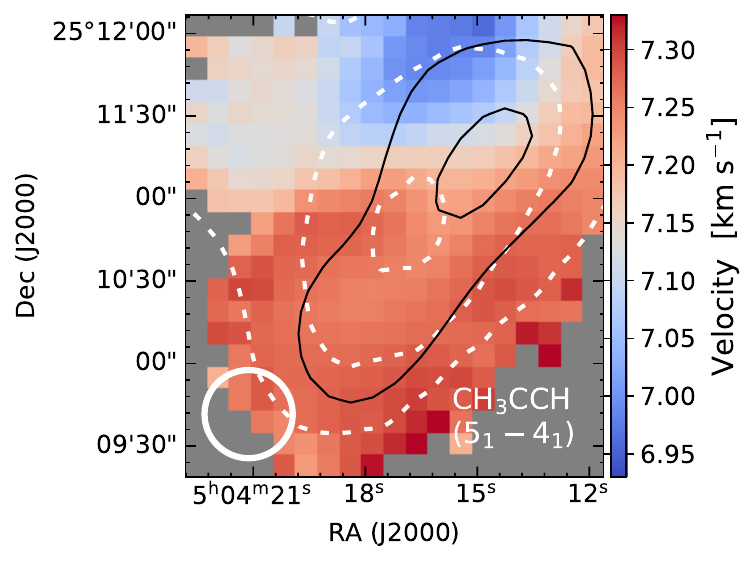}
    \includegraphics[width=0.33\textwidth]{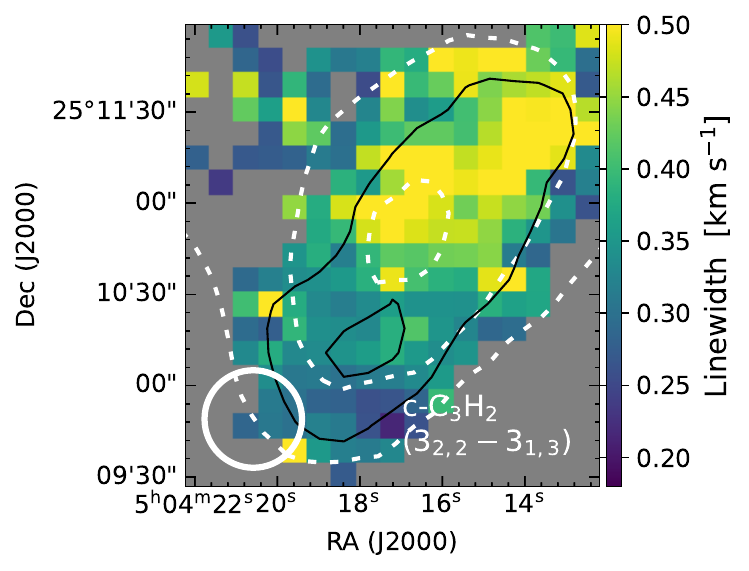}
    \includegraphics[width=0.33\textwidth]{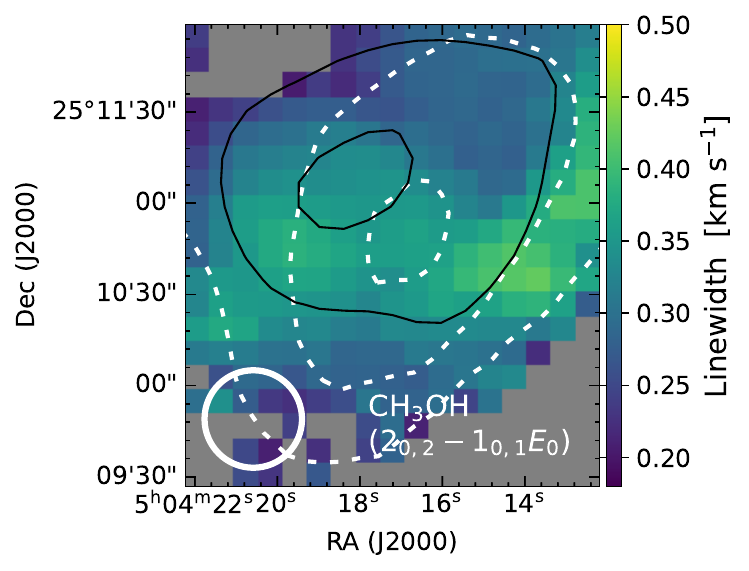}
    \includegraphics[width=0.33\textwidth]{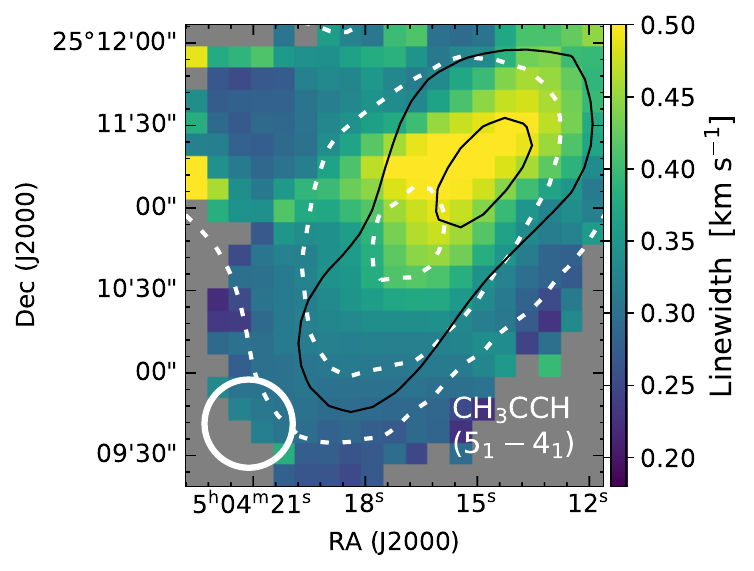}
    \caption[alt={Centroid velocities and linewidths of \cyc\,, prop\,, and \meth\, towards the prestellar core L1544.}]{Centroid velocities (\textit{top}) and linewidths (\textit{bottom}) of \cyc\, (\textit{left}), \prop\, (\textit{right}), and \meth\, (\textit{middle}) towards the prestellar core L1544. Black contours show 50\% and 90\% of the respective molecular emission peak. White contours show 30\%, 50\%, and 90\% of the H$_2$ column density peak derived from \textit{Herschel} maps \citep{Spezzano2016}. The white circle in the bottom-left corner indicates the beam size of the IRAM 30\,m telescope (32").} 
    \label{fig:VelMapsL1544}
\end{figure*}

Following this, in regions with high irradiation and therefore active gas-phase chemistry, such as the carbon-chain peak in L1544, \prop\, is quickly destroyed due to the high amounts of atomic carbon present in the gas phase.
In contrast, the north-western part of L1544 is more shielded from irradiation, allowing \prop\, to form from fresh material brought in by the filamentary flow from the north-west.
Here, the low abundance of free carbon in the gas phase protects the \prop\, from destruction.
This is further supported by the velocity and linewidth maps of \prop\,, as shown in Fig.~\ref{fig:VelMapsL1544}, along with the results for \cyc\, and \meth\,. The linewidth map of \prop\, (bottom right) shows increased linewidths near the emission peak, while the velocity map (top right) reveals a sharp velocity gradient in the same area. This suggests that the \prop\, emission peak in L1544 is the landing point or accumulation point of the incoming fresh gas. 
Similar signs of this active chemistry are observed in \cyc\,. The integrated intensity map (Fig.~\ref{fig:intmapsCH3CCH}) shows a local maximum in this region, while the linewidth map (bottom left in Fig.~\ref{fig:VelMapsL1544}) exhibits the highest linewidths not at the \cyc\, peak in the south-east but at the \prop\, peak in the north-west. The velocity map of \cyc\, (top left) also shows a gradient around this area.

The fact that \prop\, is an early-type molecule explains why, in our dataset, extended \prop\, emission is only observed in the starless cores (B68, L1521E) and the prestellar core L1544 due to the possible material accretion.
In contrast, in the other prestellar cores (HMM1, OphD, L694-2, L429), we detect some emission at the respective dust peaks, but no significant emission beyond that. 
This evolutionary trend is further supported by the \prop\, abundances at the dust peaks (shown in Fig.~\ref{fig:abundancesatdustpeak}), 
where the abundances in the starless cores are approximately one order of magnitude higher than those in the prestellar cores, except for L1544.

\section{Conclusions}\label{sec:conclusion}

We have presented an analysis of molecular differentiation using the density-based clustering algorithms DBSCAN and HDBSCAN. The clustering was applied to four different datasets, in order to compare the emission morphologies of \cyc\,, \meth\,, and \prop\, observed towards the starless cores B68 and L1521E and the prestellar core L1544.

Our main results can be summarised as follows:
\begin{itemize}
    \item The analysis with density-based clustering finds a significant chemical differentiation across the cores in our dataset. It successfully reproduces the known molecular segregation of \cyc\, and \meth\, for B68, L1521E, and L1544. 
    Furthermore, the clustering analysis identifies a segregation between \cyc\, and \prop\, in all three cores, which is not apparent from visual inspection of the emission maps. 
    \item The most relevant features in the clustering analysis are integrated intensity, velocity offset, H$_2$ column density, and H$_2$ column density gradient. 
    Distinct and recurring cluster structures in the H$_2$ column density and the gradient highlight structural and chemical patterns across the cores. Differences in the relevance of these two features for the three cores reflect the varying environmental conditions within each core. 
    The strong relation between molecular emission and velocity structure suggests that to understand anisotropic chemical structures, static chemical models are not sufficient, but  dynamical models are necessary.
    \item Increased \prop\, abundances towards the starless cores compared to prestellar cores indicate an evolutionary trend. Increased \prop\, abundances towards L1544 suggest an additional influence of environmental factors.
    In fact, in L1544, the \prop\, peak in the north-west of the core appears to trace the landing point of chemically fresh gas that is accreted to the core. Unlike the photochemically active south of the core, this area is shielded from external irradiation, which protects \prop\, from being destroyed by free carbon atoms. 
    \item The clustering analysis finds a similar behaviour between \meth\, and \prop\, relative to \cyc\, in all cores.
    This indicates that \cyc\, traces an outer layer of gas and possibly a lower-density shell compared to the other two molecules.
    In L1544, the similar clustering patterns observed for \meth\, and \prop\, may reflect the influence of accretion processes in shaping the molecular distribution.
\end{itemize}

Our results demonstrate that a successful density-based clustering approach for studying astrochemical processes does not require a large dataset covering multiple molecules across various cores. 
In fact, the results are often easier to interpret when only two or three molecules are considered. 
While this clustering method is more time-consuming than techniques such as principal component analysis, it can process much more detailed information and provide deeper insights into the core's structure.
Using the more general approach of describing a data point’s location through its H$_2$ column density and its gradient, rather than relying on spatial coordinates, also enables simple comparisons between cores.
In future studies, we aim to expand our analysis of molecular differentiation with density-based clustering to include more cores and molecules, especially those that trace other physical or chemical features. 
This will also help explore any evolutionary effects that the cores or their environment might have on the molecular distribution.

\section*{Data availability}
Figures~\href{https://zenodo.org/records/15519030}{C.3-C.14}, presenting the detailed clustering results for Cases 1-4, are published on Zenodo (\href{https://zenodo.org/records/15519030}{zenodo.org/records/15519030}). 

\begin{acknowledgements}
    We wish to thank the anonymous referee for their constructive comments.
    K.G. thanks Caroline Gieser for useful discussions. S.S. and K.G. wish to thank the Max Planck Society for the Max Planck Research Group funding. All others authors affiliated to the MPE wish to thank the Max Planck Society for financial support. 
\end{acknowledgements}

%
\bibliographystyle{aa} 
\bibliography{mybib.bib} 

\begin{thebibliography}{54}
\expandafter\ifx\csname natexlab\endcsname\relax\def\natexlab#1{#1}\fi

\bibitem[{{Alves} \& {Franco}(2007)}]{AlvesFranco2007}
{Alves}, F.~O. \& {Franco}, G.~A.~P. 2007, \aap, 470, 597

\bibitem[{{Andr{\'e}} {et~al.}(2010){Andr{\'e}}, {Men'shchikov}, {Bontemps}, {K{\"o}nyves}, {Motte}, {Schneider}, {Didelon}, {Minier}, {Saraceno}, {Ward-Thompson}, {di Francesco}, {White}, {Molinari}, {Testi}, {Abergel}, {Griffin}, {Henning}, {Royer}, {Mer{\'\i}n}, {Vavrek}, {Attard}, {Arzoumanian}, {Wilson}, {Ade}, {Aussel}, {Baluteau}, {Benedettini}, {Bernard}, {Blommaert}, {Cambr{\'e}sy}, {Cox}, {di Giorgio}, {Hargrave}, {Hennemann}, {Huang}, {Kirk}, {Krause}, {Launhardt}, {Leeks}, {Le Pennec}, {Li}, {Martin}, {Maury}, {Olofsson}, {Omont}, {Peretto}, {Pezzuto}, {Prusti}, {Roussel}, {Russeil}, {Sauvage}, {Sibthorpe}, {Sicilia-Aguilar}, {Spinoglio}, {Waelkens}, {Woodcraft}, \& {Zavagno}}]{Andre2010}
{Andr{\'e}}, P., {Men'shchikov}, A., {Bontemps}, S., {et~al.} 2010, \aap, 518, L102

\bibitem[{{Andre} {et~al.}(2000){Andre}, {Ward-Thompson}, \& {Barsony}}]{Andre2000}
{Andre}, P., {Ward-Thompson}, D., \& {Barsony}, M. 2000, in Protostars and Planets IV, ed. V.~{Mannings}, A.~P. {Boss}, \& S.~S. {Russell}, 59

\bibitem[{{Bauer} \& {Burie}(1969)}]{Bauer1969}
{Bauer}, A. \& {Burie}, J. 1969, C. R. Acad. Sci. Paris, B 268, 800

\bibitem[{{Bron} {et~al.}(2018){Bron}, {Daudon}, {Pety}, {Levrier}, {Gerin}, {Gratier}, {Orkisz}, {Guzman}, {Bardeau}, {Goicoechea}, {Liszt}, {{\"O}berg}, {Peretto}, {Sievers}, \& {Tremblin}}]{Bron2018}
{Bron}, E., {Daudon}, C., {Pety}, J., {et~al.} 2018, \aap, 610, A12

\bibitem[{Campello {et~al.}(2013)Campello, Moulavi, \& Sander}]{HDBSCAN}
Campello, R. J. G.~B., Moulavi, D., \& Sander, J. 2013, in Advances in Knowledge Discovery and Data Mining, ed. J.~Pei, V.~S. Tseng, L.~Cao, H.~Motoda, \& G.~Xu (Berlin, Heidelberg: Springer Berlin Heidelberg), 160--172

\bibitem[{{Caselli} {et~al.}(2012){Caselli}, {Keto}, {Bergin}, {Tafalla}, {Aikawa}, {Douglas}, {Pagani}, {Y{\'\i}ld{\'\i}z}, {van der Tak}, {Walmsley}, {Codella}, {Nisini}, {Kristensen}, \& {van Dishoeck}}]{Caselli2012}
{Caselli}, P., {Keto}, E., {Bergin}, E.~A., {et~al.} 2012, \apjl, 759, L37

\bibitem[{{Caselli} {et~al.}(2022){Caselli}, {Pineda}, {Sipil{\"a}}, {Zhao}, {Redaelli}, {Spezzano}, {Maureira}, {Alves}, {Bizzocchi}, {Bourke}, {Chac{\'o}n-Tanarro}, {Friesen}, {Galli}, {Harju}, {Jim{\'e}nez-Serra}, {Keto}, {Li}, {Padovani}, {Schmiedeke}, {Tafalla}, \& {Vastel}}]{Caselli2022}
{Caselli}, P., {Pineda}, J.~E., {Sipil{\"a}}, O., {et~al.} 2022, \apj, 929, 13

\bibitem[{{Caselli} {et~al.}(2002){Caselli}, {Walmsley}, {Zucconi}, {Tafalla}, {Dore}, \& {Myers}}]{Caselli2002a}
{Caselli}, P., {Walmsley}, C.~M., {Zucconi}, A., {et~al.} 2002, \apj, 565, 331

\bibitem[{{Chac{\'o}n-Tanarro} {et~al.}(2019){Chac{\'o}n-Tanarro}, {Caselli}, {Bizzocchi}, {Pineda}, {Sipil{\"a}}, {Vasyunin}, {Spezzano}, {Punanova}, {Giuliano}, \& {Lattanzi}}]{ChaconTanarro2019}
{Chac{\'o}n-Tanarro}, A., {Caselli}, P., {Bizzocchi}, L., {et~al.} 2019, \aap, 622, A141

\bibitem[{{Cleeves} {et~al.}(2014){Cleeves}, {Bergin}, {Alexander}, {Du}, {Graninger}, {{\"O}berg}, \& {Harries}}]{Cleeves2014}
{Cleeves}, L.~I., {Bergin}, E.~A., {Alexander}, C. M.~O.~D., {et~al.} 2014, Science, 345, 1590

\bibitem[{{Colombo} {et~al.}(2015){Colombo}, {Rosolowsky}, {Ginsburg}, {Duarte-Cabral}, \& {Hughes}}]{Colombo2015}
{Colombo}, D., {Rosolowsky}, E., {Ginsburg}, A., {Duarte-Cabral}, A., \& {Hughes}, A. 2015, \mnras, 454, 2067

\bibitem[{{Crapsi} {et~al.}(2005){Crapsi}, {Caselli}, {Walmsley}, {Myers}, {Tafalla}, {Lee}, \& {Bourke}}]{Crapsi2005}
{Crapsi}, A., {Caselli}, P., {Walmsley}, C.~M., {et~al.} 2005, \apj, 619, 379

\bibitem[{{Crapsi} {et~al.}(2007){Crapsi}, {Caselli}, {Walmsley}, \& {Tafalla}}]{Crapsi2007}
{Crapsi}, A., {Caselli}, P., {Walmsley}, M.~C., \& {Tafalla}, M. 2007, \aap, 470, 221

\bibitem[{{Drozdovskaya} {et~al.}(2022){Drozdovskaya}, {Coudert}, {Margul{\`e}s}, {Coutens}, {J{\o}rgensen}, \& {Manigand}}]{Drozdovskaya2022}
{Drozdovskaya}, M.~N., {Coudert}, L.~H., {Margul{\`e}s}, L., {et~al.} 2022, \aap, 659, A69

\bibitem[{{Drozdovskaya} {et~al.}(2021){Drozdovskaya}, {Schroeder I}, {Rubin}, {Altwegg}, {van Dishoeck}, {Kulterer}, {De Keyser}, {Fuselier}, \& {Combi}}]{Drozdovskaya2021}
{Drozdovskaya}, M.~N., {Schroeder I}, I. R.~H.~G., {Rubin}, M., {et~al.} 2021, \mnras, 500, 4901

\bibitem[{{Ester} {et~al.}(1996){Ester}, {Kriegel}, {Sander}, \& {Xu}}]{DBSCAN}
{Ester}, M., {Kriegel}, H.-P., {Sander}, J., \& {Xu}, X. 1996, Proceedings of the Second International Conference on Knowledge Discovery and Data Mining, KDD-96 (AAAI Press), 226

\bibitem[{{Fotopoulou}(2024)}]{Fotopoulou2024}
{Fotopoulou}, S. 2024, Astronomy and Computing, 48, 100851

\bibitem[{{Galli} {et~al.}(2019){Galli}, {Loinard}, {Bouy}, {Sarro}, {Ortiz-Le{\'o}n}, {Dzib}, {Olivares}, {Heyer}, {Hernandez}, {Rom{\'a}n-Z{\'u}{\~n}iga}, {Kounkel}, \& {Covey}}]{Galli2019}
{Galli}, P.~A.~B., {Loinard}, L., {Bouy}, H., {et~al.} 2019, \aap, 630, A137

\bibitem[{{Galli} {et~al.}(2018){Galli}, {Loinard}, {Ortiz-L{\'e}on}, {Kounkel}, {Dzib}, {Mioduszewski}, {Rodr{\'\i}guez}, {Hartmann}, {Teixeira}, {Torres}, {Rivera}, {Boden}, {Evans}, {Brice{\~n}o}, {Tobin}, \& {Heyer}}]{Galli2018}
{Galli}, P. A.~B., {Loinard}, L., {Ortiz-L{\'e}on}, G.~N., {et~al.} 2018, \apj, 859, 33

\bibitem[{{Ginsburg} {et~al.}(2019){Ginsburg}, {Koch}, {Robitaille}, {Beaumont}, {Adamginsburg}, {ZuHone}, {Sipocz}, {Patra}, {Jones}, {Lim}, {Rosolowsky}, {Stern}, {Earl}, {De Val-Borro}, {Jrobbfed}, {Shuokong}, {Kepley}, {Sokolov}, {Badger}, {Maret}, {Garrido}, {Booker}, \& {Tollerud}}]{Spectralcube}
{Ginsburg}, A., {Koch}, E., {Robitaille}, T., {et~al.} 2019, {radio-astro-tools/spectral-cube: v0.4.4}

\bibitem[{{Hirota} {et~al.}(2002){Hirota}, {Ito}, \& {Yamamoto}}]{Hirota2002}
{Hirota}, T., {Ito}, T., \& {Yamamoto}, S. 2002, \apj, 565, 359

\bibitem[{{Keto} \& {Caselli}(2008)}]{KetoCaselli2008}
{Keto}, E. \& {Caselli}, P. 2008, \apj, 683, 238

\bibitem[{{Keto} {et~al.}(2015){Keto}, {Caselli}, \& {Rawlings}}]{Keto2015}
{Keto}, E., {Caselli}, P., \& {Rawlings}, J. 2015, \mnras, 446, 3731

\bibitem[{{Lada} {et~al.}(2003){Lada}, {Bergin}, {Alves}, \& {Huard}}]{Lada2003}
{Lada}, C.~J., {Bergin}, E.~A., {Alves}, J.~F., \& {Huard}, T.~L. 2003, \apj, 586, 286

\bibitem[{{Lee} {et~al.}(2001){Lee}, {Myers}, \& {Tafalla}}]{Lee2001}
{Lee}, C.~W., {Myers}, P.~C., \& {Tafalla}, M. 2001, \apjs, 136, 703

\bibitem[{{Lin} {et~al.}(2022){Lin}, {Spezzano}, {Sipil{\"a}}, {Vasyunin}, \& {Caselli}}]{Lin2022}
{Lin}, Y., {Spezzano}, S., {Sipil{\"a}}, O., {Vasyunin}, A., \& {Caselli}, P. 2022, \aap, 665, A131

\bibitem[{{Mangum} \& {Shirley}(2015)}]{Mangum2015}
{Mangum}, J.~G. \& {Shirley}, Y.~L. 2015, \pasp, 127, 266

\bibitem[{McInnes {et~al.}(2017)McInnes, Healy, \& Astels}]{HDBSCANpython}
McInnes, L., Healy, J., \& Astels, S. 2017, The Journal of Open Source Software, 2

\bibitem[{Moulavi {et~al.}(2014)Moulavi, Jaskowiak, Campello, Zimek, \& Sander}]{DBCV}
Moulavi, D., Jaskowiak, P.~A., Campello, R. J. G.~B., Zimek, A., \& Sander, J. 2014, Density-Based Clustering Validation, 839--847

\bibitem[{{M{\"u}ller} {et~al.}(2001){M{\"u}ller}, {Thorwirth}, {Roth}, \& {Winnewisser}}]{Mueller2001}
{M{\"u}ller}, H.~S.~P., {Thorwirth}, S., {Roth}, D.~A., \& {Winnewisser}, G. 2001, \aap, 370, L49

\bibitem[{{Nagy} {et~al.}(2019){Nagy}, {Spezzano}, {Caselli}, {Vasyunin}, {Tafalla}, {Bizzocchi}, {Prudenzano}, \& {Redaelli}}]{Nagy2019}
{Nagy}, Z., {Spezzano}, S., {Caselli}, P., {et~al.} 2019, \aap, 630, A136

\bibitem[{{Ohashi} {et~al.}(1999){Ohashi}, {Lee}, {Wilner}, \& {Hayashi}}]{Ohashi1999}
{Ohashi}, N., {Lee}, S.~W., {Wilner}, D.~J., \& {Hayashi}, M. 1999, \apjl, 518, L41

\bibitem[{{Okoda} {et~al.}(2021){Okoda}, {Oya}, {Abe}, {Komaki}, {Watanabe}, \& {Yamamoto}}]{Okoda2021}
{Okoda}, Y., {Oya}, Y., {Abe}, S., {et~al.} 2021, \apj, 923, 168

\bibitem[{{Okoda} {et~al.}(2020){Okoda}, {Oya}, {Sakai}, {Watanabe}, \& {Yamamoto}}]{Okoda2020}
{Okoda}, Y., {Oya}, Y., {Sakai}, N., {Watanabe}, Y., \& {Yamamoto}, S. 2020, \apj, 900, 40

\bibitem[{Pedregosa {et~al.}(2011)Pedregosa, Varoquaux, Gramfort, Michel, Thirion, Grisel, Blondel, Prettenhofer, Weiss, Dubourg, Vanderplas, Passos, Cournapeau, Brucher, Perrot, \& Duchesnay}]{scikit-learn}
Pedregosa, F., Varoquaux, G., Gramfort, A., {et~al.} 2011, Journal of Machine Learning Research, 12, 2825

\bibitem[{{Pety}(2005)}]{Pety2005}
{Pety}, J. 2005, in SF2A-2005: Semaine de l'Astrophysique Francaise, ed. F.~{Casoli}, T.~{Contini}, J.~M. {Hameury}, \& L.~{Pagani}, 721

\bibitem[{{Punanova} {et~al.}(2018){Punanova}, {Caselli}, {Feng}, {Chac{\'o}n-Tanarro}, {Ceccarelli}, {Neri}, {Fontani}, {Jim{\'e}nez-Serra}, {Vastel}, {Bizzocchi}, {Pon}, {Vasyunin}, {Spezzano}, {Hily-Blant}, {Testi}, {Viti}, {Yamamoto}, {Alves}, {Bachiller}, {Balucani}, {Bianchi}, {Bottinelli}, {Caux}, {Choudhury}, {Codella}, {Dulieu}, {Favre}, {Holdship}, {Jaber Al-Edhari}, {Kahane}, {Laas}, {LeFloch}, {L{\'o}pez-Sepulcre}, {Ospina-Zamudio}, {Oya}, {Pineda}, {Podio}, {Quenard}, {Rimola}, {Sakai}, {Sims}, {Taquet}, {Theul{\'e}}, \& {Ugliengo}}]{Punanova2018}
{Punanova}, A., {Caselli}, P., {Feng}, S., {et~al.} 2018, \apj, 855, 112

\bibitem[{{Redaelli} {et~al.}(2021){Redaelli}, {Sipil{\"a}}, {Padovani}, {Caselli}, {Galli}, \& {Ivlev}}]{Redaelli2021}
{Redaelli}, E., {Sipil{\"a}}, O., {Padovani}, M., {et~al.} 2021, \aap, 656, A109

\bibitem[{{Semenov} {et~al.}(2010){Semenov}, {Hersant}, {Wakelam}, {Dutrey}, {Chapillon}, {Guilloteau}, {Henning}, {Launhardt}, {Pi{\'e}tu}, \& {Schreyer}}]{Semenov2010}
{Semenov}, D., {Hersant}, F., {Wakelam}, V., {et~al.} 2010, \aap, 522, A42

\bibitem[{{Sipil{\"a}} {et~al.}(2015){Sipil{\"a}}, {Caselli}, \& {Harju}}]{Sipila2015}
{Sipil{\"a}}, O., {Caselli}, P., \& {Harju}, J. 2015, \aap, 578, A55

\bibitem[{{Soler} {et~al.}(2013){Soler}, {Hennebelle}, {Martin}, {Miville-Desch{\^e}nes}, {Netterfield}, \& {Fissel}}]{Soler2013}
{Soler}, J.~D., {Hennebelle}, P., {Martin}, P.~G., {et~al.} 2013, \apj, 774, 128

\bibitem[{{Spezzano} {et~al.}(2016){Spezzano}, {Bizzocchi}, {Caselli}, {Harju}, \& {Br{\"u}nken}}]{Spezzano2016}
{Spezzano}, S., {Bizzocchi}, L., {Caselli}, P., {Harju}, J., \& {Br{\"u}nken}, S. 2016, \aap, 592, L11

\bibitem[{{Spezzano} {et~al.}(2017){Spezzano}, {Caselli}, {Bizzocchi}, {Giuliano}, \& {Lattanzi}}]{Spezzano2017}
{Spezzano}, S., {Caselli}, P., {Bizzocchi}, L., {Giuliano}, B.~M., \& {Lattanzi}, V. 2017, \aap, 606, A82

\bibitem[{{Spezzano} {et~al.}(2020){Spezzano}, {Caselli}, {Pineda}, {Bizzocchi}, {Prudenzano}, \& {Nagy}}]{Spezzano2020}
{Spezzano}, S., {Caselli}, P., {Pineda}, J.~E., {et~al.} 2020, \aap, 643, A60

\bibitem[{{Tafalla} \& {Santiago}(2004)}]{TafallaSantiago2004}
{Tafalla}, M. \& {Santiago}, J. 2004, \aap, 414, L53

\bibitem[{{Thaddeus} {et~al.}(1985){Thaddeus}, {Vrtilek}, \& {Gottlieb}}]{Thaddeus1985}
{Thaddeus}, P., {Vrtilek}, J.~M., \& {Gottlieb}, C.~A. 1985, \apjl, 299, L63

\bibitem[{{Valdivia-Mena} {et~al.}(2023){Valdivia-Mena}, {Pineda}, {Segura-Cox}, {Caselli}, {Schmiedeke}, {Choudhury}, {Offner}, {Neri}, {Goodman}, \& {Fuller}}]{ValdiviaMena2023}
{Valdivia-Mena}, M.~T., {Pineda}, J.~E., {Segura-Cox}, D.~M., {et~al.} 2023, \aap, 677, A92

\bibitem[{{Wakelam} {et~al.}(2015){Wakelam}, {Loison}, {Herbst}, {Pavone}, {Bergeat}, {B{\'e}roff}, {Chabot}, {Faure}, {Galli}, {Geppert}, {Gerlich}, {Gratier}, {Harada}, {Hickson}, {Honvault}, {Klippenstein}, {Le Picard}, {Nyman}, {Ruaud}, {Schlemmer}, {Sims}, {Talbi}, {Tennyson}, \& {Wester}}]{WakelamKIDA2015}
{Wakelam}, V., {Loison}, J.~C., {Herbst}, E., {et~al.} 2015, \apjs, 217, 20

\bibitem[{{W}es {M}c{K}inney(2010)}]{pandas}
{W}es {M}c{K}inney. 2010, in {P}roceedings of the 9th {P}ython in {S}cience {C}onference, ed. {S}t\'efan van~der {W}alt \& {J}arrod {M}illman, 56 -- 61

\bibitem[{{Williams} {et~al.}(1999){Williams}, {Myers}, {Wilner}, \& {Di Francesco}}]{Williams1999}
{Williams}, J.~P., {Myers}, P.~C., {Wilner}, D.~J., \& {Di Francesco}, J. 1999, \apjl, 513, L61

\bibitem[{{Xu} \& {Lovas}(1997)}]{XuLovas1997}
{Xu}, L.-H. \& {Lovas}, F. 1997, J. Phys. Chem. Ref. Data, 26, 17

\bibitem[{{Yan} {et~al.}(2022){Yan}, {Yang}, {Su}, {Sun}, {Zhou}, {Xu}, {Wang}, {Zhang}, \& {Chen}}]{Yan2022}
{Yan}, Q.-Z., {Yang}, J., {Su}, Y., {et~al.} 2022, \aj, 164, 55

\bibitem[{{Yun} \& {Lee}(2023)}]{YunLee2023}
{Yun}, H.-S. \& {Lee}, J.-E. 2023, \apj, 958, 113

\end{thebibliography}
%

\FloatBarrier

\begin{appendix}
\onecolumn

\section{Integrated intensity maps}\label{app:intintmaps}

\begin{figure}[h!]
    \centering
    \includegraphics[width=0.33\textwidth]{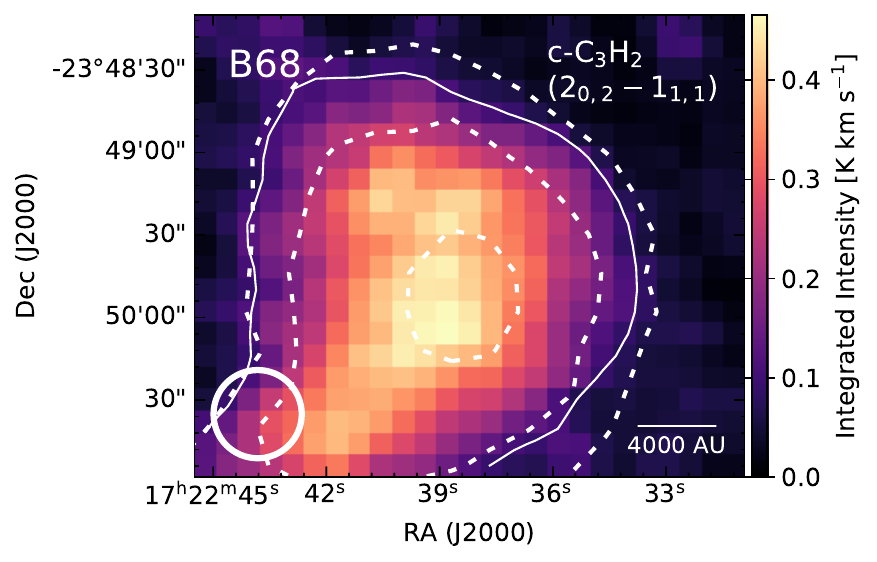}
    \includegraphics[width=0.33\textwidth]{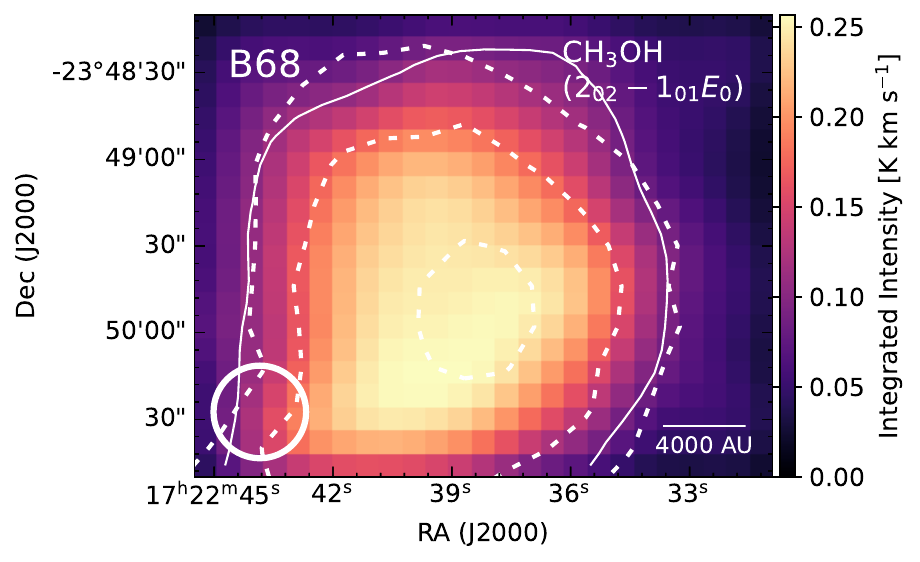}
    \includegraphics[width=0.33\textwidth]{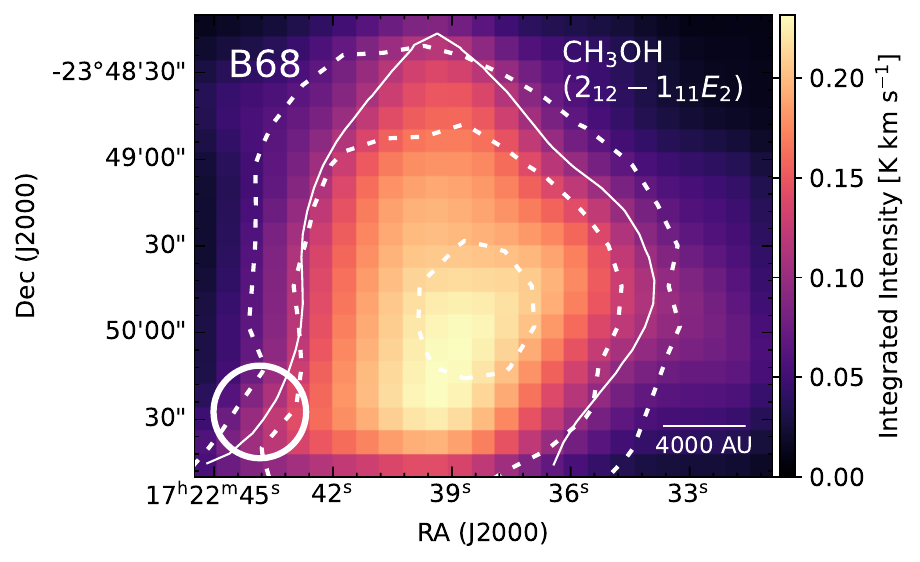}
    \includegraphics[width=0.33\textwidth]{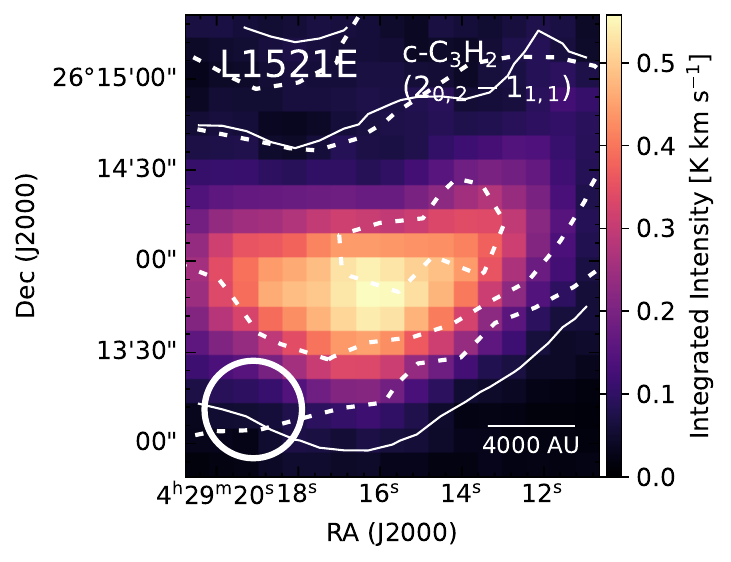}
    \includegraphics[width=0.33\textwidth]{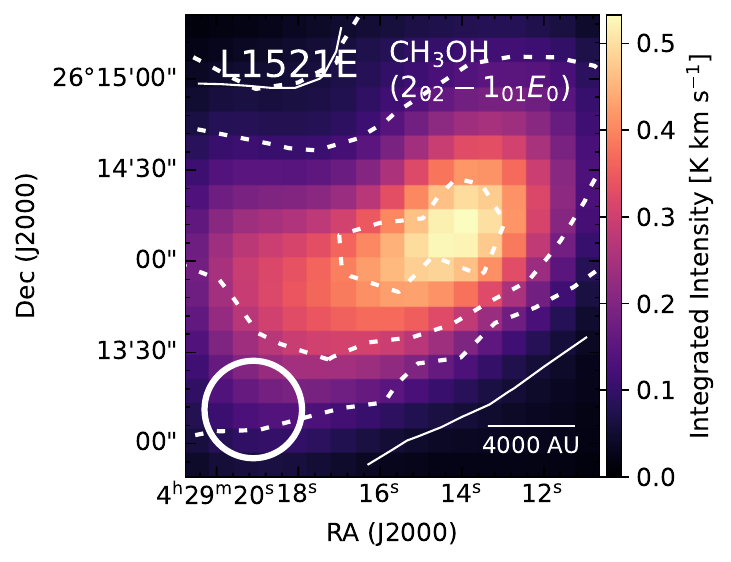}
    \includegraphics[width=0.33\textwidth]{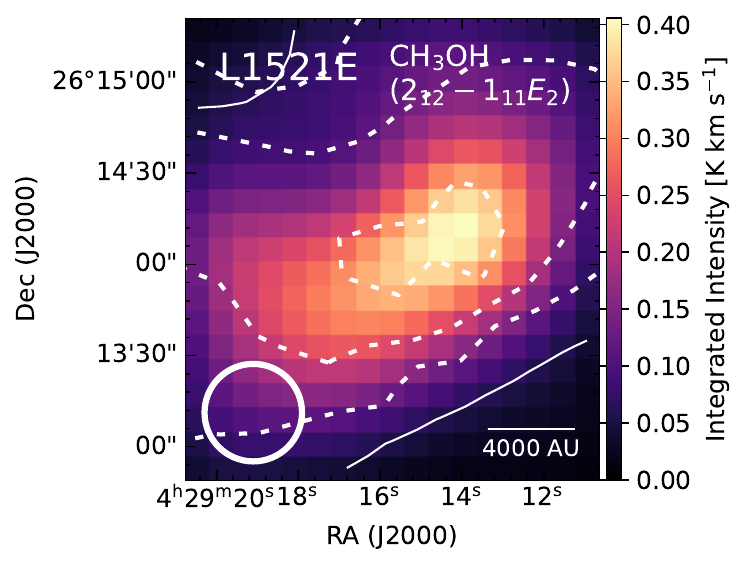}
    \includegraphics[width=0.33\textwidth]{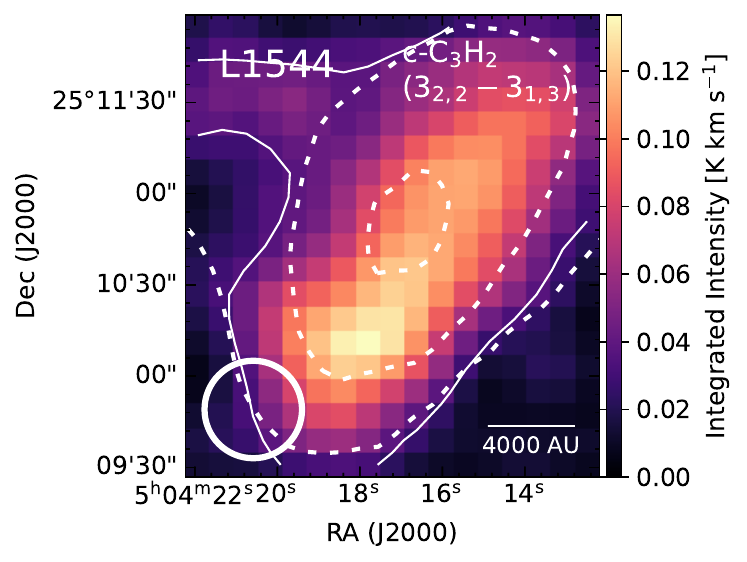}
    \includegraphics[width=0.33\textwidth]{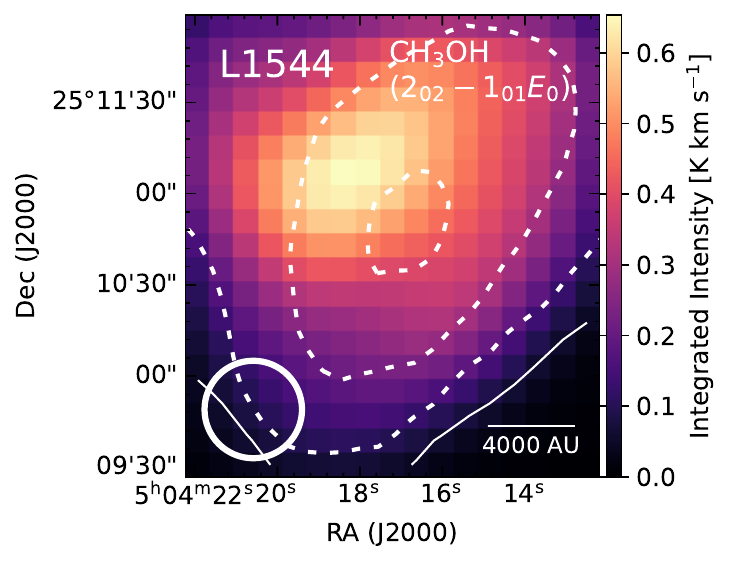}
    \includegraphics[width=0.33\textwidth]{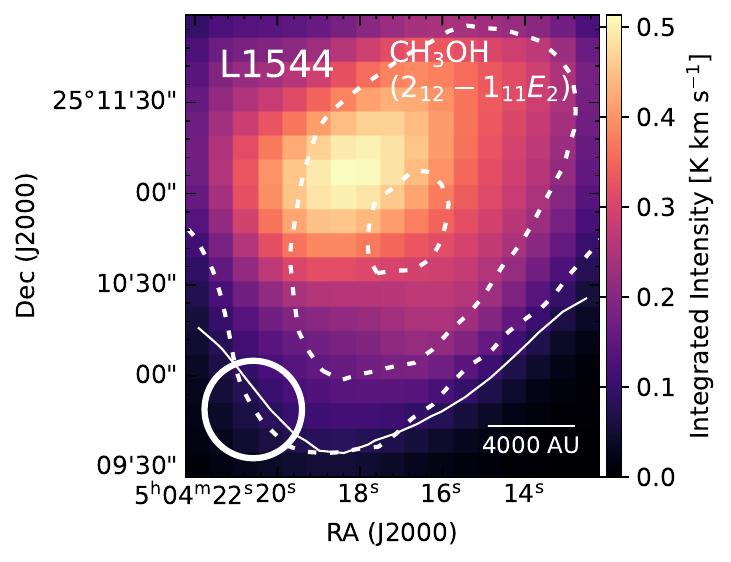}
    \caption[alt={Integrated intensity maps of \cyc\, and \meth\, observed towards B68, L1521E, and L1544.}]{Integrated intensity maps of \cyc\, and \meth\, observed towards B68 \citep{Spezzano2020}, L1521E \citep{Nagy2019}, and L1544 \citep{Spezzano2016}. The solid line contours indicate the 3$\sigma$ level of the integrated intensity, except for \meth\, in L1544, where they indicate the 9$\sigma$ level. The dashed line contours represent 90\%, 50\%, and 30\% of the H$_2$ column density peak derived from \textit{Herschel} maps \citep{Spezzano2020}. The white circle in the bottom-left corner indicates the beam size of the IRAM 30\,m telescope (32"). }
    \label{fig:intmapsC3H2CH3OH}
\end{figure}

\begin{figure}[h!]
    \includegraphics[width=0.33\textwidth]{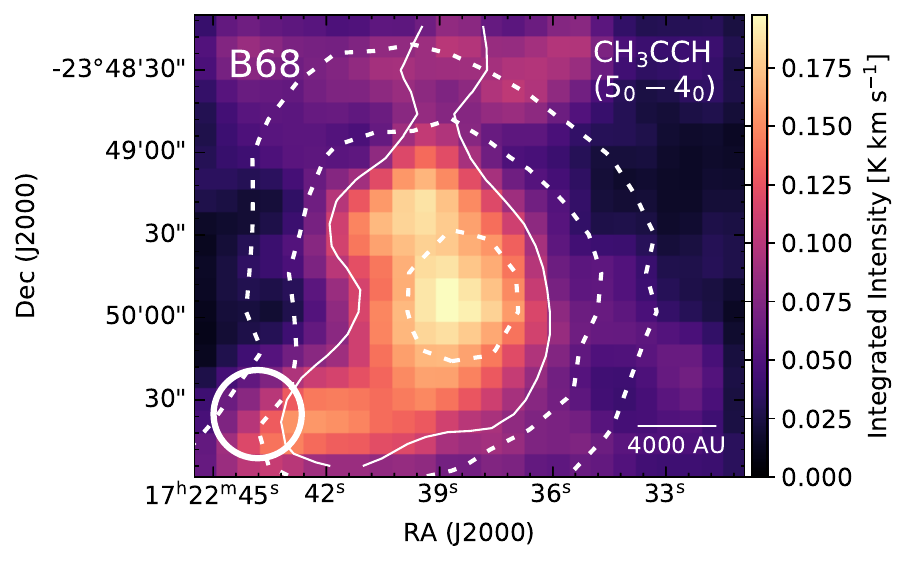}
    \includegraphics[width=0.33\textwidth]{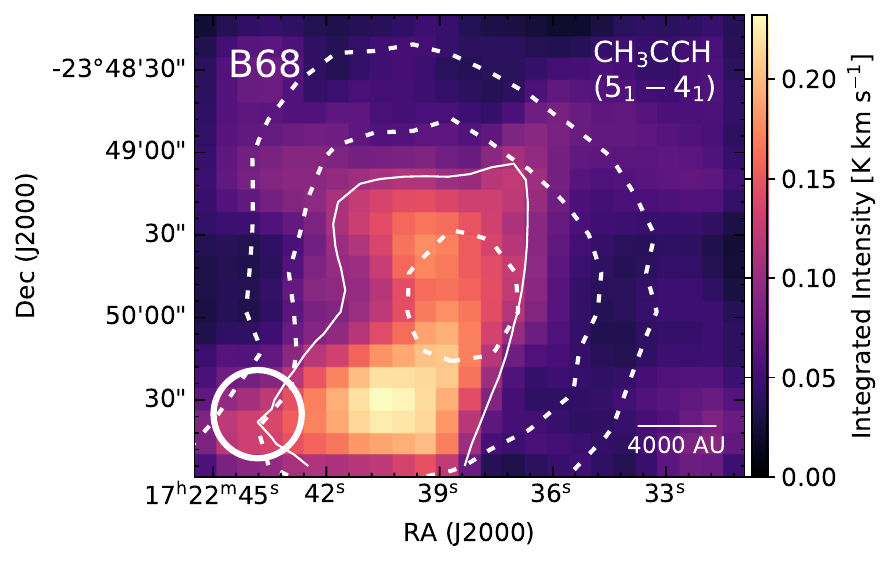}
    \includegraphics[width=0.33\textwidth]{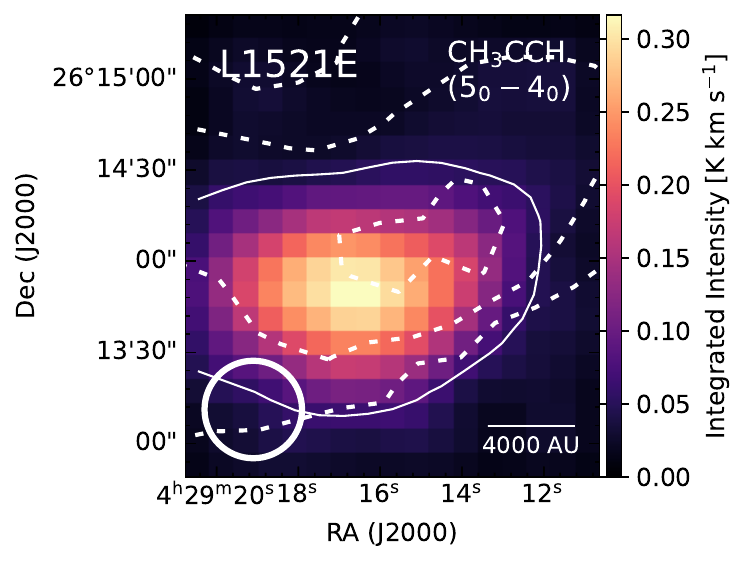}
    \includegraphics[width=0.33\textwidth]{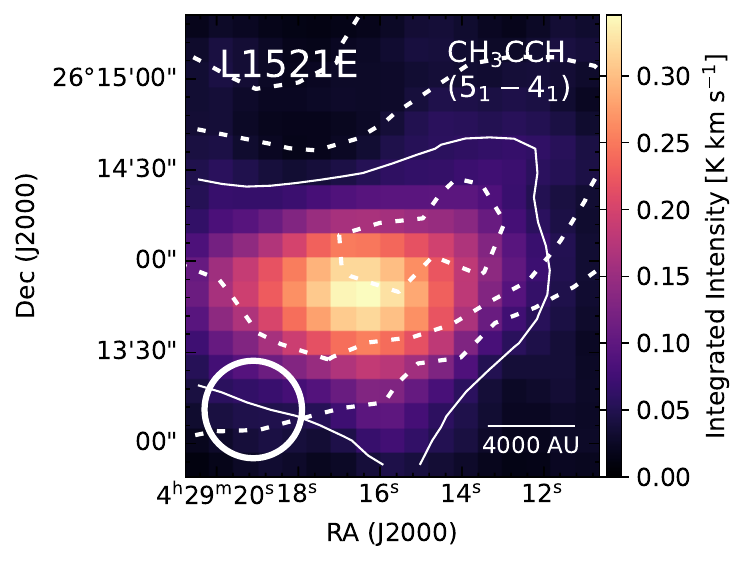}
    \includegraphics[width=0.33\textwidth]{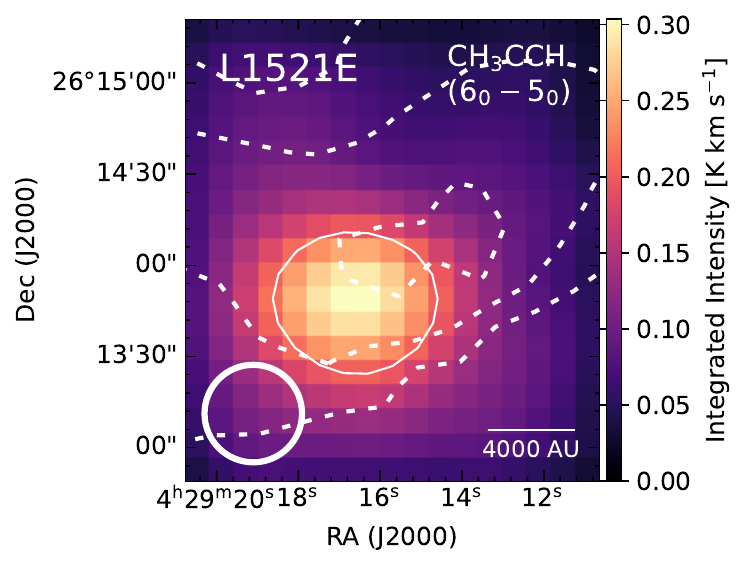}
    \includegraphics[width=0.33\textwidth]{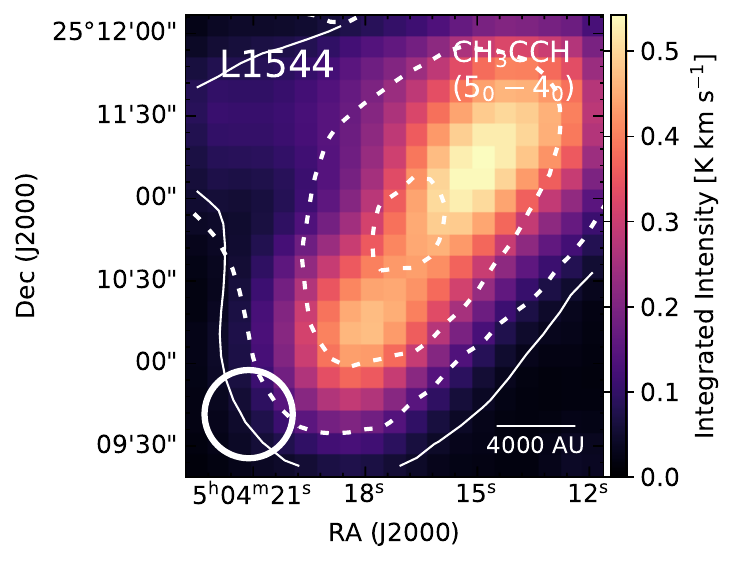}
    \includegraphics[width=0.33\textwidth]{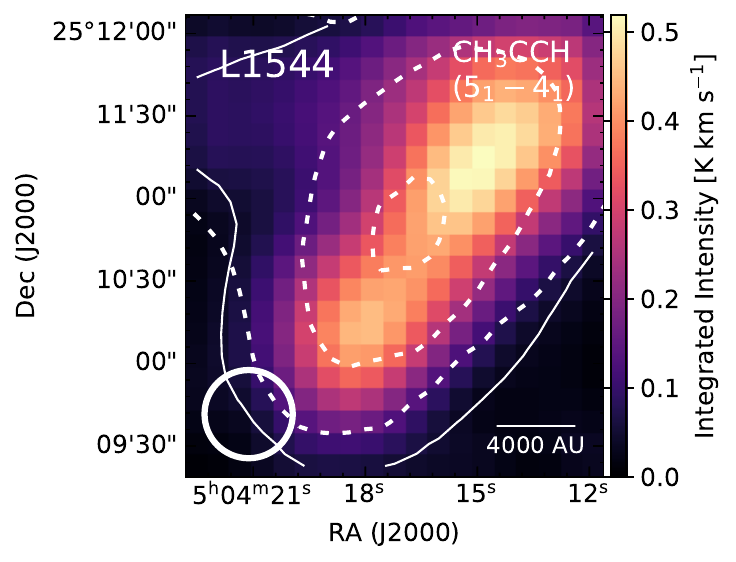}
    \includegraphics[width=0.33\textwidth]{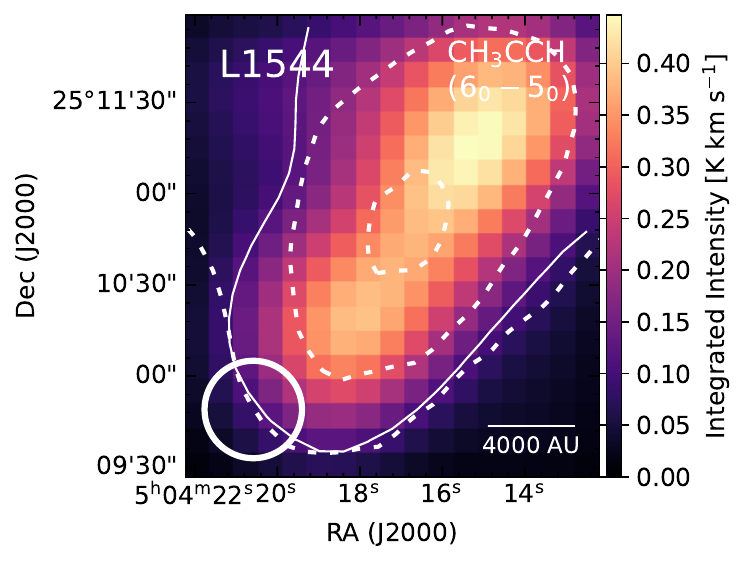}
    \includegraphics[width=0.33\textwidth]{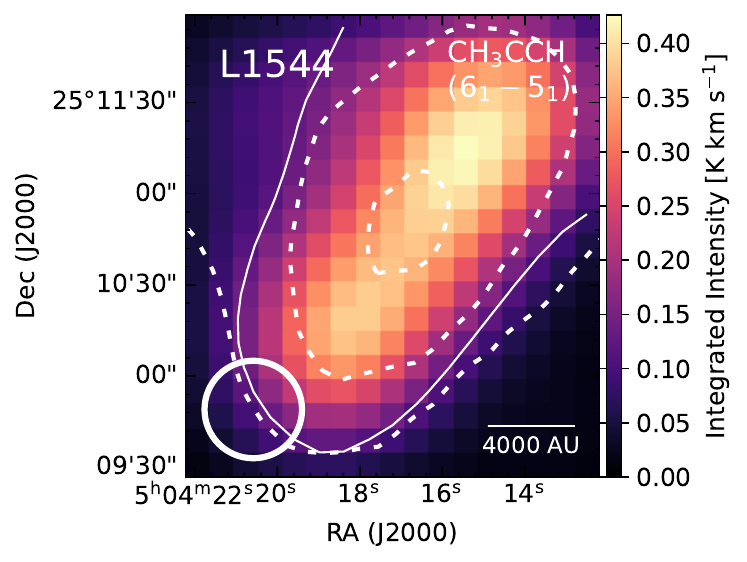}
    \caption[alt={Integrated intensity maps of \prop\, observed towards B68, L1521E, and L1544.}]{Integrated intensity maps of \prop\, observed towards B68, L1521E \citep{Nagy2019}, and L1544 \citep{Spezzano2017}. The solid line contours indicate the 3$\sigma$ level of the integrated intensity for B68 and L1521E, and the 6$\sigma$ level for L1544. The dashed line contours represent 90\%, 50\%, and 30\% of the H$_2$ column density peak derived from \textit{Herschel} maps \citep{Spezzano2020}. The white circle in the bottom-left corner indicates the beam size of the IRAM 30\,m telescope (32").}
    \label{fig:intmapsCH3CCH}
\end{figure}

The integrated intensity maps observed towards B68, L1521E, and L1544 are shown in Fig.~\ref{fig:intmapsC3H2CH3OH} (\cyc\, and \meth\,), and in Fig.~\ref{fig:intmapsCH3CCH} (\prop\,).

\FloatBarrier

\section{H$_2$ column density gradient maps} \label{app:NH2gradient}

\begin{figure*}[h!]
    \centering
    \includegraphics[width=0.9\textwidth]{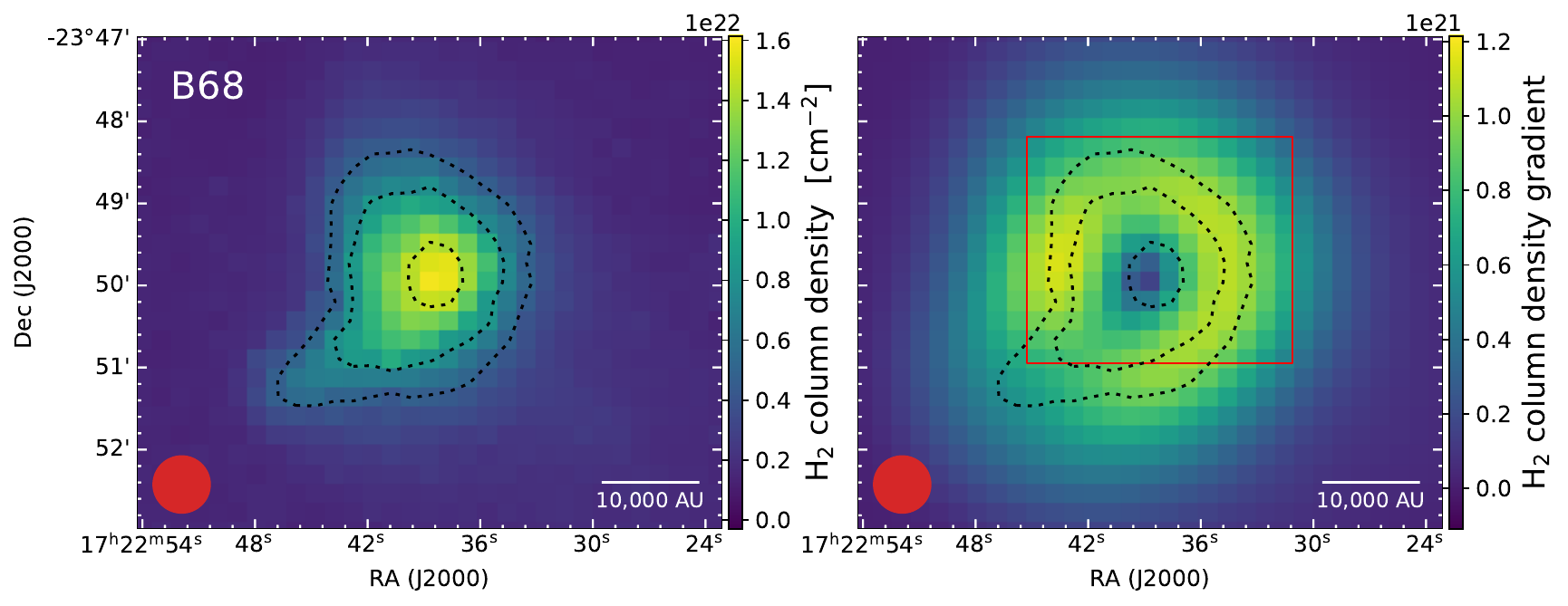}
    \includegraphics[width=0.9\textwidth]{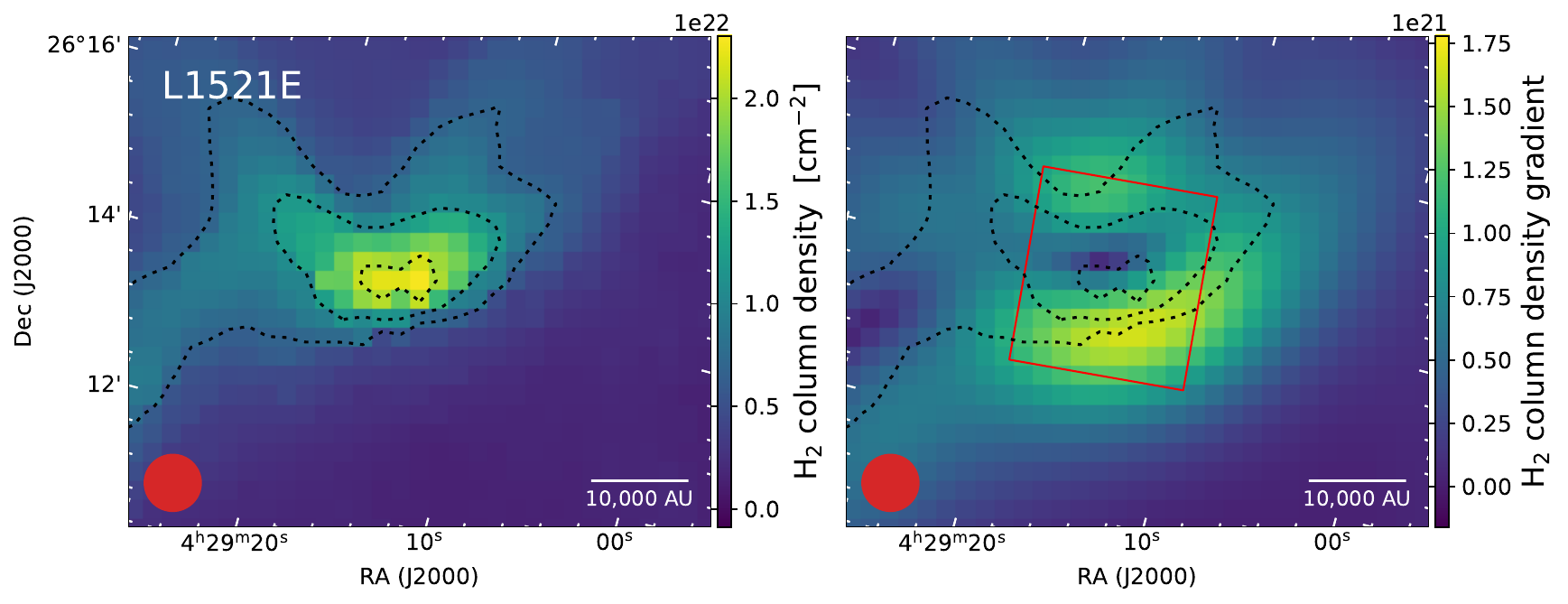}
    \includegraphics[width=0.9\textwidth]{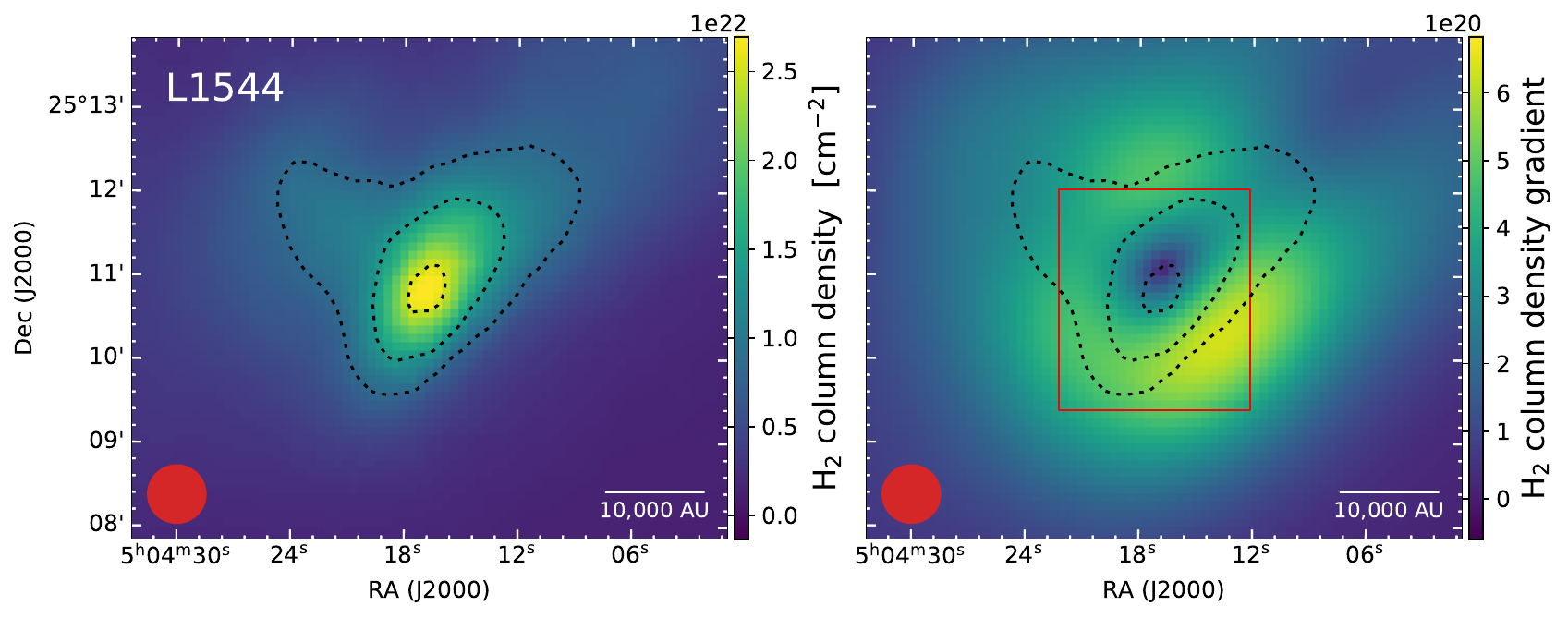}
    \caption[alt={H$_2$ column density maps derived from Herschel SPIRE maps and the corresponding H$_2$ column density gradient maps for B68, L1521E, and L1544.}]{H$_2$ column density maps (\textit{left}) derived from \textit{Herschel} SPIRE maps \citep{Spezzano2016,Spezzano2020} and the corresponding H$_2$ column density gradient maps (\textit{right}) for B68 (\textit{top}), L1521E (\textit{middle}), and L1544 (\textit{bottom}). The dashed line contours represent 90\%, 50\%, and 30\% of the H$_2$ column density peak. The red rectangle marks the location and size of the emission maps observed towards each core. The red circle in the bottom-left corner indicates the \textit{Herschel} beam size (40").}
    \label{fig:NH2maps}
\end{figure*}

Figure~\ref{fig:NH2maps} shows the H$_2$ column density gradient maps derived from \textit{Herschel} SPIRE maps \citep[see][]{Spezzano2016,Spezzano2020} and the corresponding H$_2$ column density maps, for B68, L1521E, and L1544, respectively.

\FloatBarrier
\section{Detailed clustering results}\label{app:clusteringresults}

\begin{figure*}[h!]
    \centering
    \includegraphics[width=0.49\textwidth]{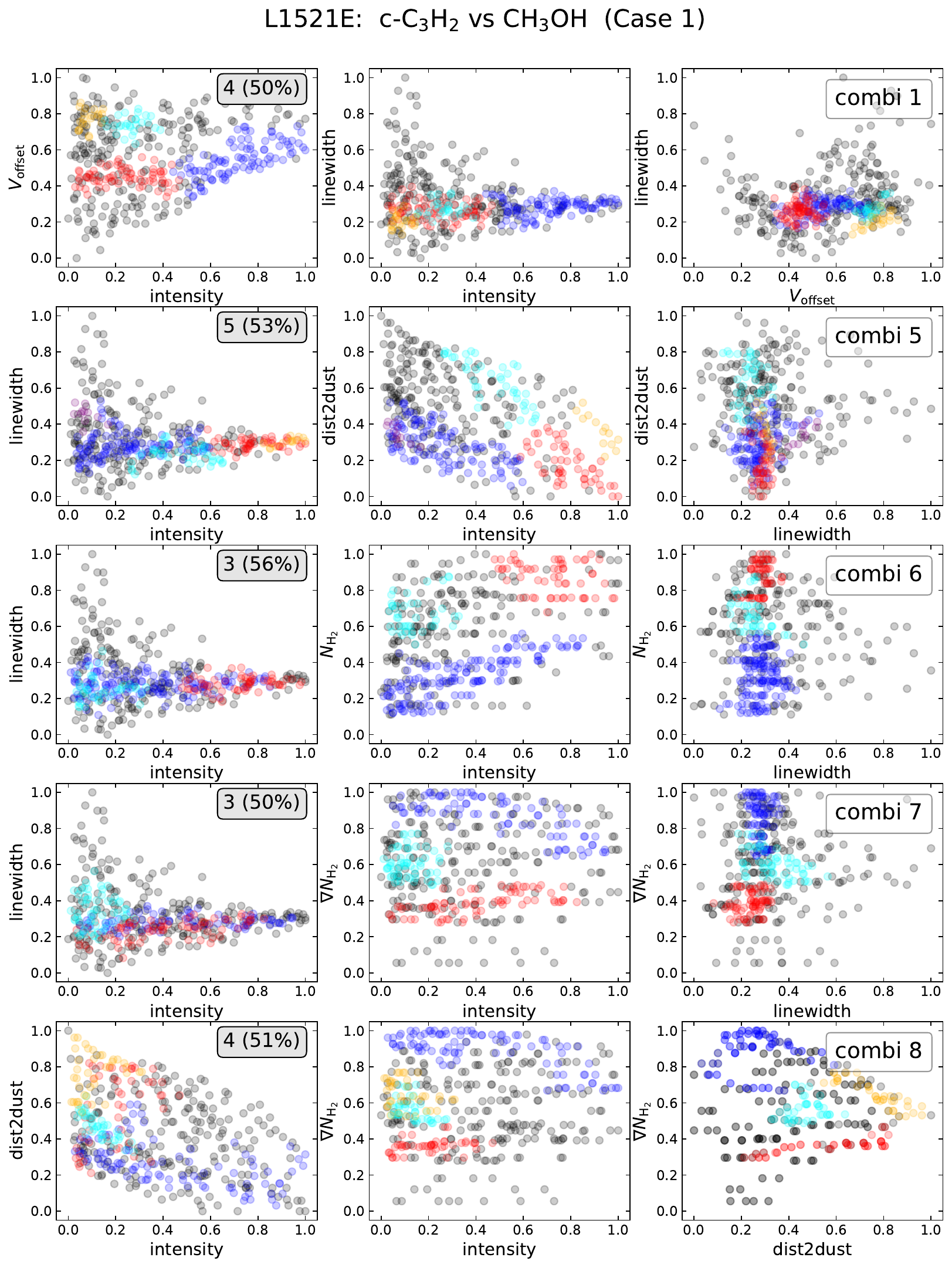}
    \includegraphics[width=0.49\textwidth]{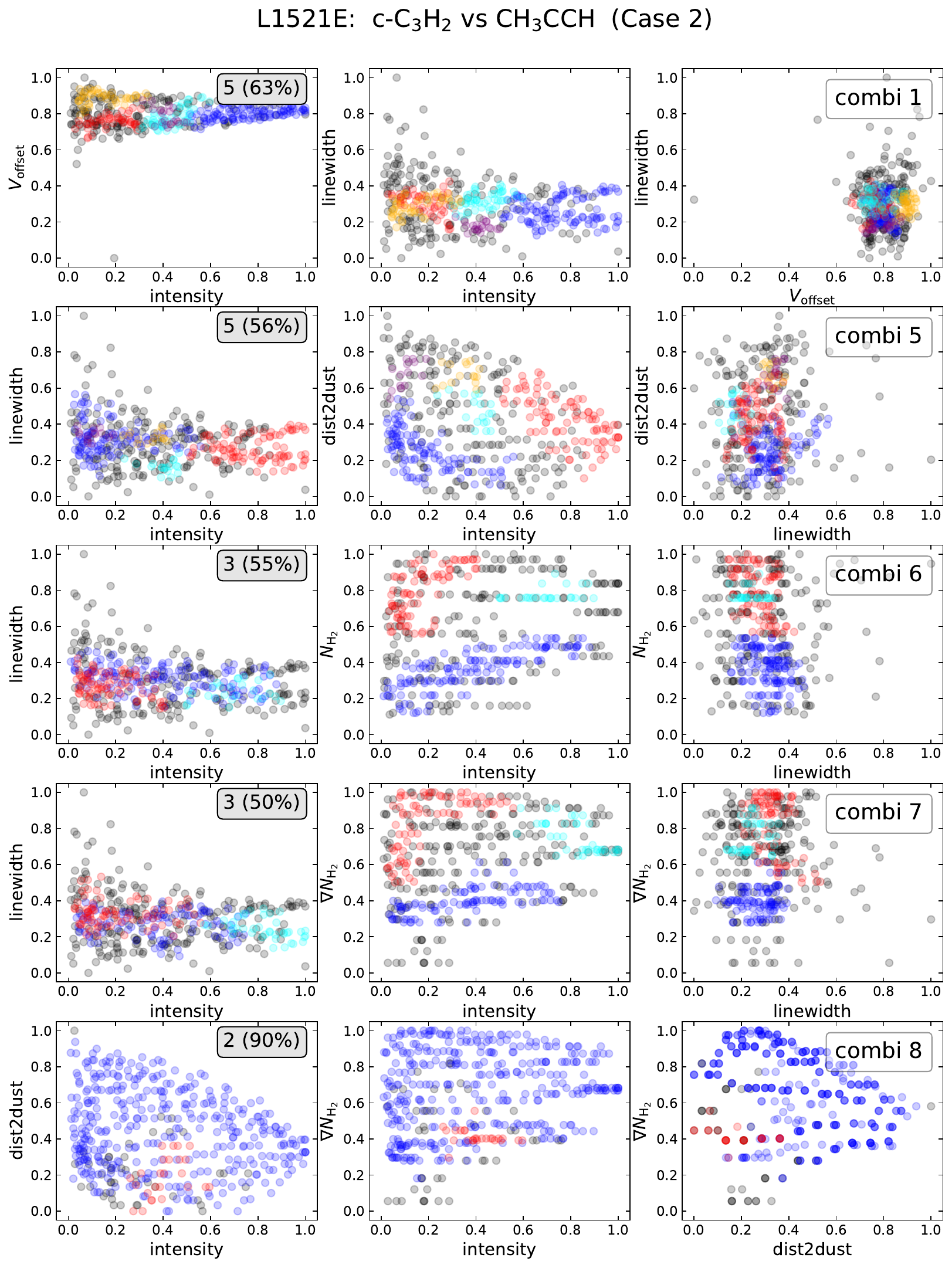}
    \includegraphics[width=0.45\textwidth]{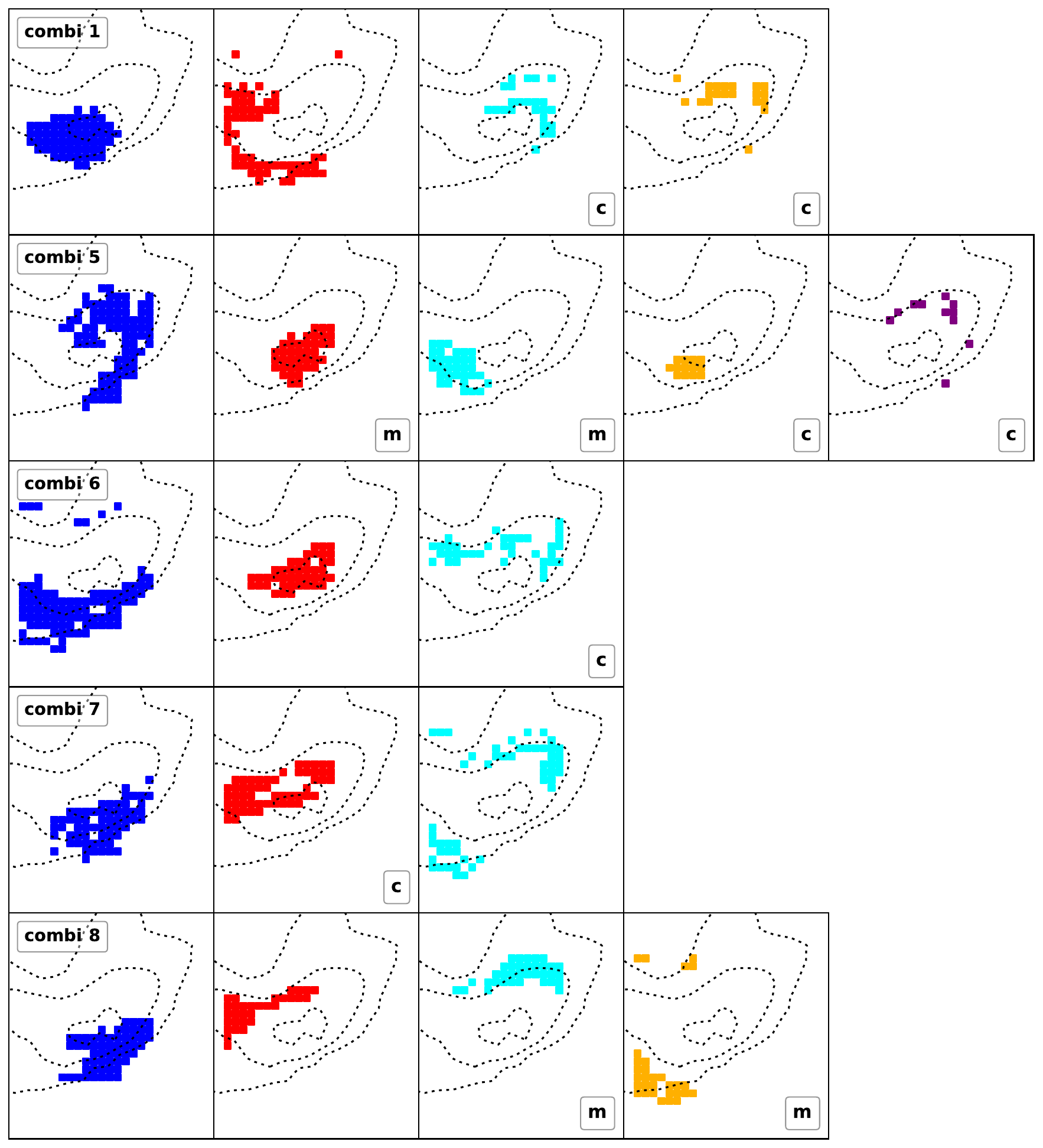}
    \includegraphics[width=0.45\textwidth]{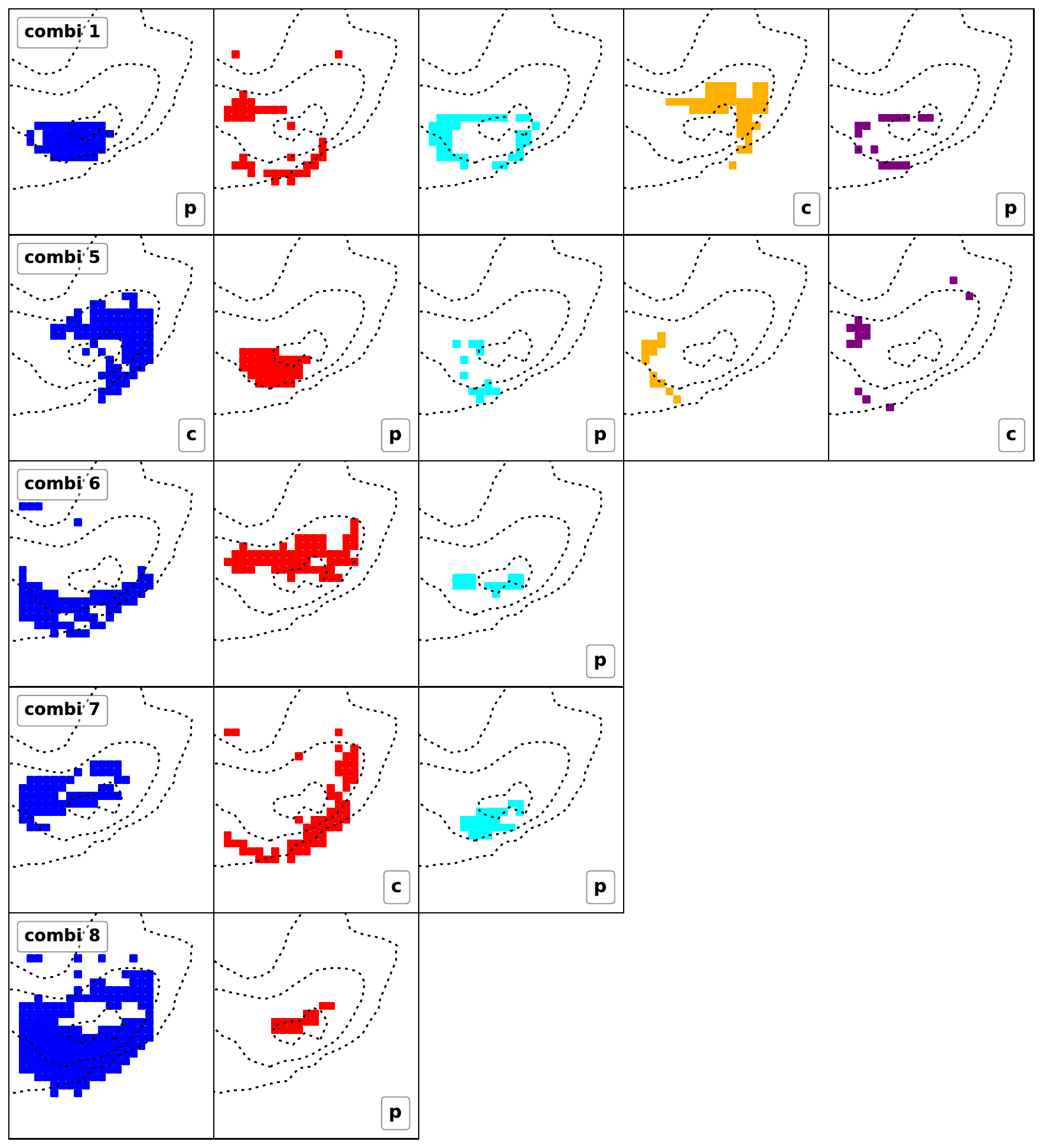}
    \caption[alt={Clustering results for the starless core L1521E for the dataset of Case 1 and Case 2, for feature combinations 2, 3, 4, 9, and 10. The top part shows the distribution of the resulting clusters in the input features. The bottom part shows the corresponding spatial distribution of each cluster across the core.}]{Clustering results for the starless core L1521E for the dataset of Case 1 (\textit{left}) and the dataset of Case 2 (\textit{right}), for feature combinations 2, 3, 4, 9, and 10. Each row represents a different combination of features (see Table.~\ref{Tab:FeatureCombinations}): combination 2 (intensity, $V_\mathrm{offset}$, and dist2dust), combination 3 (intensity, $V_\mathrm{offset}$, and \nht\,), combination 4 (intensity, $V_\mathrm{offset}$, and \ngrad\,), combination 9 (intensity, dist2dust, and \nht\,), and combination 10 (intensity, \nht\,, \ngrad\,).
    \textit{Top}: Distribution of the resulting clusters in the input features. The annotations provide information on how many clusters were found by the algorithm (two to five) and what percentage of data points are assigned to the clusters. Noise points (=points not assigned to any cluster) are plotted in black. The colours of the clusters are ordered by cluster size: the biggest cluster is given in blue, followed by red, cyan, yellow, and purple.
    \textit{Bottom}: Corresponding spatial distribution of each cluster across the core. The annotations indicate if a cluster contains more than $60\%$ of one molecule (c: \cyc\,; m: \meth\,; p: \prop\,). The dashed line contours represent 30\%, 50\%, 90\% of the H$_2$ column density peak derived from \textit{Herschel} maps \cite{Spezzano2020}.}
    \label{fig:ClusteringL1521E}
\end{figure*}

\begin{figure*}[h!]
    \centering
    \includegraphics[width=0.49\textwidth]{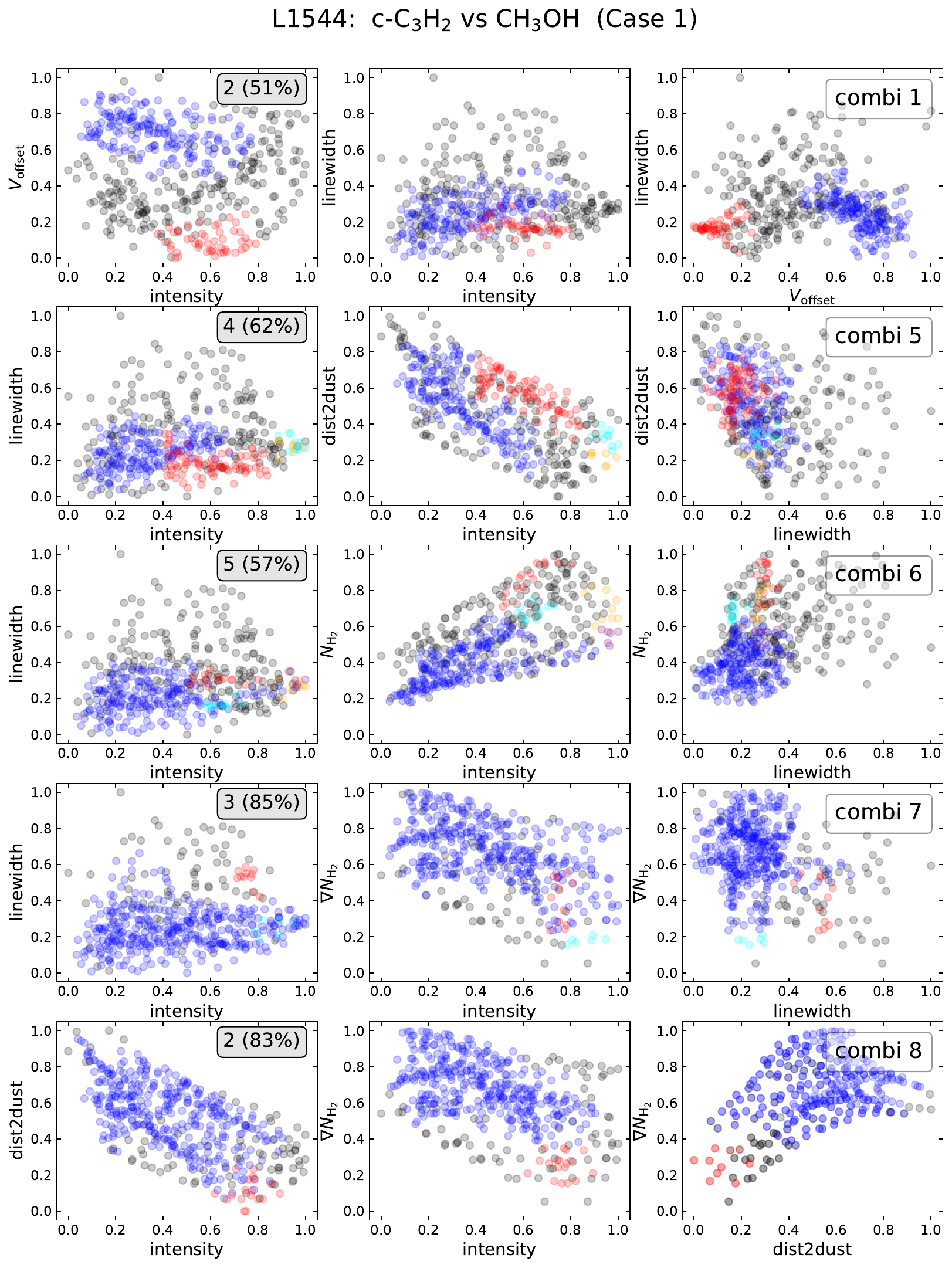}
    \includegraphics[width=0.49\textwidth]{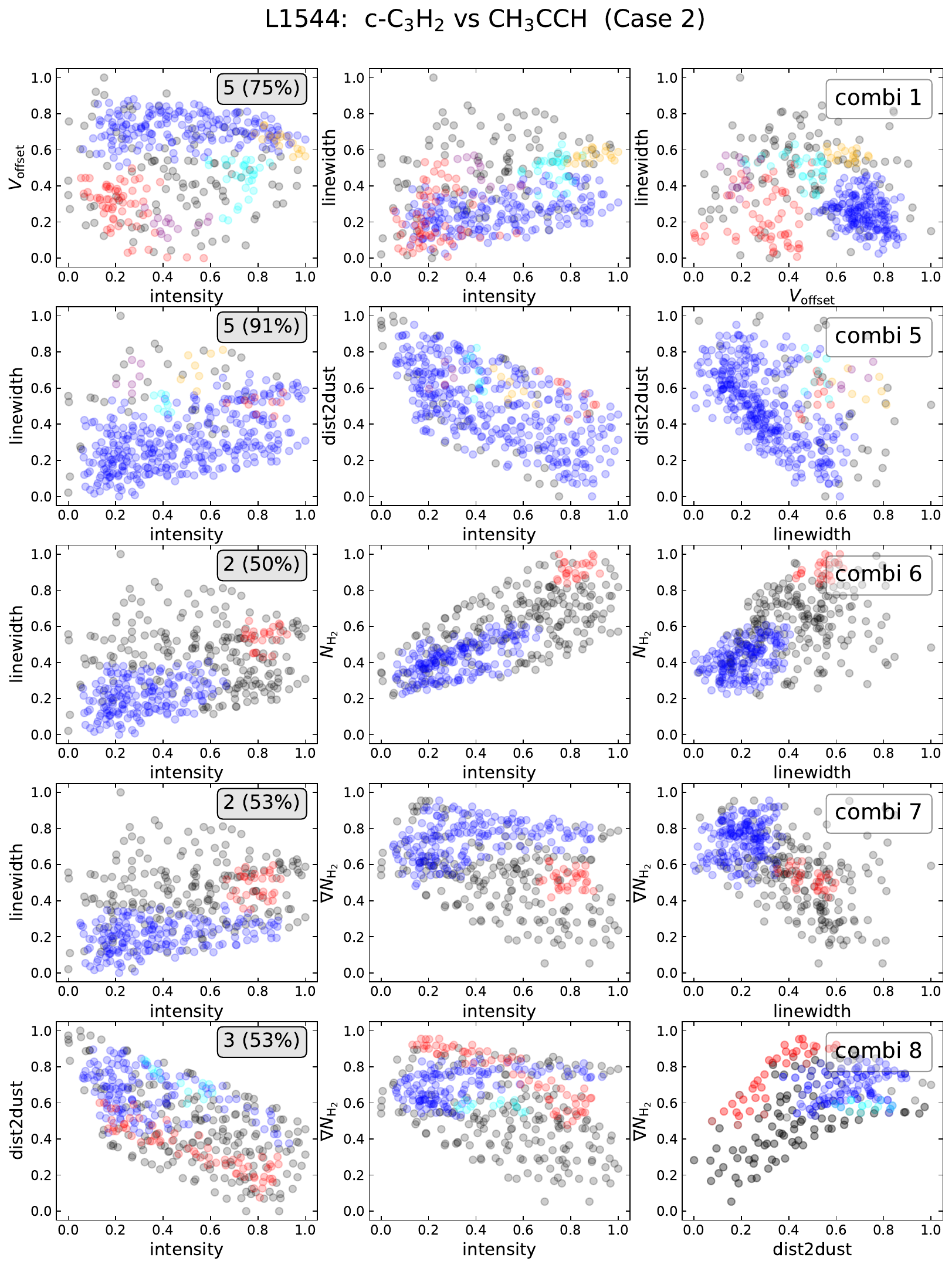}
    \includegraphics[width=0.49\textwidth]{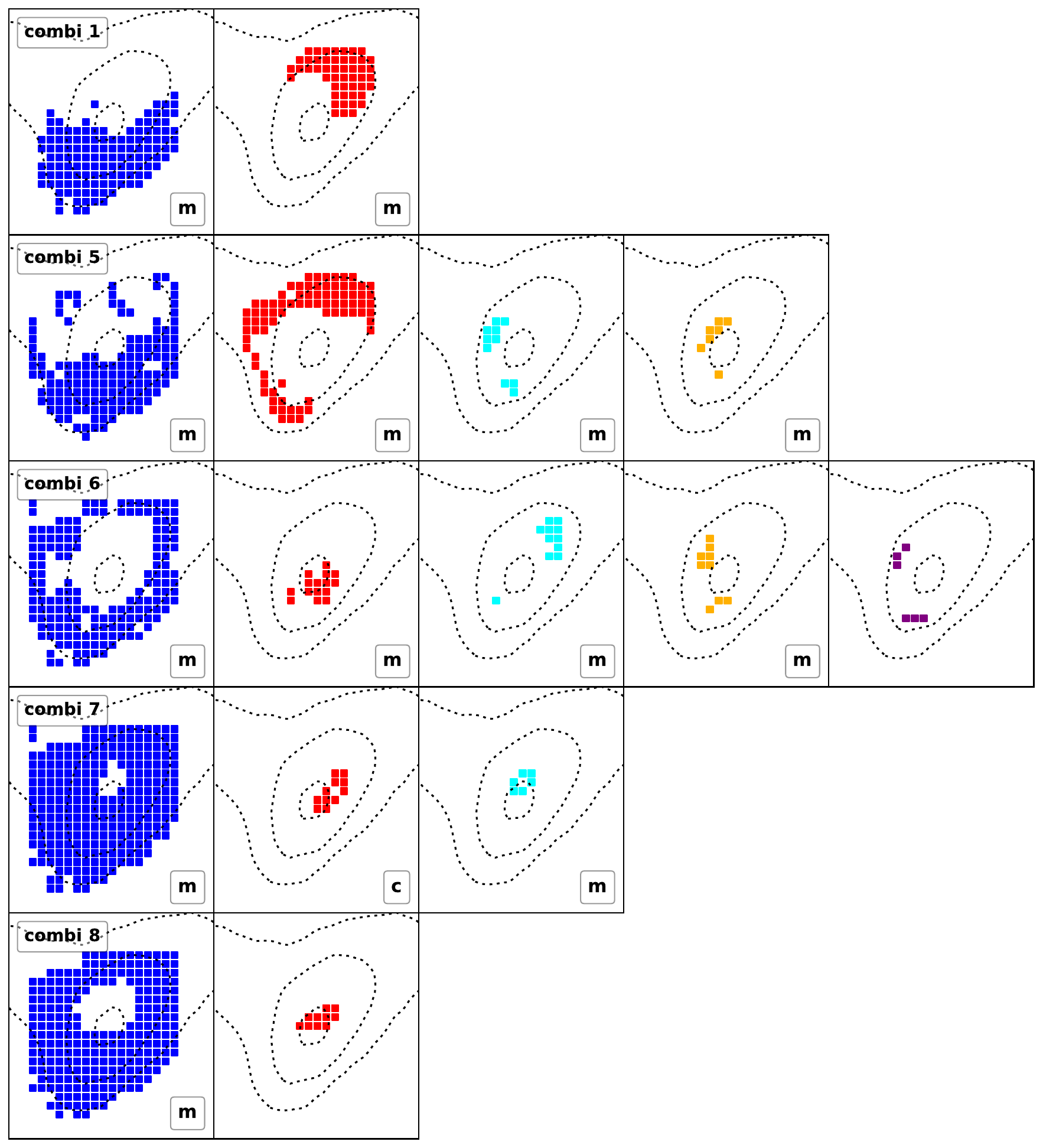}
    \includegraphics[width=0.49\textwidth]{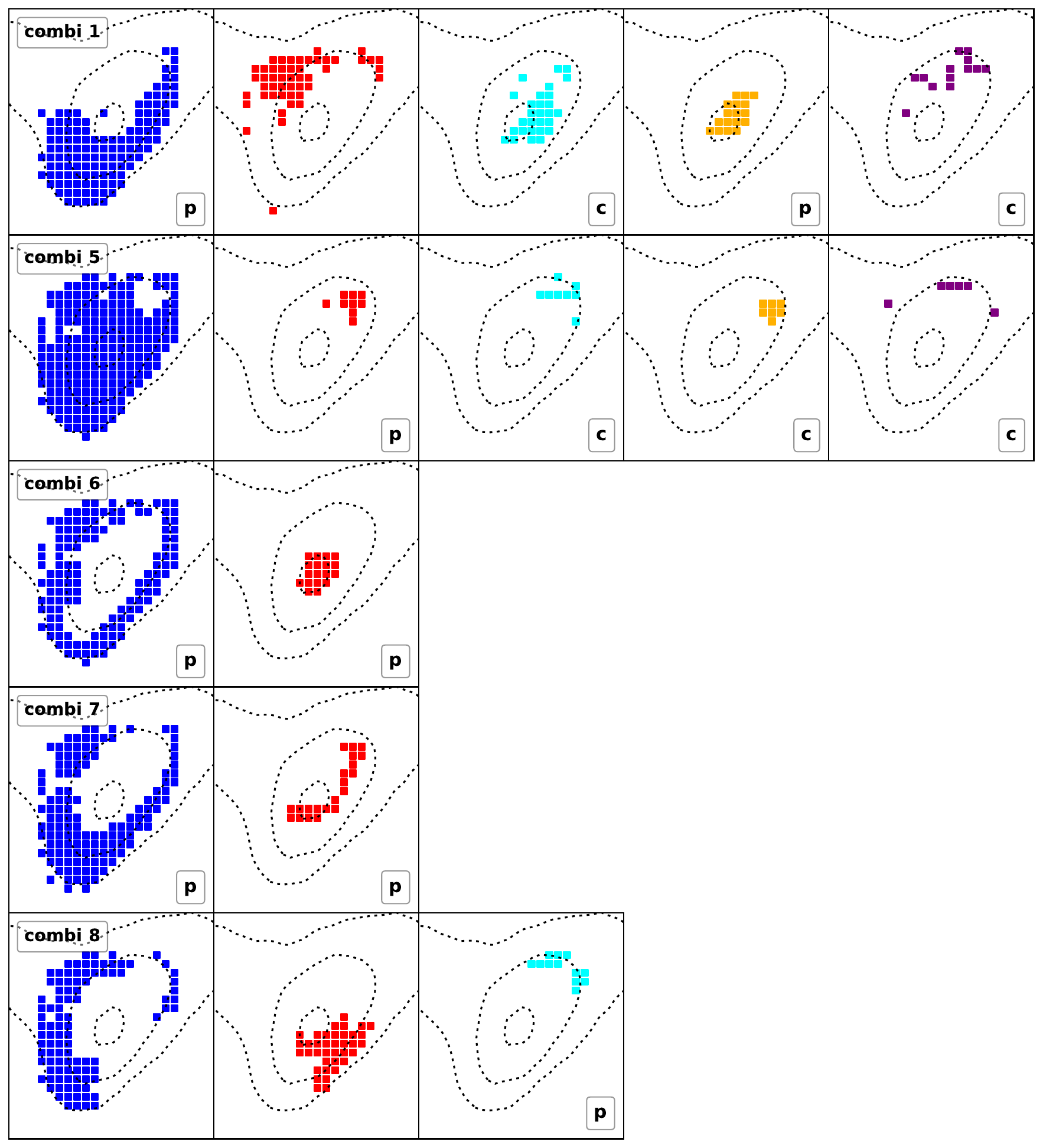}
    \caption[alt={Clustering results for the prestellar core L1544 for the dataset of Case 1 and Case 2, for feature combinations 2, 3, 4, 9, and 10. The top part shows the distribution of the resulting clusters in the input features. The bottom part shows the corresponding spatial distribution of each cluster across the core.}]{Same as in Fig.~\ref{fig:ClusteringL1521E} but for the prestellar core L1544.}
    \label{fig:ClusteringL1544}
\end{figure*}

The clustering results for Case 1 and Case 2 are visualised in Fig.~\ref{fig:ClusteringL1521E} and Fig.~\ref{fig:ClusteringL1544} for feature combinations 2, 3, 4, 9, 10 for L1521E and L1544, respectively.
The remaining results for Cases 1 and 2, together with the results for Cases 3 and 4 are published on Zenodo, in Figs.~\href{https://zenodo.org/records/15519030}{C.3-C.14}. 
The molecular ratio in a cluster and the number of data points assigned to it are given in Table~\ref{Tab:ClusterContent} for all combinations and Cases.

\begin{landscape}
{\scriptsize
\begin{longtable}{l | r r r r | r r r r | r r r r}
    \caption{Cluster content for all Cases and feature combinations.}
    \label{Tab:ClusterContent}
    
    \\\hline\hline
     & \multicolumn{4}{c}{B68} & \multicolumn{4}{c}{L1521E} & \multicolumn{4}{c}{L1544} \\
    \hline
    \endfirsthead
    \caption{continued.}\\
    \hline\hline
     & \multicolumn{4}{c}{B68} & \multicolumn{4}{c}{L1521E} & \multicolumn{4}{c}{L1544} \\
    \hline
    \endhead
    \hline
    \endfoot
     combi 1 & Case 1 (502) & Case 2 (443) & Case 3 (439) & Case 4 (693) & Case 1 (436) & Case 2 (437) & Case 3 (481) & Case 4 (677) & Case 1 (445) & Case 2 (404) & Case 3 (477) & Case 4 (663) \\
      & \#\hspace{2mm}| \hspace{2mm}\%\hspace{3mm} 
      & \#\hspace{2mm}| \hspace{2mm}\%\hspace{3mm} 
      & \#\hspace{2mm}| \hspace{2mm}\%\hspace{3mm} 
      & \#\hspace{2mm}| \hspace{3mm}\%\hspace{6mm} 
      & \#\hspace{2mm}| \hspace{2mm}\%\hspace{3mm} 
      & \#\hspace{2mm}| \hspace{2mm}\%\hspace{3mm} 
      & \#\hspace{2mm}| \hspace{2mm}\%\hspace{3mm} 
      & \#\hspace{2mm}| \hspace{3mm}\%\hspace{6mm} 
      & \#\hspace{2mm}| \hspace{2mm}\%\hspace{3mm}
      & \#\hspace{2mm}| \hspace{2mm}\%\hspace{3mm}
      & \#\hspace{2mm}| \hspace{2mm}\%\hspace{3mm}
      & \#\hspace{2mm}| \hspace{3mm}\%\hspace{6mm} \\\hline
     cluster 1 & 250 | 44/56 & 212 | 55/45 & 156 | \textbf{74/26}  & 304 | 32/27/42 
     & 97 | 52/48 & 109 | \textbf{33/67} & 104 | \textbf{38/62}  & 191 | 35/30/35 
     & 178 | 39/61 & 184 | \textbf{38/62} & 168 | \textbf{67/33}  & 298 | 26/31/43 \\
     cluster 2 &  52 | 54/46 &  26 | \textbf{85/15} &  135 | \textbf{45/55} & 39 | 38/41/21 
     & 71 | \textbf{58/42} &  49 | \textbf{55/45} & 96 | \textbf{29/71}  & 158 | 34/32/34 
     &  49 | \textbf{0/100} &  59 | \textbf{53/47} & 49 | \textbf{96/\;\,4}  & 47 | \textbf{\;\;0/\;\,4/96} \\ 
     cluster 3 &             &   7 | 57/43 &  20 | \textbf{75/25} & 22 | \textbf{45/\;\,9/45} & 27 | \textbf{63/37} &  48 | \textbf{58/42} &  16 | \textbf{69/31} & 34 | \textbf{62/29/\;\,9} &             &  28 | \textbf{82/18} & 18 | \textbf{39/61}  & 30 | \textbf{37/43/20} \\  
     cluster 4 &             &          &     &  11 | \textbf{27/\;\,0/73} & 21 | \textbf{90/10} &  48 | \textbf{83/17} &  15 | 47/53 & 20 | \textbf{\,0/100/ 0} &             &  17 | \textbf{0/100} & 17 | \textbf{0/100}  & 17 | \textbf{\;0/100/ 0} \\ 
     cluster 5 &             &             &  &            &  20 | \textbf{0/100} &  & 10 | \textbf{90/10} & 17 | \textbf{\;\;0/94/\;\,6} &  &  13 | \textbf{62/38} & &  \\
    \hline
     combi 2 & Case 1 & Case 2 & Case 3 & Case 4 & Case 1 & Case 2 & Case 3 & Case 4 & Case 1 & Case 2 & Case 3 & Case 4 \\\hline
     cluster 1 &  99 | 46/54 & 330 | 55/45 & 172 | \textbf{72/28}  & 334 | 34/27/39 & 120 | \textbf{57/43} & 141 | \textbf{35/65} & 359 | 54/46  & 183 | \textbf{39/27/33} & 218 | 46/64 & 261 | 41/59 & 132 | \textbf{66/34}  & 202 | \textbf{35/17/48} \\ 
     cluster 2 &  66 | \textbf{21/79} &   6 | \textbf{67/13} & 21 | \textbf{86/14}  & 41 | \textbf{37/\;\,7/56} &  60 | \textbf{23/77} & 101 | \textbf{64/36} & 19 | \textbf{0/100}  & 164 | 24/39/37 &  36 | \;\;\textbf{8/92} &  80 | \textbf{59/41} & 45 | \textbf{73/27}  & 78 | \textbf{23/72/\;\,5} \\
     cluster 3 &  45 | 42/58 &             &  20 | \textbf{0/100} &  &  42 | 50/50 &             & 17 | \textbf{0/100} &   &  32 | 41/59 &   & 41 | \textbf{88/12}  & 67 | \textbf{27/24/49} \\
     cluster 4 &  32 | \textbf{81/19} &             & 16 | \textbf{0/100}  &  &             &             &   &  &  19 | \textbf{16/84} &   & 38 | \textbf{11/89}  & 63 | \textbf{11/\;\,8/81} \\
     cluster 5 &  30 | \textbf{73/27} &             &   &  &             &             &   &  &  18 | \textbf{17/83} &   & 24 | \textbf{0/100}  & 23 | \textbf{0/100/\;\,0} \\
    \hline
     combi 3 & Case 1 & Case 2 & Case 3 & Case 4 & Case 1 & Case 2 & Case 3 & Case 4 & Case 1 & Case 2 & Case 3 & Case 4 \\\hline
     cluster 1 & 114 | 43/57 & 264 | 53/47 &  264 | 58/42 & 469 | 34/28/38 & 83 | 43/57 & 154 | \textbf{38/62} & 173 | 48/52  & 260 | 28/39/33 & 159 | 42/58 & 224 | 47/53 & 111 | \textbf{68/32}  & 182 | 24/35/41 \\
     cluster 2 & 107 | 48/52 &  33 | \textbf{67/33} & 21 | \textbf{100/0}  & 47 | \textbf{49/\;\,6/45} & 51 | \textbf{65/35} & 106 | \textbf{54/46} & 34 | \textbf{71/29}  & 86 | \textbf{41/20/40} & 151 | \textbf{28/72} &  50 | \textbf{22/78} &  110 | \textbf{65/35} & 63 | \textbf{38/52/10} \\
     cluster 3 &  50 | 42/58 &             & 11 | \textbf{82/18}  &  & 50 | 40/60 &  24 | \textbf{29/71} & 31 | \textbf{\;\,6/94}  & 30 | \textbf{37/57/\;\,7} &  29 | \textbf{69/31} &  19 | \;\;\textbf{95/5} &  52 | \textbf{\;\,4/96} & 61 | 25/31/44 \\
     cluster 4 &  50 | \textbf{24/76} &             &  10 | \textbf{90/10} &  & 34 | \textbf{56/44} &             &  25 | \textbf{40/60} &  &  23 | \;\;\textbf{91/9}  &  & 36 | \textbf{0/100}  & 24 | \textbf{\;\,8/17/75} \\
     cluster 5 &             &             &   &  & 20 | \textbf{30/70} &             & 23 | \textbf{39/61}  &  &             &  & 22 | \textbf{100/0}  & 12 | \textbf{0/100/\;\,0} \\
    \hline
     combi 4 & Case 1 & Case 2 & Case 3 & Case 4 & Case 1 & Case 2 & Case 3 & Case 4 & Case 1 & Case 2 & Case 3 & Case 4 \\\hline
     cluster 1 & 362 | 48/52 & 391 | 60/40 &  96 | \textbf{83/17} & 473 | 40/19/41 & 96 | \textbf{54/46} & 120 | 45/55 & 206 | 55/45  & 169 | \textbf{30/46/24} & 146 | 36/64 & 160 | \textbf{37/63} &  104 | \textbf{83/17} & 630 | 28/31/41 \\
     cluster 2 &  32 | \textbf{41/59} &  20 | \textbf{50/50} & 59 | 54/46  & 57 | \textbf{32/16/53} & 75 | \textbf{37/63} &  61 | 46/54 & 102 | \textbf{36/64}  & 85 | 34/36/29 &  55 | \textbf{65/35} &  59 | \textbf{56/44} & 52 | \textbf{79/21}  & 14 | \textbf{\,0/100/ 0} \\ 
     cluster 3 &  20 | 50/50 &  10 | \textbf{0/100} & 37 | \textbf{30/70}  & 40 | \textbf{25/50/25} & 45 | 40/60 &  33 | \textbf{36/64} & 47 | \textbf{32/68}  & 69 | \textbf{28/48/25} &  27 | \textbf{0/100} &   9 | \textbf{0/100} & 43 | \textbf{77/23}  & 11 | \textbf{27/73/\;\,0} \\  
     cluster 4 &  18 | 56/44 &   8 | 50/50 &  21 | \textbf{100/0} & 26 | \textbf{46/23/31} &            &  29 | \textbf{38/62} &   & 54 | 26/44/30 &  20 | 50/50 &   9 | \textbf{78/22} & 29 | \textbf{0/100}  &  \\ 
     cluster 5 &  17 | \textbf{29/71} &             &  19 | 53/47 & 20 | \textbf{20/80/\;\;0} &            &             &   & 43 | \textbf{28/16/56} &             &   6 |\textbf{67/33} & 19 | \textbf{0/100}  &  \\
    \hline
     combi 5 & Case 1 & Case 2 & Case 3 & Case 4 & Case 1 & Case 2 & Case 3 & Case 4 & Case 1 & Case 2 & Case 3 & Case 4 \\\hline
     cluster 1 & 470 | 49/51 & 182 | 56/44 & 216 | \textbf{85/15}  & 638 | 36/26/38 & 112 | \textbf{54/46} & 109 | \textbf{66/34} &  141 | 48/52 & 180 | \textbf{41/23/36} & 178 | \textbf{33/67} & 339 | 43/57 & 415 | 61/39  & 367 | \textbf{19/32/48} \\
     cluster 2 &  24 | \textbf{83/17} &  38 | \textbf{50/50} & 13 | \textbf{0/100}  & 11 | \textbf{82/\;\,0/18} &  51 | \textbf{27/73} &  89 | \textbf{31/69} & 56 | \textbf{80/20}  & 106 | 22/41/38 &  80 | \textbf{19/81} &   9 | \textbf{0/100} & 24 | \textbf{0/100}  & 16 | \textbf{44/56/\;\,0} \\ 
     cluster 3 &             &             & 11 | \textbf{0/100} & 7 | \textbf{100/ 0/ 0} &  42 | 38/62 &  18 | \textbf{0/100} & 47 | \textbf{40/60}  & 39 | \textbf{31/\;\,8/62} &  10 | \textbf{30/70} &   8 | \textbf{75/25} &   & 14 | \textbf{14/21/64} \\  
     cluster 4 &             &             & 10 | \textbf{0/100}  &  &  13 | \textbf{100/0} &  15 | 47/53 &   & 22 | 27/36/36 &   7 | \textbf{14/86} &   7 | \textbf{100/0} &   &  \\ 
     cluster 5 &             &            &   10 | \textbf{90/10} &  &  11 | \textbf{100/0} &  14 | \textbf{64/36} &   &  &             &   6 | \textbf{100/0} &   &  \\
     \hline
     combi 6 & Case 1 & Case 2 & Case 3 & Case 4 & Case 1 & Case 2 & Case 3 & Case 4 & Case 1 & Case 2 & Case 3 & Case 4 \\\hline
     cluster 1 & 204 | \textbf{38/62} & 174 | 61/39 & 106 | \textbf{73/27}  & 333 | \textbf{34/21/45} & 142 | 49/51 & 137 | 47/53 &  159 | 47/53 & 239 | 31/34/35 & 211 | \textbf{27/73} & 179 | \textbf{35/65} & 318 | 57/43  & 377 | \textbf{19/36/46} \\
     cluster 2 &  84 | 48/52 &  35 | \textbf{66/34} & 52 | \textbf{71/29}  & 23 | \textbf{43/\;\;9/48} &  59 | 44/56 &  79 | \textbf{56/44} & 32 | \textbf{81/19}  & 57 | \textbf{51/23/26} &  15 | \;\;\textbf{7/93} &  24 | \textbf{38/62} & 13 | \textbf{0/100}  & 24 | \textbf{33/67/ \;0} \\ 
     cluster 3 &             &  18 | \textbf{33/67} & 41 | \textbf{78/22}  &  &  41 | \textbf{71/29} &  26 | \textbf{27/73} & 32 | \textbf{62/38}  & 31 | \textbf{23/52/26} &  11 | \textbf{\;\;9/91} &             & 8 | \textbf{0/100}  & 16 | \textbf{31/19/50} \\  
     cluster 4 &             &            & 32 | \textbf{22/78}   &  &             &             & 31 | 42/58  & 20 | \textbf{50/\;\;5/45} &   9 | 33/67 &             &   &  \\ 
     cluster 5 &             &             &   &  &             &             & 27 | \textbf{0/100}  & 16 | \textbf{25/\;\;6/69} &   6 | 50/50 &             &  &  \\
    \hline
     combi 7 & Case 1 & Case 2 & Case 3 & Case 4 & Case 1 & Case 2 & Case 3 & Case 4 & Case 1 & Case 2 & Case 3 & Case 4 \\\hline
     cluster 1 & 266 | 44/56 & 224 | \textbf{68/32} &  240 | \textbf{72/28} & 415 | 37/19/43 & 84 | 48/52 & 101 | 48/52 & 88 | 42/58  & 124 | 34/35/31 & 362 | \textbf{32/68} & 185 | \textbf{38/62} &  250 | 63/37 & 477 | 21/32/46 \\
     cluster 2 &  29 | \textbf{0/100} &  70 | 51/49 & 41 | \textbf{78/22}  & 40 | \textbf{\;\,0/20/80} & 74 | \textbf{62/38} &  79 | \textbf{62/38} & 82 | 46/54  & 87 | \textbf{39/13/48} &  11 | \textbf{100/0} &  30 | \textbf{37/63} & 24 | \textbf{42/58}  & 15 | \textbf{33/67/\;\,0} \\ 
     cluster 3 &  16 | \textbf{100/0} &  25 | 52/48 &  18 | 56/44 & 35 | \textbf{74/26/\;\,0} & 60 | 50/50 &  40 | \textbf{17/83} & 40 | \textbf{78/22}  & 61 | 34/28/38 &   6 | \textbf{0/100} &             & 18 | \textbf{0/100}  &  \\  
     cluster 4 &             &  13 | \textbf{\;\;8/92} & 17 | \textbf{0/100}  & 28 | 36/29/36 &            &             & 35 | \textbf{26/74}  & 37 | \textbf{35/19/46} &             &             & 12 | \textbf{0/100}  &  \\ 
     cluster 5 &             &             &  \textbf{10 | 0/100} &  &            &             &   & 32 | 22/38/41 &             &             & 11 | \textbf{64/36}  &  \\
    \hline
     combi 8 & Case 1 & Case 2 & Case 3 & Case 4 & Case 1 & Case 2 & Case 3 & Case 4 & Case 1 & Case 2 & Case 3 & Case 4 \\\hline
     cluster 1 & 435 | 49/51 & 398 | 60/40 & 273 | \textbf{69/31}  & 149 | \textbf{32/17/50} & 94 | 46/54 & 368 | 45/55 & 184 | 51/49  & 470 | 28/33/39 & 351 | 36/64 & 136 | 40/60 &  390 | 59/41 & 372 | \textbf{20/32/48} \\
     cluster 2 &  19 | 47/53 &  22 | \textbf{45/55} & 40 | \textbf{90/10}  & 134 | \textbf{49/\;\,6/45} & 48 | 48/52 &  25 | \textbf{32/68} & 37 | \textbf{97/\;\,3}  & 32 | \textbf{25/59/16} &  18 | 50/50 &  64 | 50/50 & 17 | 59/41  & 35 | \textbf{14/37/49} \\ 
     cluster 3 &             &   8 | 50/50 & 38 | \textbf{39/61}  & 63 | \textbf{35/14/51} & 41 | \textbf{15/85} &             & 26 | \textbf{23/77}  &  &             &  14 | \textbf{21/79} &   & 35 | \textbf{37/46/17} \\  
     cluster 4 &             &   8 | \textbf{0/100} &  17 | \textbf{0/100} &  & 38 | \textbf{34/66} &  &  26 | 38/62            &  &             &             &   & 13 | \textbf{15/23/62} \\ 
     cluster 5 &             &             &   &  &            &             & 18 | 50/50  &  &             &             &   &  \\
    \hline
     combi 9 & Case 1 & Case 2 & Case 3 & Case 4 & Case 1 & Case 2 & Case 3 & Case 4 & Case 1 & Case 2 & Case 3 & Case 4 \\\hline
     cluster 1 & 135 | \textbf{39/61}  & 224 | 54/46 & 153 | \textbf{73/27}  & 168 | \textbf{33/17/51} & 131 | 39/61 & 98 | \textbf{55/45} & 116 | \textbf{72/28}  & 186 | 33/32/35 & 183 | 34/66 & 168 | \textbf{38/62} & 261 | 52/48  & 382 | 26/34/41 \\
     cluster 2 & 122 | \textbf{37/63}  &  30 | \textbf{73/27} &  68 | \textbf{75/25} & 101 | \textbf{30/19/51} &  60 | \textbf{33/67} & 65 | \textbf{35/65} & 75 | 43/57  & 116 | 26/40/34 &  38 | \textbf{55/45} &  57 | 44/56 & 42 | \textbf{36/64}  & 54 | \textbf{37/35/28} \\
     cluster 3 & 26 | \textbf{35/65}   &             &   & 50 | 30/26/44 &  22 | 45/55 & 28 | \textbf{54/46} & 24 | 54/46  & 44 | \textbf{41/27/32} &  23 | 39/61 &  &   &  \\
     cluster 4 &              &             &   & 20 | \textbf{35/ \;5/60} &  18 | \textbf{17/83} & 25 | \textbf{32/68} & 23 | \textbf{0/100}  & 27 | \textbf{15/26/59} &             &  &   &  \\
     cluster 5 &              &             &   & 14 | \textbf{57/29/14} &             & 23 | \textbf{30/70} & 10 | \textbf{30/70} &  &             &  & & \\
    \hline
     combi 10 & Case 1 & Case 2 & Case 3 & Case 4 & Case 1 & Case 2 & Case 3 & Case 4 & Case 1 & Case 2 & Case 3 & Case 4 \\\hline
     cluster 1 & 146 | 51/49 & 248 | \textbf{65/35} & 287 | \textbf{68/32}  & 142 | \textbf{45/12/43} & 273 | 42/58 & 76 | 47/53 & 106 | \textbf{72/28}  & 344 | \textbf{28/28/44} & 319 | \textbf{34/66} & 138 | 40/60 & 391 | 56/44  & 486 | 25/33/42 \\
     cluster 2 &  89 | 48/52 &  33 | 58/42 & 60 | \textbf{40/60}  & 123 | 38/24/38 &  45 | 42/58 & 47 | \textbf{57/43} & 68 | 49/51  & 56 | \textbf{21/54/25} &  20 | 50/50 &  31 | 42/58 & 15 | \textbf{87/13}  & 25 | \textbf{36/52/12} \\ 
     cluster 3 &  44 | 45/55 &  29 | \textbf{34/66} &  30 | \textbf{33/67} & 78 | 31/38/31 &  31 | 48/52 & 39 | \textbf{36/64} & 35 | \textbf{34/66}  & 50 | \textbf{24/52/24} &  13 | \textbf{100/0} &  30 | 50/50 &  10 | 60/40 &  \\  
     cluster 4 &  22 | 45/55 &  15 | \textbf{13/87} & 21 | \textbf{0/100}  & 33 | 33/30/36 &  24 | 50/50 & 31 | \textbf{55/45} & 15 | \textbf{40/60}  & 40 | \textbf{28/57/15} &  11 | \textbf{0/100} &  15 | 53/47 & 8 | \textbf{0/100}  &  \\ 
     cluster 5 &             &             & 18 | \textbf{83/17}  &  &  20 | \textbf{60/40} & 25 | \textbf{32/68} & 15 | \textbf{40/60}  & 23 | \textbf{22/48/30} &   7 | \textbf{100/0} &  13 | \textbf{15/85} & 6 | 50/50  &  \\
    \hline
\end{longtable}
\tablefoot{ Given are number of points per cluster and the percentage of \cyc\, over \meth\, (Case 1), \cyc\, over \prop\, (Case 2), and \cyc\, over \meth\, over \prop\, (Case 4). Given in brackets is the total number of data points per dataset. Ratios that differ from the initial input ratio by $\geq10\%$ are given in boldface. The clusters are ordered by size and enumerated from 1 to 5. The corresponding colours in Fig.~\ref{fig:ClusteringB68} and Figs.~\ref{fig:ClusteringL1521E}-~\ref{fig:ClusteringL1544} are: 1=blue, 2=red, 3=cyan, 4=yellow, 5=purple.}
}
\end{landscape}

\FloatBarrier

\section{Spectra at dust peak} \label{app:dustpeakspectra}
\begin{figure*}[h!]
    \centering
    \includegraphics[width=0.8\textwidth]{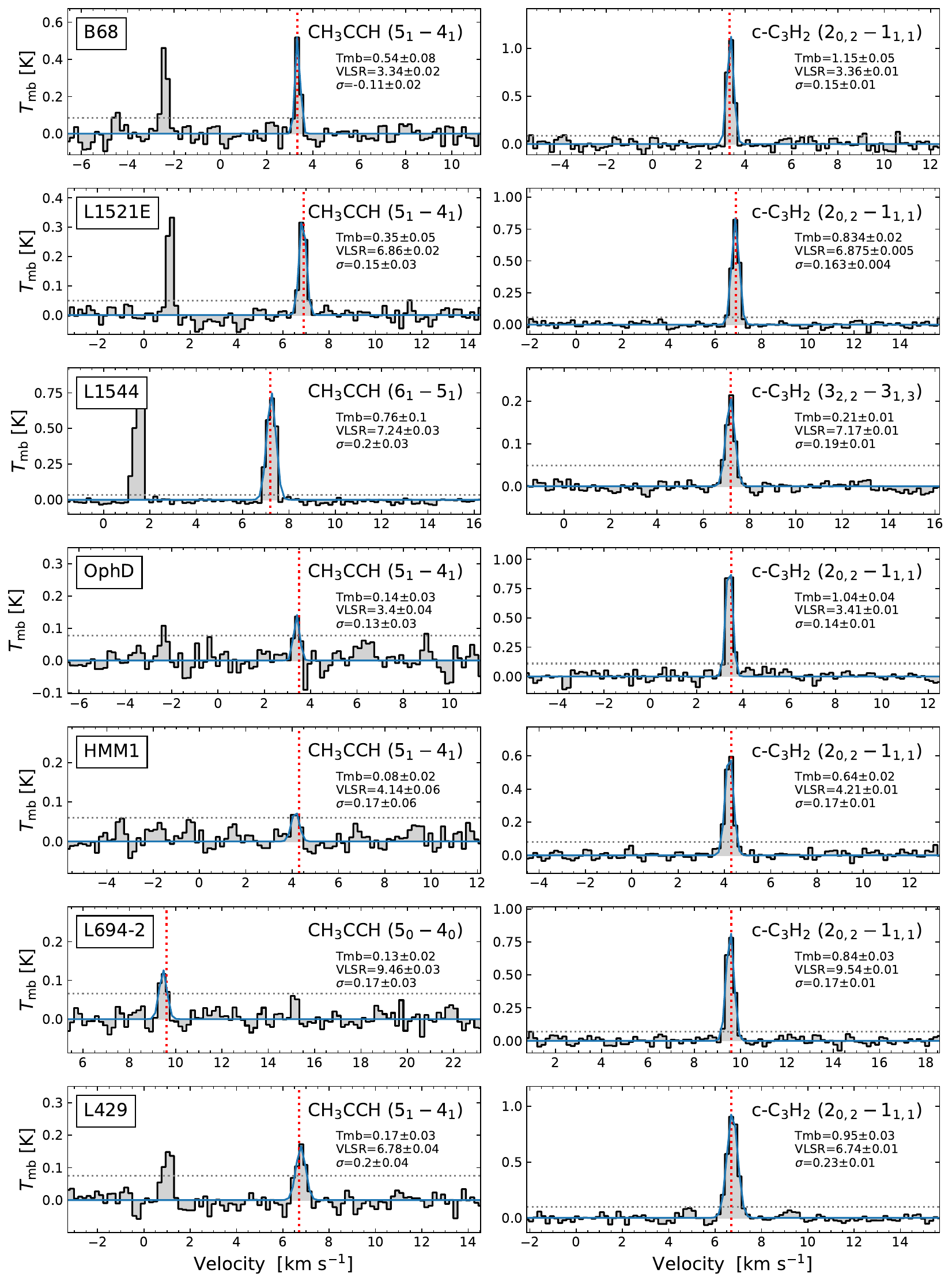}
    \caption[alt={Spectra of \prop\, and \cyc\, at the dust peak of the seven cores.}]{Spectra of \prop\, (\textit{left}) and \cyc\, (\textit{right}) at the dust peak of each core (black) extracted within a circular aperture of radius 8" and the corresponding Gaussian fit (cyan). The 3$\sigma$ level is indicated by the grey dotted line. The systemic velocity with respect to the line chosen for analysis is shown by the red dotted line. The Gaussian fit parameters are annotated for each line.}
    \label{fig:DustpeakSpectra}
\end{figure*}

The spectra of \cyc\, and \prop\, observed towards the dust peaks of B68, L1521E, L1544, OphD, HMM1, L694-2, and L429 are shown in Fig.~\ref{fig:DustpeakSpectra}.
The observed data cubes are convolved with the \textit{Herschel} beam size (40"), then a circular aperture with radius 8" is used to extract the spectra.

\end{appendix}

\end{document}